\documentclass[pra,10pt,twocolumn,amsmath,amssymb,superscriptaddress,noshowpacs]{revtex4-1}

\usepackage{ifthen}

\usepackage{amsmath}
\usepackage{amssymb}
\usepackage[T1]{fontenc}
\usepackage{epsfig}
\usepackage{multirow}

\usepackage[colorlinks=true,linkcolor=blue,citecolor=blue,urlcolor=blue]{hyperref}
\usepackage{geometry}
\usepackage{times, newtxmath}

\usepackage{xcolor} % Only for "\comment"

\usepackage{ucs}
\usepackage{bbm}

\newcommand{\ket}[1]{|#1\rangle}
\newcommand{\ketbra}[1]{| #1\rangle \langle #1|}

\newcommand{\be}{\begin{equation}}
\newcommand{\ee}{\end{equation}}
\newcommand{\eea}{\end{eqnarray}}
\newcommand{\bea}{\begin{eqnarray}}

\newcommand{\va}[1]{\ensuremath{(\Delta#1)^2}}

\newcommand{\ex}[1]{\ensuremath{\left\langle{#1}\right\rangle}}
\newcommand{\exs}[1]{\ensuremath{\langle{#1}\rangle}}

\newcommand{\eins}{\mathbbm{1}}
\newcommand{\qed}{\ensuremath{\hfill \blacksquare}}

\newcommand{\kommentar}[1]{}

\newcommand{\forget}[1]{}

\newcommand{\APPENDIX}[1]{Appendix~\ref{#1}}
\newcommand{\EQ}[1]{Eq.~\eqref{#1}}
\newcommand{\EQS}[1]{Eqs.~\eqref{#1}}
\newcommand{\EQL}[1]{Equation~\eqref{#1}}

\newcommand{\SEC}[1]{Sec.~\ref{#1}}
\newcommand{\FIG}[1]{Fig.~\ref{#1}}
\newcommand{\FIGL}[1]{Figure~\ref{#1}}
\newcommand{\REF}[1]{Ref.~\cite{#1}}

\newcommand{\TABLE}[1]{Table~\ref{#1}}
\newcommand{\comment}[1]{}

\setcitestyle{citesep={,\kern-0.24em}}

\begin{document}

\title{Optimal witnessing of the quantum Fisher information with few measurements}

\author{Iagoba Apellaniz}
\affiliation{Department of Theoretical Physics, University of the Basque Country
UPV/EHU, %P.O. Box 644,
E-48080 Bilbao, Spain}
\author{Matthias Kleinmann}
\affiliation{Department of Theoretical Physics, University of the Basque Country
UPV/EHU, %P.O. Box 644,
E-48080 Bilbao, Spain}
\author{Otfried G\"uhne}
\affiliation{Naturwissenschaftlich-Technische Fakult\"at, Universit\"at Siegen, Walter-Flex-Str. 3, 57068 Siegen, Germany}
\author{G\'eza T\'oth}
\email{toth@alumni.nd.edu}
\homepage{http://www.gtoth.eu}
\affiliation{Department of Theoretical Physics, University of the Basque Country
UPV/EHU, %P.O. Box 644,
E-48080 Bilbao, Spain}
\affiliation{IKERBASQUE, Basque Foundation for Science, E-48013 Bilbao, Spain}
\affiliation{Wigner Research Centre for Physics, Hungarian Academy of Sciences,
%P.O. Box 49,
H-1525 Budapest, Hungary}

\makeatletter
\def\Dated@name{Published: }
\makeatother

%\pacs{03.67.Mn, 03.65.Ud,42.50.St}


% 03.65.Ud Entanglement and quantum nonlocality
% 03.67.Mn Entanglement measures, witnesses, and other characterizations
% 42.50.St Nonclassical interferometry, subwavelength lithography
% 03.67.Bg Entanglement production and manipulation (for entanglement in Bose-Einstein condensates, see 03.75.Gg)
% 03.75.Gg	Entanglement and decoherence in Bose-Einstein condensates
% 03.75.Dg	Atom and neutron interferometry

\begin{abstract}

We show how to verify the metrological usefulness of quantum states based on the expectation values of an arbitrarily chosen set of observables. In particular, we estimate the quantum Fisher information as a figure of merit of metrological usefulness. Our approach gives a tight lower bound on the quantum Fisher information for the given incomplete information. We apply our method to the results of various multiparticle quantum states prepared in experiments with photons and trapped ions, as well as to spin-squeezed states and Dicke states realized in cold gases. Our approach can be used for detecting and quantifying metrologically useful entanglement in very large systems, based on a few operator expectation values. We also gain new insights into the difference between metrological useful multipartite entanglement and entanglement in general.

\vspace{2ex}
\noindent
DOI: \href{https://doi.org/10.1103/PhysRevA.95.032330}{10.1103/PhysRevA.95.032330}

%So far, there is not a similar general scheme
%for entanglement detection in the literature.

\end{abstract}
\date{28 March 2017}

\maketitle

\section{Introduction}\label{sec:Introduction}

Entanglement lies at the heart of many problems in
quantum mechanics and has attracted increasing attention
in recent years.
There are now efficient methods to detect it with a moderate experimental effort \cite{Horodecki2009Quantum,Guhne2009Entanglement}.
However, in spite of intensive research,
many of the intriguing properties of entanglement are not fully understood.
One such puzzling fact is that,
while entanglement is a sought after resource,
not all entangled states are useful
for some particular quantum information processing task.
For instance, it has been realized recently that entanglement is needed
in very general metrological tasks
to achieve a high precision \cite{Pezze2009Entanglement}.
Remarkably, this is
true
even in the case of millions of particles,
which is especially important
for characterizing the entanglement properties of cold atomic ensembles
\cite{Louchet-Chauvet2010Entanglement-assisted,Appel2009Mesoscopic,Riedel2010Atom-chip-based,Gross2010Nonlinear,Lucke2011Twin,Strobel2014Fisher}.
However,
there are highly
entangled pure states that are useless for metrology \cite{Hyllus2010Not}.

In the light of the these results, besides verifying that\break a quantum state is entangled, we should also show that it
is useful for metrology.
This is possible if we know the quantum Fisher information
$\mathcal{F}_{\rm Q}[\varrho,J_l]$ for the state.  Here $\varrho$ is a density matrix of an ensemble of $N$ two-level systems\break (i.e., qubits),  $J_l=\tfrac{1}{2}\sum_n \sigma_l^{(n)}$ for $l=x,y,z$ are the angular momentum components and $\sigma_l^{(n)}$ are the Pauli spin matrices acting on qubit $n.$

The quantum Fisher information
is a central quantity of quantum metrology. It is connected to the task of estimating the phase $\theta$
for the unitary dynamics of a linear interferometer $U=$\break$\exp(-iJ_l\theta),$
assuming that we start from $\varrho$ as the initial state.\break
It
provides a tight bound for
the precision of phase estimation
as  \cite{Giovannetti2004Quantum-Enhanced,*Paris2009QUANTUM,*Demkowicz-Dobrzanski2014Quantum,*Pezze2014Quantum,Helstrom1976Quantum,*Holevo1982Probabilistic,*Braunstein1994Statistical,
*Petz2008Quantum,*Braunstein1996Generalized}
\be\label{eq:cr}
 %$
\va{\theta}\geqslant {1}/{\mathcal{F}_{\rm Q}[\varrho,J_l]}.
%\frac{1}{\mathcal{F}_{\rm Q}[\varrho,J_l]}.
\ee
%$
It has been shown that if
$\mathcal{F}_{\rm Q}[\varrho,J_l]$ is larger than the value achieved by product states \cite{Pezze2009Entanglement}, then the state $\varrho$ is entangled.
Higher values of the quantum Fisher information indicate\break even multipartite entanglement \cite{{Hyllus2012Fisher,*Toth2012Multipartite}}; this fact has been used to analyze the results of several experiments
\cite{Lucke2011Twin,Krischek2011Useful,Strobel2014Fisher}.

In this paper, we suggest estimating the quantum Fisher\break information based on a few measurements \cite{*[{In another context, the measurement of the quantum Fisher information has recently been considered in systems in thermal equilibrium. }] [{;}]  Hauke2016Measuring,*Shitara2016Determining}. Our method can be called ``witnessing the quantum Fisher information'' since our estimation scheme is based on measuring operator expectation values similarly to how entanglement witnesses work \cite{Horodecki2009Quantum,Guhne2009Entanglement}.
Our findings are expected
to simplify the experimental determination of metrological sensitivity since the\break proposed set of a few
measurements is much easier to carry out than the direct determination of the metrological sensitivity,\break which has been applied in several experiments
 \cite{Lucke2011Twin,Strobel2014Fisher,Pezze2016Witnessing,Frowis2016Detecting}.
The archetypical criterion in this regard is \cite{Pezze2009Entanglement}
\be\label{eq:archetypical}
\mathcal{F}_{\rm Q}[\varrho,J_y]\geqslant \frac{\exs{J_z}^2}{\va{J_x}},
\ee
which is expected to
work best for states that are almost\break completely polarized in the $z$ direction and spin-squeezed in the $x$ direction.
Apart from spin-squeezed states, there are conditions similar to \EQ{eq:archetypical} for symmetric states close to Dicke states \cite{Dicke1954Coherence,Zhang2014Quantum,Frowis2014Tighter,Apellaniz2015Verifying}
and for two-mode squeezed states
\cite{Oudot2015Two-mode}.

After finding criteria for various systems, it is
crucial to develop a general method that provides an {\it optimal} lower bound
on the quantum Fisher information in a wide class of cases,
especially for the states most relevant for experiments
such as spin-squeezed states
\cite{Kitagawa1993Squeezed,*Wineland1994Squeezed},
Greenberger-Horne-Zeilinger (GHZ) states \cite{Greenberger1990Bells},
and symmetric Dicke states \cite{Dicke1954Coherence}. It seems that such a method
would
involve a numerical minimization over all density matrix elements constrained
for some operator expectation values, which would be impossible
except in very small systems.

In this paper, we demonstrate that tight lower bounds on the quantum Fisher information can still be computed efficiently. Remarkably, our method works for thousands of particles.
We show how to obtain a bound on the quantum Fisher information from
fidelity measurements for GHZ states \cite{Bouwmeester1999Observation,Pan2000Experimental,Zhao2003Experimental,Lu2007Experimental,Gao2010Experimental,Leibfried2004Toward,Sackett2000Experimental,Monz201114-Qubit} and for symmetric Dicke states \cite{Kiesel2007Experimental,Wieczorek2009Experimental,Prevedel2009Experimental,Krischek2011Useful,Chiuri2012Experimental,Schindler2013Quantum}. We also discuss  how to obtain such bounds based on collective measurements for spin-squeezed states of thousands of atoms
\cite{Gross2012Spin,*Ma2011Quantum,*Hald1999Spin,*Echaniz2005Conditions,*Sewell2012Magnetic,Riedel2010Atom-chip-based,Gross2010Nonlinear}
and for symmetric Dicke states prepared recently in cold gases
\cite{Lucke2014Detecting,Lucke2011Twin,Hamley2012Spin-nematic,Luo2017Deterministic}.
We stress that the method is very general, and needs only the expectation values of a set of operators chosen by the experimenter.
Then it provides a tight lower bound on the quantum Fisher information.

Due to the relation between the quantum Fisher information and entanglement mentioned above, our method can also be
used for entanglement detection and quantification based on an arbitrary set of operator expectation values in very large systems.
So far, methods that can be used for large systems, such as spin-squeezing inequalities \cite{Sorensen2001Many-particle,Korbicz2005Spin,Toth2007Optimal}, work only for a specific set of observables.
In addition, methods that can quantify entanglement based on the expectation values of an arbitrary set of observables,
such as semidefinite programming \cite{Doherty2002Distinguishing,Wunderlich2009Quantitative,Toth2015Evaluating},
work only for small systems.

The paper is organized as follows. In \SEC{sec:Estimation},
we show how to bound the quantum Fisher information based
on the knowledge of some operator expectation values.
In \SEC{sec:examples}, we test our method on theoretical examples
in small systems. In \SEC{sec:calcexp}, we present calculations for experimental data.
Finally, in \SEC{sec:scaling}, we discuss how the quantum Fisher information
is expected to scale with the particle number in the limit of large particle numbers.

\section{Estimation of the quantum Fisher information}
\label{sec:Estimation}

In this section, first we review some important properties of the quantum Fisher information. Then we present our method for estimating it based on a few measurements.

\subsection{Entanglement quantification with the quantum Fisher information}

In \SEC{sec:Introduction}, we mentioned briefly, how quantum Fisher information connects quantum metrology and entanglement theory. In more detail,
the bounds on the quantum Fisher information make it possible
to detect metrologically useful entanglement.
It has been shown that if
\be\label{eq:FQent}
    \mathcal{F}_{\rm Q}[\varrho,J_l]> (k-1)N,
\ee
where $k$ is an integer, then
the state has a better metrological performance than any state with at most $(k-1)$--particle entanglement, hence it possesses
 at least $k$-particle metrologically useful entanglement
\cite{Pezze2009Entanglement,Hyllus2012Fisher,*Toth2012Multipartite}.
We can immediately see that a perfect $N$-particle GHZ
state possesses metrologically useful $N$-particle entanglement.
Based on the ideas above, it is possible to use the quantum Fisher information for
entanglement detection \cite{Lucke2011Twin,Krischek2011Useful,Strobel2014Fisher}.

Let us analyze the condition, \EQ{eq:FQent}, further.
A simple calculation shows that for a tensor product of $(k-1)$--particle GHZ states the two sides of \EQ{eq:FQent}
are equal.
% where we have to assume also that $(k-1)$ is a divisor of $N.$
Hence, a state is detected by \EQ{eq:FQent} if it performs better than a state in which all particles
are in GHZ states of $(k-1)$ particles. For instance, if in an experiment with $10\;000$ particles we detect five-particle metrologically useful entanglement,
then the state is better metrologically than a tensor product of $2500$ four-particle GHZ states.
Based on this example, it is easy to see that the requirements for metrologically useful $k$-particle entanglement are much stricter
than for general $k$-particle entanglement.

\subsection{Estimation of a general function of $\varrho$}

First, we review a method that can be used to find a lower bound on
a convex function $g(\varrho)$ based on only a single operator expectation value $w=\exs{W}_\varrho={\rm Tr}(W\varrho).$  Theory tells us that a
tight lower bound can be obtained as \cite{Rockafellar2015Convex,Guhne2007Estimating,Eisert2007Quantitative}
\be\label{eq:LowerBoundOnF}
g(\varrho)\geqslant \mathcal{B}(w):=\sup_{r}\left[ r w-\hat{g}\left( r W\right)\right],
\ee
%where $w={\rm Tr}(\varrho W),$
where $\hat{g}$ is the Legendre transform, in this context defined as
\be\label{eq:LegendreTransform}
\hat{g}(W)=\sup_{\varrho}[\langle W\rangle_{\varrho}-g(\varrho)].
\ee
\EQL{eq:LowerBoundOnF} has been applied to entanglement measures \cite{Guhne2007Estimating,Eisert2007Quantitative}. Since those are defined as convex roofs
over all possible decompositions of the density matrix,
it is sufficient to carry out the optimization in \EQ{eq:LowerBoundOnF}
for pure states only. However, still an optimization over a general pure state, i.e.,
over many variables, has to be carried out, which is practical only for small systems.

Based on this method,
we would like to estimate the quantum
Fisher information, which is strongly connected to entanglement,
while it also has a clear physical meaning in
metrological applications.
As the first step, we note that
$\mathcal{F}_{\rm Q}[\varrho,J_l]$ can be obtained as a closed formula with $\varrho$ and $J_l$
\cite{{Helstrom1976Quantum,*Holevo1982Probabilistic,*Braunstein1994Statistical,
*Petz2008Quantum,*Braunstein1996Generalized}},
however, this is a highly nonlinear expression which would make the computation of the Legendre transform
very demanding.
A key point in our approach is using a very recent finding showing that $\mathcal{F}_{\rm Q}[\varrho,J_l]$
is the convex roof of $4\va{J_y}$
\cite{Toth2013Extremal,*Yu2013Quantum},
and hence
the optimization may be carried out only for pure states.
With this, however, we are still facing an optimization problem
that cannot be solved numerically for system sizes relevant
for quantum metrology.

We now arrive at our first main result.
We show that, for the quantum Fisher information, \EQ{eq:LegendreTransform}
can be rewritten as an optimization over a {\it single} real parameter.

{\it Observation 1.} The quantum Fisher information can be estimated using the Legendre transform
\be\label{eq:LegendreTransformFQ_H}
\hat{\mathcal{F}}_{\rm Q}(W)=\sup_{\mu} \left\{\lambda_{\max}\left[W-4(J_l-\mu)^2\right]\right\},
\ee
where $\lambda_{\rm max}(A)$ denotes the maximal eigenvalue of $A.$

{\it Proof.}
Based on the previous discussion, we can rewrite the right-hand side of \EQ{eq:LegendreTransform} for our case as
\be
\label{eq:LegendreTransform2_ExpSquare}
\hat{\mathcal{F}}_{\rm Q}(W)=\sup_{\Psi}[\langle W-4J_l^2\rangle_{\Psi}+4\ex{J_l}^2_\Psi].
\ee
\EQL{eq:LegendreTransform2_ExpSquare} is quadratic in operator expectation values. It can be rewritten as an optimization linear in operator expectation values as
\be
\hat{\mathcal{F}}_{\rm Q}(W)=\sup_{\Psi,\mu}[\langle W-4J_l^2\rangle_{\Psi}+8\mu \ex{J_l}_\Psi-4\mu^2],
\ee
which can be reformulated as
\EQ{eq:LegendreTransformFQ_H}.
 At the extremum, the derivative with respect to $\mu$ must be 0, hence
at the optimum $\mu=\ex{J_l}_\Psi.$
This also means that we have to test $\mu$ values in the interval $-N/2 \leqslant \mu\leqslant N/2$ only.
$\qed$

In this paper, we use \EQ{eq:LegendreTransformFQ_H} to calculate the Legendre transform
\footnote{An alternative definition is $\hat{\mathcal{F}}_{\rm Q}(W)=\sup_{\nu} \exs{W}_{\psi_\nu}-4\va{J_l}_{\psi_\nu},$ where \unexpanded{$\ket{\psi_\nu}$} is the eigenstate with the maximal eigenvalue of the operator $W-4J_l^2-\nu J_l.$ In certain cases, this form can be calculated numerically more easily than \EQ{eq:LegendreTransformFQ_H}.}.
The full optimization problem to be solved consists of \EQ{eq:LegendreTransformFQ_H} and \EQ{eq:LowerBoundOnF} with the substitutions $g(\varrho)={\mathcal{F}}_{\rm Q}[\varrho,J_l]$ and $\hat{g}(W)=\hat{\mathcal{F}}_{\rm Q}(W).$

We want to stress the generality of our findings
beyond the linear interferometers covered in this article.
For nonlinear interferometers \cite{Luis2004Nonlinear,Boixo2007Generalized,Choi2008Bose-Einstein,Roy2008Exponentially,Napolitano2011Interaction-based,Hall2012Does}, the phase $\theta$ must be estimated in a unitary dynamics $U=\exp(-iA\theta),$ where $A$ is not a sum of single spin operators and, hence,
is different from the angular momentum components.
Using  Observation 1, we can obtain
lower bounds for the corresponding quantum Fisher information
$\mathcal{F}_{\rm Q}[\varrho,A]$ if we replace
 $J_l$ with $A$ in \EQ{eq:LegendreTransformFQ_H}.

\subsection{Measuring several observables}

We now consider the estimation of the quantum Fisher information
based on several expectation values.
We can generalize the method described by \EQS{eq:LowerBoundOnF} and \eqref{eq:LegendreTransform} for measuring several observables $W_k$ as
\cite{Guhne2007Estimating}
\be\label{eq:multir}
\mathcal{F}_{\rm Q}[\varrho,J_y]\geqslant \sup_{r_1,r_2,...,r_K}\left[\sum_{k=1}^K r_k w_k-\hat{\mathcal{F}}_{\rm Q}\left(\sum_{k=1}^K r_k W_k\right)\right],
\ee
where $w_k=\exs{W_k}_{\varrho}.$
As we can see, we now have several\break parameters $r_k.$
Combining \EQ{eq:multir}
with the Legendre transform \eqref{eq:LegendreTransformFQ_H},
we arrive at the formula
\be\label{eq:FQsupM}
\mathcal{F}_{\rm Q}[\varrho,J_l]\geqslant \sup_{\{r_k\}}\left[ \sum_k r_k w_k-\sup_{\mu} \lambda_{\max}\left(M\right)\right],
\ee
where
\be
M=\sum_k r_k W_k-4(J_l-\mu)^2.
\ee
Since $\hat{\mathcal{F}}_{\rm Q}(\sum r_kW_k)$ is a convex function in $r_k$, in \EQ{eq:FQsupM} the quantity to be maximized
in $r_k$ is concave \cite{Rockafellar2015Convex}. Thus, we can easily find the maximum with the gradient method.
If we do not find the optimal $r_k$, then we underestimate
the real bound. Hence, we will still have a valid lower bound.
This does not hold for the optimization over $\mu.$
The function to be optimized is not a convex function of
$\mu,$ and not finding the optimal $\mu$ leads to overestimating the bound.
Thus, great care must be taken
when optimizing over $\mu.$

\section{Examples}

\label{sec:examples}

In this section, we show how to use our method to estimate the quantum Fisher information based on
fidelity measurements, as well as collective measurements.

\subsection{Exploiting symmetries}
\label{sec:symmetries}

When making calculations for quantum systems with
an increasing number of qubits, we soon run into difficulties
when computing
the largest eigenvalue of \EQ{eq:LegendreTransformFQ_H}.
The reason is that for $N$ qubits, we need
to handle $2^N\times 2^N$ matrices, hence we are limited to systems
of $10$--$15$ qubits.

We can obtain bounds for much larger particle numbers,
if we restrict ourselves to the symmetric subspace
\cite{Sorensen2001Entanglement}.
This approach can give optimal bounds for  many systems, such as Bose-Einstein condensates of two-state atoms, which
are in a symmetric multiparticle state.
The bound computed for the symmetric subspace might not be correct for general states.

Finally, it is important to note that if the operator $W$ is permutationally invariant and the eigenstate with the maximal eigenvalue of the matrix in \EQ{eq:LegendreTransformFQ_H} is nondegenerate,
then the two bounds coincide,  as shown in \APPENDIX{sec:symm_subspace_appendix}.

\subsection{Fidelity measurements}
\label{sec:Fidelity_measurements}

Let us examine the case where $W$ is a projector onto
a pure quantum state. First, we consider GHZ states
\cite{Greenberger1990Bells}.
We choose $W=\ketbra{\rm GHZ},$ hence  $\exs{W}$ is equal to $F_{\rm GHZ},$
the fidelity with respect to the GHZ state.
 Based on knowing  $F_{\rm GHZ},$ we would like to estimate  $\mathcal{F}_{\rm Q}[\varrho,J_z]$.

 \comment{They have already been realized experimentally many times
\cite{Bouwmeester1999Observation,Pan2000Experimental,Zhao2003Experimental,Lu2007Experimental,Gao2010Experimental,Leibfried2004Toward,Sackett2000Experimental,Monz201114-Qubit}.
}

\begin{figure}

\centerline{
\epsfxsize1.7in \epsffile{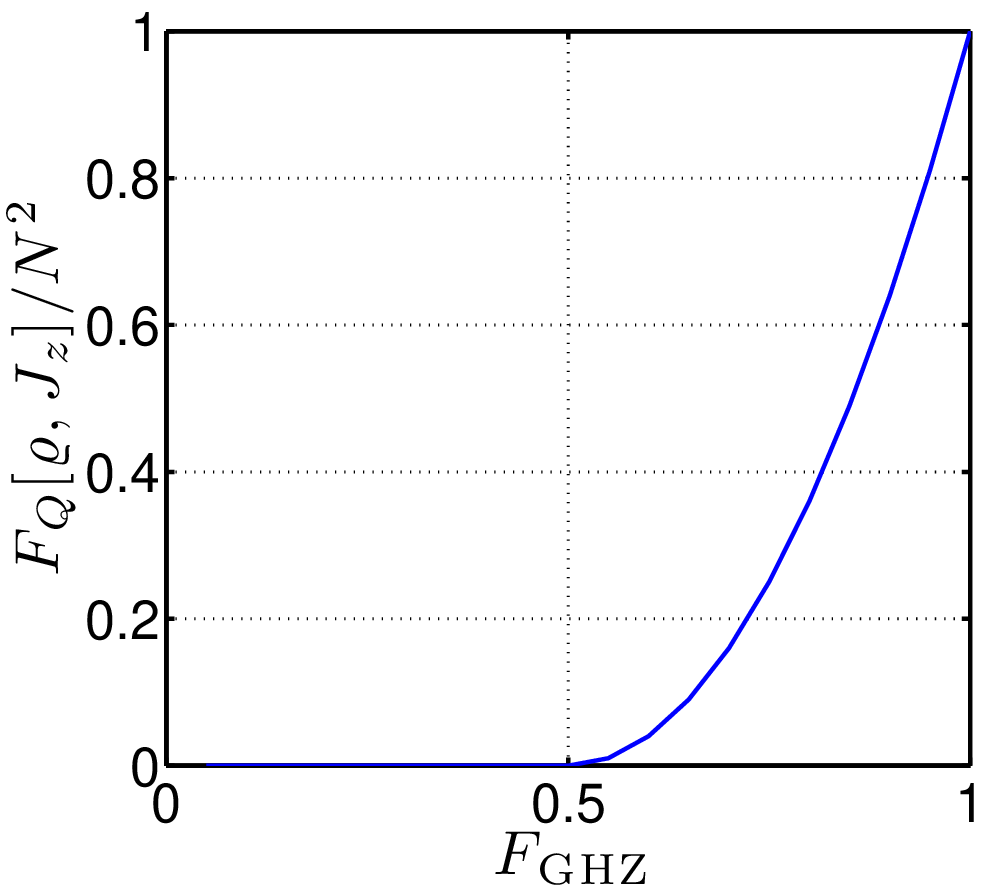}
\epsfxsize1.7in \epsffile{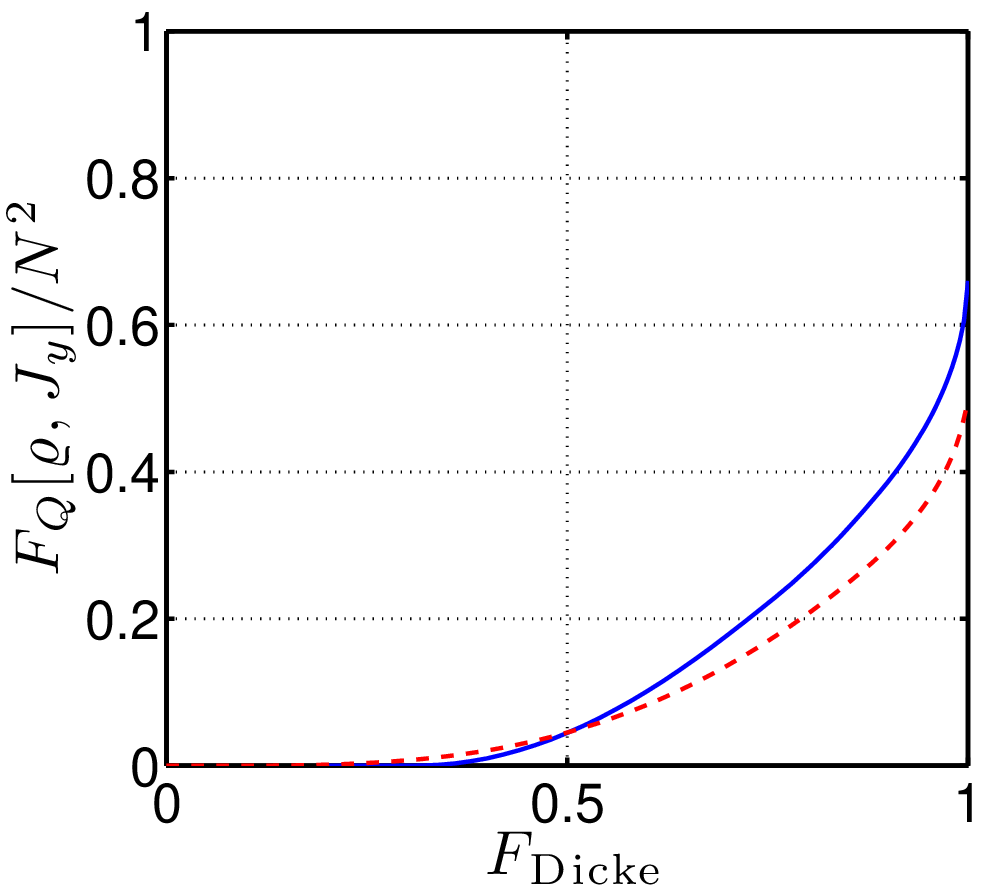}
 }
\centerline{\hskip0.8cm (a) \hskip3.8cm (b)}

\caption{(a) Fidelity vs. lower bound on the quantum Fisher information for GHZ states of $N$ qubits. The quantum Fisher information is 0 if the fidelity is less than $0.5.$
(b) The same, but for Dicke states for with $N=6$ (solid line) and $N=40$ (dashed line).}
\label{fig:GHZ}
\end{figure}

{\it Observation 2.} A sharp lower bound on the quantum Fisher information with the fidelity $F_{\rm GHZ}$ is given by
\be\label{eq:FQFGHZ}
\frac{\mathcal{F}_{\rm Q}[\varrho,J_z]}{N^2}\geqslant \left\{\begin{array}{ll}
 (1-2F_{\rm GHZ})^2 & \text{ if }\; F_{\rm GHZ}> 1/2,\\
0& \text{ if }\; F_{\rm GHZ}\leqslant 1/2.
\end{array}\right.
\ee
The proof is based on carrying out the optimization described above analytically and can be found in \APPENDIX{sec:proofobs2_app} \cite{*[{A not tight lower bounds on the quantum Fisher based on the fidelity has been presented in }] [{. }] Augusiak2016Asymptotic}.
\EQL{eq:FQFGHZ} is plotted in \FIG{fig:GHZ}(a). Note that the bound on the quantum Fisher information normalized by $N^2$ in \EQ{eq:FQFGHZ} is independent of the number of particles. Moreover, the bound is 0 for $F_{\rm GHZ}\leqslant 0.5.$ This is consistent with the fact that for the product state $\ket{111...11}$ we have $F_{\rm GHZ}=1/2,$ while $\mathcal{F}_{\rm Q}[\varrho,J_z]=0.$

Next, let us consider symmetric Dicke states.
An $N$-qubit symmetric Dicke state is given as
\begin{equation}\label{eq:DickeNm}
\ket{{\rm D}_N^{(m)}}=\binom{N}{m}^{-\frac{1}{2}}\sum_k \mathcal{P}_k (\ket{1}^{\otimes m}\otimes\ket{0}^{\otimes (N-m)}),
\end{equation}
where the summation is over all the different permutations of the product state having $m$ particles in the $\ket{1}$ state and $(N-m)$ particles in the $\ket{0}$ state.

From the point of view of metrology, we are interested mostly in the symmetric Dicke state for even $N$ and $m=\frac{N}{2}.$ This state is known to be highly entangled \cite{Toth2007Detection,Toth2009Practical} and allows for Heisenberg limited interferometry \cite{Holland1993Interferometric}.
In the following, we omit the superscript giving the number of  $\ket{1}$'s and use the notation
\begin{equation}\label{eq:Dicke}
\ket{{\rm D}_N}\equiv\ket{{\rm D}_N^{(\frac{N}{2})}}.
\end{equation}
The witness operator that can be used for noisy Dicke states is $W=\ketbra{{\rm D}_N},$
hence for the expectation value of the witness it is just the fidelity with respect to Dicke states, i.e., $\exs{W}=F_{\rm Dicke}.$
In \FIG{fig:GHZ}(b), we plot the results for Dicke states of various numbers of qubits. Now the normalized curve is not the same for all particle numbers.
$F_{\rm Dicke}=1$ corresponds to ${\mathcal F}_{\rm Q}[\varrho,J_y]=N(N+2)/2.$
At this point note that for the examples presented above, the quantum Fisher information scales
as $O(N^2)$ if the quantum state has been prepared perfectly,
where $O(x)$ is the usual Landau notation used to describe the
asymptotic behavior of a quantity for large $x$
\cite{{Hyllus2012Fisher,*Toth2012Multipartite}}.

Note that estimating
$\mathcal{F}_{\rm Q}[\varrho,J_y]$ based on $F_{\rm Dicke}$
was possible for $40$ qubits in \FIG{fig:GHZ}(b), since we carried out the calculations
for the symmetric subspace. For our case, the witness operator $W$ is permutationally invariant and it has a nondegenerate  eigenstate corresponding to the maximal eigenvalue. Hence, based on the arguments in \SEC{sec:symmetries}  the bound is valid even for general, i.e., nonsymmetric states.
Further calculations for the large-$N$ limit are given in \APPENDIX{sec:fisherdicke_appendix}.

\subsection{Spin-squeezed states}
\label{sec:spinsqueezed_th}

In the case of spin-squeezing, the quantum state has a large spin in the $z$ direction, but a decreased variance in the $x$ direction. By measuring $\exs{J_z}$ and $\va{J_x}$ we can estimate the quantum Fisher information by \EQ{eq:archetypical}. However, this formula does not necessarily give the best lower bound for all values of the collective observables. With our approach we can find the best bound.

\begin{figure}

\centerline{
   \epsfxsize1.95in
   \epsffile{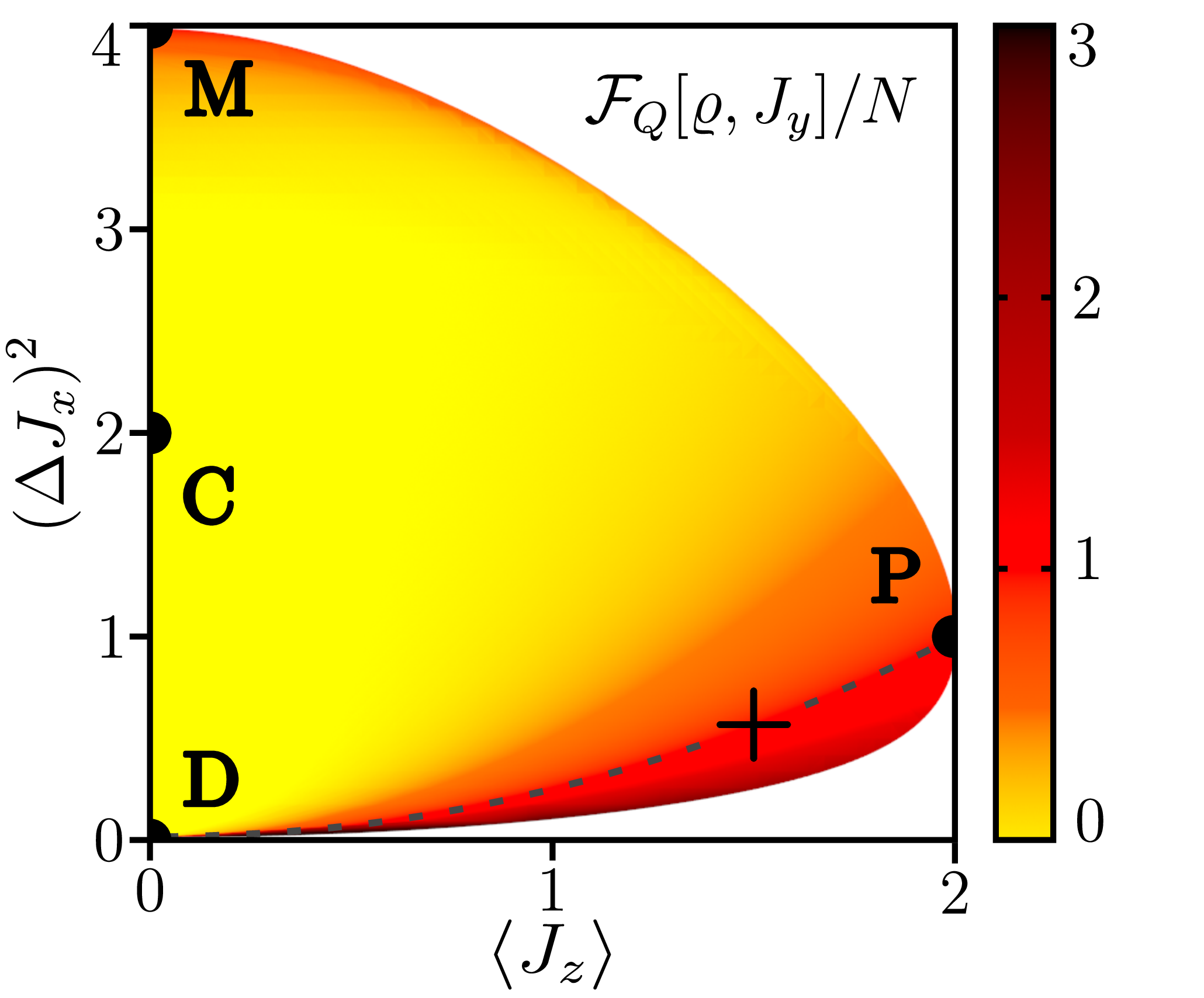}
   \hskip-0.2cm
      \epsfxsize1.5in \epsffile{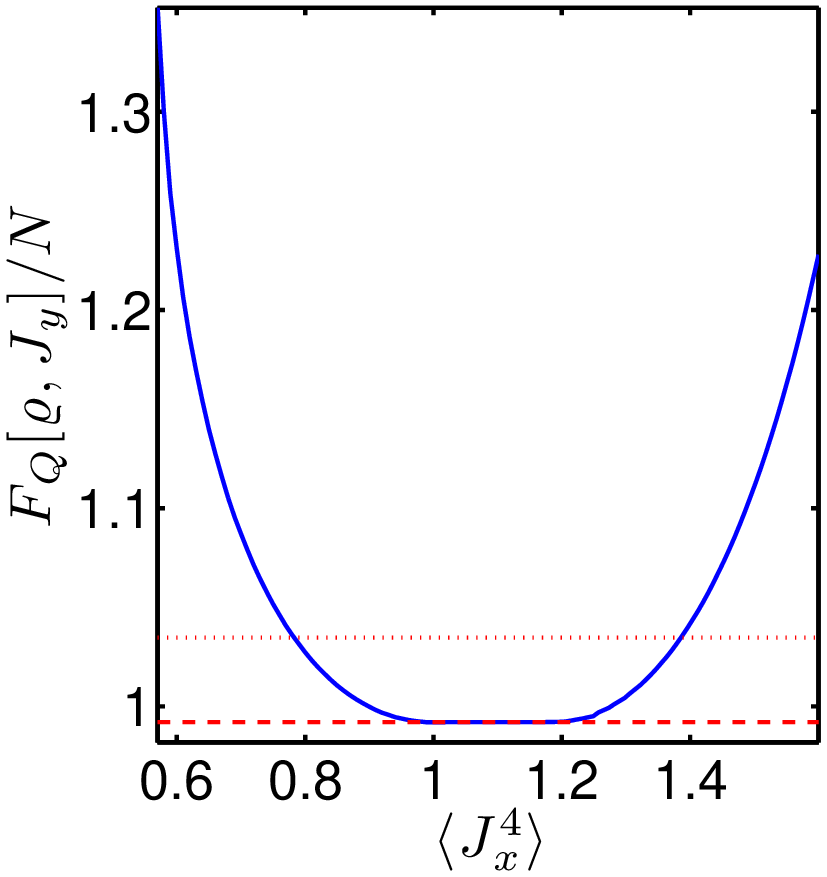}
 }
\centerline{\hskip0.65cm (a) \hskip4.4cm (b)}

\caption{(a) Optimal lower bound on the quantum Fisher information $\mathcal{F}_{\rm Q}[\varrho,J_y]$ based on collective measurements for spin-squeezing with $N=4.$  The mean spin points in the $z$ direction.
Below the dashed line we have $\mathcal{F}_{\rm Q}[\varrho,J_y]/N>1.$
For the description of points P, D, M, and C, see the text.
(b) Lower bound on $\mathcal{F}_{\rm Q}[\varrho,J_y]$ for $\ex{J_z}= 1.5$ and
$\va{J_x}=0.567,$ as a function of $\exs{J_x^4}.$
The corresponding point in (a) is denoted by a cross.
Dashed horizontal line: Lower bound without constraining $\exs{J_x^4}.$
Dotted horizontal line: Lower bound for states in the symmetric subspace.
As shown, an additional constraint or assuming symmetry improves the bound.
}
\label{fig:spin-squeezing}
\end{figure}

To give a concrete example, we choose $W_1=J_z,$ $W_2=J_x^2$, and $W_3=J_x$ for the operators to be measured.
We change $w_1$ and $w_2$ in some interval. We also require that $w_3=0,$
%XXX .
since we assume that the mean spin points in the $z$ direction \footnote{Due to symmetries of the problem, when minimizing ${\mathcal F}_{\rm Q}[\varrho,J_y]$ with the constrains on $\exs{J_z}$ and $\exs{J_x^2}$, we do not have to add explicitly the constraint $\exs{J_x}=0.$ Optimization with only the first two constraints will give the same bound (see \APPENDIX{sec:assuming_Jx0_appendix}).}. This is reasonable
since in most spin-squeezing experiments we know the direction of the mean spin.

Our results are shown in \FIG{fig:spin-squeezing}(a).
We chose $N=4$ particles since for small $N$ the main features of the plot are clearly visible. The white areas correspond to nonphysical combinations of expectation values. States at the boundary can be obtained as  ground states of $H_{{\rm bnd}}^{(\pm)}(\mu)=\pm J_x^2-\mu J_z$ (\APPENDIX{sec:physicalboundry_appendix}).
In \FIG{fig:spin-squeezing}(a), the state fully polarized in the $z$ direction, an initial state for spin-squeezing experiments, corresponds to point P. The Dicke state, \eqref{eq:Dicke}, corresponds to point
D \footnote{Outside the symmetric subspace, there are other states with $\exs{J_z}=\exs{J_x^2}=0,$ which also correspond to point D. For example, such a state is the multiparticle singlet. However, usual spin-squeezing procedures remain in the symmetric subspace, thus we discuss only the Dicke state.}.
Spin-squeezing makes $\va{J_x}$ decrease, while
 $\exs{J_z}$ also decreases somewhat. Hence, at least for small squeezing,
it corresponds to moving down from point P towards point D on the boundary of the plot, while the metrological usefulness is increasing.
Below the dashed line $\mathcal{F}_{\rm Q}[\varrho,J_y]/N>1,$
hence the state possesses metrologically useful entanglement \cite{Pezze2009Entanglement}.
The equal mixture of $\ket{000..00}_x$ and
$\ket{111..11}_x$ corresponds to point
M,
with
${\mathcal F}_{\rm Q}[\varrho_M,J_y]=N.$
Finally, the completely mixed state corresponds to point C. It cannot be used for metrology, hence $\mathcal{F}_{\rm Q}[\varrho_C,J_y]=0.$

We now  compare the difference between
our bound and \EQ{eq:archetypical}.
First, we consider the experimentally relevant region for which $\va{J_x}<1.$ We find that for points
that are away from the boundary at least by $0.01$ on the vertical axis,
the difference between the two bounds for ${\mathcal F}_{\rm Q}[\varrho,J_y]$ is smaller than $2\times10^{-6}.$
For points at the boundary the difference is somewhat larger but still small;
the relative difference is less than $2\%$ (see \APPENDIX{sec:QFI_at_lboundry_appendix}).
Hence, \EQ{eq:archetypical} practically coincides with the optimal bound for $\va{J_x}<1.$
We now consider the region in \FIG{fig:spin-squeezing}(a) for which $\va{J_x}>1.$ The difference between the two bounds is now larger. It is largest at
point M, for which the bound, \eqref{eq:archetypical}, is 0.
Hence, for measurement values corresponding to points close to M, our method could improve formula \eqref{eq:archetypical}.
It is important from the point of view of applying our method
to spin-squeezing experiments that the bound, \eqref{eq:archetypical}, can be substantially improved
even for $\va{J_x}<1,$ if we assume a bosonic symmetry
or we measure an additional quantity, such as
$\exs{J_x^4}$ as shown in \FIG{fig:spin-squeezing}(b).

\subsection{Dicke states}
\label{sec:dickeexp}

In this section, we use our method to find
lower bounds on the quantum Fisher information
for states close to the Dicke states, \eqref{eq:Dicke}, based on collective measurements.
We discuss what operators have to be measured
to estimate the metrological usefulness
of the state. In \SEC{sec:dickeexp_large}, we test our approach for
a realistic system with very many particles.

In order to estimate the metrological usefulness of states created in such experiments, we choose to measure $W_1=J_x^2,$ $W_2=J_y^2,$ and $W_3=J_z^2$
since the expectation values of these operators uniquely define the ideal Dicke state,
and they have already been used for entanglement detection
\cite{Lucke2014Detecting}.
 In cold gas experiments it is common that the state created is invariant under transformations of the type
$U_z(\phi)=\exp(-iJ_z\phi)$ \cite{Apellaniz2015Verifying}. For such states $\exs{J_x^2}=\exs{J_y^2},$ which we also use as a constraint in our optimization.

\begin{figure}
% \vskip-1cm
\centerline{
{\epsfxsize5.5cm \epsffile{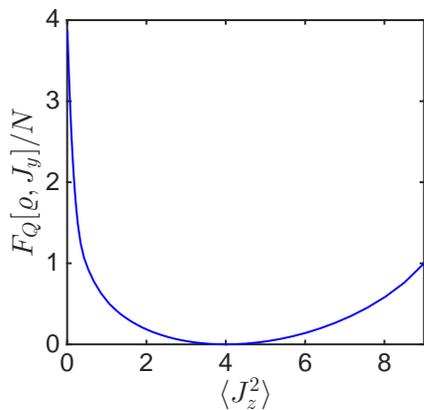}
}
}
\vskip-0.5cm
\caption{%(Color online)
Optimal lower bound on the quantum Fisher information for symmetric states close to Dicke states for $N=6.$
}
\label{fig:dicke-squeezing}
\end{figure}

Let us demonstrate how our method works in an example for small systems.
\FIGL{fig:dicke-squeezing} shows the results for $N=6$ particles for symmetric states for which
\be\label{eq:WJN}
\exs{J_x^2+J_y^2+J_z^2}=\tfrac{N}{2}\left(\tfrac{N}{2}+1\right)=:\mathcal{J}_N.
\ee It can be seen that the lower bound on the quantum Fisher information is the largest
for $\exs{J_z^2}=0.$ It reaches the value corresponding to
the ideal Dicke state, $N(N+2)/2=24.$
It is remarkable that the state is also useful for metrology
if $\exs{J_z^2}$ is very large.
In this case $\exs{J_x^2}$ and $\exs{J_y^2}$ are smaller than $\exs{J_z^2},$
and this cigar-shaped uncertainty ellipse can be used for metrology.

\section{Calculations for experimental data}
\label{sec:calcexp}

In this section, we use our method to find tight lower bounds on the quantum Fisher information based on experimental data. First,
%Our results can be summarized  as follows
%, giving the details in the Appendix.
% \CITESUPP,
we determine the quantum Fisher information for several experiments in photons and trapped ions
creating GHZ states and Dicke states, in which
the fidelity has been measured \cite{Kiesel2007Experimental,Chiuri2012Experimental,Krischek2011Useful,Wieczorek2009Experimental,Prevedel2009Experimental,Zhao2003Experimental,Zhao2004Experimental,Huang2011Multi-partite,Gao2010Experimental,Leibfried2004Toward,Sackett2000Experimental,Leibfried2005Creation,Monz201114-Qubit}. Our method is much simpler than obtaining the
quantum Fisher information from the density matrix \cite{Krischek2011Useful} or estimating it
from a metrological procedure \cite{Lucke2011Twin}. Second, we obtain a bound on
the quantum Fisher information for a spin-squeezing experiment with thousands of particles \cite{Gross2010Nonlinear}.
Based on numerical examples, we see that the bound, \eqref{eq:archetypical},
is close to optimal even if the state is not completely polarized.
Assuming symmetry or knowing additional expectation values
can improve the bound \eqref{eq:archetypical}.
Finally, we also obtain the bound for the quantum Fisher information for a recent experiment with Dicke states \cite{Lucke2014Detecting}.
The
estimate of the precision based on considering the particular
case where $\exs{J_z^2}$ is measured
for parameter estimation \cite{Apellaniz2015Verifying} is close to the optimal bound computed by our method.

\subsection{Few-particle experiments}

We now estimate the quantum Fisher information
based on the fidelity with respect to Dicke states and GHZ states
for several experiments with photons and trapped cold ions,
following the ideas in \SEC{sec:Fidelity_measurements}.

Our results are summarized in \TABLE{tab:exp}.
For the experiments aiming to create  Dicke states, the lower bound on $\mathcal{F}_{\rm Q}[\varrho,J_y]/N^2$ is shown,
while for the experiments with GHZ states we estimate $\mathcal{F}_{\rm Q}[\varrho,J_z]/N^2.$
In \cite{Chiuri2012Experimental, Gao2010Experimental} several logical qubits are stored in a particle, but in the rest of the experiments only a single qubit.
Reference \cite{Monz201114-Qubit} describes experiments with $2$--$14$ ions, of which we analyze the $8$-qubit and $10$-qubit GHZ sates. Finally, for the experiment in \REF{Zhao2004Experimental}
we used the fidelity estimated using reasonable assumptions discussed in that paper, while
the worst-case fidelity is lower.

We can compare our estimate to the quantum Fisher information of the state
for the experiment in \REF{Krischek2011Useful}, where
the quantum Fisher information  for the
density matrix was obtained as
$\mathcal{F}_{\rm Q}[\varrho,J_y]/N^2=(10.326 \pm0.093)/N^2=(
0.6454\pm0.0058).$
As reported in \TABLE{tab:exp}, this value is larger than the one we obtained,
however, it was calculated by knowing the entire density matrix, while our
bound is obtained from the fidelity alone.

\begin{table}[ht]
\begin{center}
\begin{tabular}{ l l c c c }
\hline
\hline
         & Targeted & & & \\
Physical & quantum  & & & \\
system   & state    & Fidelity & $\frac{\mathcal{F}_{\rm Q}}{N^2}\geqslant$ & Ref. No.\\[1ex]
\hline
Photons  & $\ket{{\rm D}_4}$ & $0.844\pm 0.008$ & $0.358\pm0.011$ & \cite{Kiesel2007Experimental}\\
 & & $0.78\pm 0.005$ & $0.281\pm0.059$ & \cite{Chiuri2012Experimental} \\
 & & $0.8872\pm0.0055$ & $0.420\pm0.009$ & \cite{Krischek2011Useful} \\
 & & $0.873\pm 0.005$ & $0.351\pm0.006$ & \cite{Toth2010Permutationally} \\
 & $\ket{{\rm D}_6}$ & $0.654\pm 0.024$ & $0.141\pm0.019$ & \cite{Wieczorek2009Experimental} \\
 & & $0.56\pm 0.02$ & $0.0761\pm0.012$ & \cite{Prevedel2009Experimental} \\
Photons & $\ket{{\rm GHZ}_4}$ & $0.840\pm0.007$ & $0.462\pm0.019$ & \cite{Zhao2003Experimental} \\
 & $\ket{{\rm GHZ}_5}$  & $0.68$ & $0.130$ & \cite{Zhao2004Experimental} \\
 & $\ket{{\rm GHZ}_8}$  & $0.59\pm0.02$ & $0.032\pm0.016$ & \cite{Huang2011Multi-partite} \\
 & $\ket{{\rm GHZ}_{8}}$  & $0.776\pm0.006$ &
% 0.3047 +/- 0.013392
 $0.305\pm0.013$ & \cite{Gao2010Experimental} \\
 &$\ket{{\rm GHZ}_{10}}$  & $0.561 \pm 0.019$ & $0.015\pm0.011$ & \cite{Gao2010Experimental} \\
Trapped ions & $\ket{{\rm GHZ}_3}$  & $0.89\pm0.03$ & $0.608\pm0.097$ & \cite{Leibfried2004Toward} \\
 &$\ket{{\rm GHZ}_4}$  & $0.57\pm0.02$ & $0.020\pm0.013$ & \cite{Sackett2000Experimental} \\
 &$\ket{{\rm GHZ}_6}$  & $\geqslant 0.509\pm0.004$ & $0.0003\pm0.0003$ & \cite{Leibfried2005Creation} \\
 &$\ket{{\rm GHZ}_{8}}$  & $0.817\pm0.004$ & $0.402\pm0.010$ & \cite{Monz201114-Qubit} \\
 &$\ket{{\rm GHZ}_{10}}$  & $0.626\pm0.006$ & $0.064\pm0.006$ & \cite{Monz201114-Qubit} \\[0.2ex]
\hline
\end{tabular}
\end{center}
\caption{Fidelity values and the corresponding bounds on the quantum Fisher information for several experiments with Dicke states and GHZ states. For experiments targeting Dicke states, bounds on $\mathcal{F}_{\rm Q}[\varrho,J_y]/N^2$ are listed. The maximal value of this quantity is
$0.75$ and $0.67$ for $N=4$ and $N=6,$ respectively.
For experiments with GHZ states, bounds on $\mathcal{F}_{\rm Q}[\varrho,J_z]/N^2$ are shown,
and, in this case, the maximal value is $1.$
}
\label{tab:exp}
\end{table}%

\subsection{Many-particle experiments}

So far, we have studied the quantum state of few particles. Next we turn to experiments with very many particles, in which a fidelity measurement is not practical. In such systems, the quantum Fisher information must be
estimated based on collective measurements.

By far the most relevant quantum states in many-particle experiments are spin-squeezed states, which can be used to increase the precision in magnetometry and in atomic clocks \cite{Sorensen2001Many-particle}. We also discuss Dicke states, since they have been realized in several experiments \cite{Lucke2014Detecting,Lucke2011Twin,Hamley2012Spin-nematic,Luo2017Deterministic}.
Dicke states realized in cold gases are the focus of our attention, since they can be used for high-precision interferometry \cite{Holland1993Interferometric}.

\subsubsection{Spin-squeezing experiment}
\label{sec:spinsqueezingexp}

We now use our method to find
lower bounds on the quantum Fisher information
for a recent spin-squeezing experiment in cold gases,
following the ideas in \SEC{sec:spinsqueezed_th}.
With it we show that the lower bound given in \EQ{eq:archetypical} is close to optimal
in this case.
We also demonstrate that we can carry out calculations for
real systems.

In particular, for our calculations
we use the data from the spin-squeezing experiment in \REF{Gross2010Nonlinear}.
The particle number is
$N=2300,$  and the spin-squeezing parameter, defined as
\be
\xi_s^2=N\frac{\va{J_x}}{\exs{J_z}^2},
\label{eq:xis2}
\ee
has the value
$\xi_s^2=-8.2{\rm dB}=10^{-8.2/10}=0.1514.$
The spin length $\exs{J_z}$ has been close to maximal.
In our calculations, we choose
\be
\exs{J_z}=\alpha \tfrac{N}{2},
\label{eq:alpha}
\ee
where we test our method with various values for $\alpha.$
For each $\alpha,$ we use a value for $\va{J_x}$ such that we get the experimentally obtained spin-squeezing parameter, \eqref{eq:xis2}.
Moreover, we assume that $\exs{J_x}=0$, as the $z$direction was the direction of the mean spin in the experiment.
Based on \EQ{eq:archetypical}, the bound for the quantum Fisher information
is obtained as
\be\label{eq:PS_exp}
\frac{\mathcal{F}_{\rm Q}[\varrho_{N},J_y]}{N}\geqslant \frac{1}{\xi_s^2}= 6.605 .
\ee
where $\varrho_N$ is the state of the system in the sxperiment satisfying \EQS{eq:xis2} and \eqref{eq:alpha}.

We carry out the calculations for symmetric states.
This way we obtain a lower bound on the quantum Fisher information, which we denote $\mathcal{B}_{\rm sym}(\exs{J_z}_{\varrho_{N}},\exs{J_x^2}_{\varrho_{N}}).$
As mentioned in \SEC{sec:Fidelity_measurements}, we could obtain
a bound for the quantum Fisher information that is
valid even for general, not necessarily symmetric states
if the matrix in \EQ{eq:LegendreTransformFQ_H} had nondegenerate eigenvalues.
This is not the case for the spin-squeezing problem.
However, we still know that the bound obtained
with our calculations restricted to the symmetric subspace
cannot be smaller than the optimal bound for general states, $\mathcal{B}(\exs{J_z}_{\varrho_{N}},\exs{J_x^2}_{\varrho_{N}}).$
On the other hand, we know that bound \eqref{eq:archetypical}
cannot be larger than the optimal bound for general states.
These relations can be summarized as
\bea\label{eq:bounds}
\mathcal{B}_{\rm sym}(\exs{J_z}_{\varrho_{N'}},\exs{J_x^2}_{\varrho_{N}})
&\ge&
\mathcal{B}(\exs{J_z}_{\varrho_{N}},\exs{J_x^2}_{\varrho_{N}})\nonumber\\
&\ge&
\frac{\exs{J_z}^2_{\rm \varrho_{N}}}{\va{J_x}_{\varrho_{N}}},%\nonumber\\
\eea
where on the right-hand side of \EQ{eq:bounds} there is just the bound in
\EQ{eq:archetypical}.

Our calculations lead to
\be\label{eq:bounds2}
\mathcal{B}_{\rm sym}(\exs{J_z}_{\varrho_{N}},\exs{J_x^2}_{\varrho_{N}})
=
6.605
\ee
for almost completely polarized spin-squeezed states with $\alpha=0.85,$ as well as for not fully polarized ones with $\alpha=0.5.$ That is, based on numerics, the left-hand side and the right-hand side of  \EQ{eq:bounds} seem to be equal. This implies that the lower bound, \eqref{eq:archetypical}, for the quantum Fisher information is optimal for the system.
In \APPENDIX{sec:spinsq_exp_appendix},  the details of the calculations are given, and we also show examples where we can improve the bound, \eqref{eq:archetypical}, with our approach, if symmetry is assumed.

\subsubsection{ Experiment creating Dicke states}
\label{sec:dickeexp_large}

We now present our calculations for an experiment aimed at creating Dicke states in cold gases \cite{Lucke2014Detecting}.
The basic ideas are similar to the ones explained in \SEC{sec:dickeexp} for small systems.
The experimental data are
$N=7900, \exs{J_z^2}_N=112\pm 31, \exs{J_x^2}_N=\exs{J_y^2}_N=6\times10^6\pm 0.6\times 10^6$
\cite{Apellaniz2015Verifying}. Applying some simple transformations, we can obtain a lower bound on $\mathcal{F}_{\rm Q}[\varrho_n, J_y]$
for this very large number of particles, even for general, nonsymmetric systems.

For many particles we can make calculations directly only in the symmetric subspace. Thus, we transform the collective quantities such that
they are compatible with symmetric states, i.e.,
they have to fulfill
\be\label{eq:sym_cond}
\exs{J_x^2}_{{\rm sym},N}+\exs{J_y^2}_{{\rm sym},N}+\exs{J_z^2}_{{\rm sym},N}=
\mathcal{J}_N,
\ee
where $\mathcal{J}_N$ is given in \EQ{eq:WJN}.
This can be done by multiplying all the second moments by the same number as
\be\label{eq:symconst}
\exs{J_l^2}_{{\rm sym},N}=\gamma\exs{J_l^2}_N,
\ee
where $l=x,y,z,$ and we defined the coefficient
\be\label{eq:gamma}
\gamma=\frac{\mathcal{J}_N}{\exs{J_x^2+J_y^2+J_z^2}_N}.
\ee
For a symmetric state, $\gamma=1$.
In practice, $\gamma\le1$, but close to 1.
From this we can see that there are no symmetric states that are compatible with the experimentally observed expectation values.
This is the reason why we needed to apply the transformation \eqref{eq:symconst}.

Based on the ideas of \SEC{sec:dickeexp}, we calculate
the lower bound on the quantum Fisher information\break for symmetric systems, which we denote
$\mathcal{B}_{{\rm sym},N}(\exs{J_x^2}_{{\rm sym},N},\exs{J_y^2}_{{\rm sym},N},\exs{J_z^2}_{{\rm sym},N}).$

Finally, to obtain the results for the original, non-symmetric case,
we need the following observation.

{\it Observation 3.} For the bounds for original system and symmetric system, respectively, the inequality
\be\label{eq:FQbound_ineq}
\mathcal{B}_N
\leqslant \tfrac{1}{\gamma}\mathcal{B}_{{\rm sym},N}
\ee
holds, where $\gamma$ is given in \EQ{eq:gamma}. Here, for brevity we have omitted the arguments
of $\mathcal{B}_N$ and $\mathcal{B}_{{\rm sym},N}.$

{\it Proof.}  For our proof we need to know that
for an $N$-qubit singlet state $\varrho_{{\rm singlet},N}$
the relations
$\exs{J_l^2}_{\varrho_{{\rm singlet},N}}=0$ hold for $l=x,z,y.$
Due to the well-known inequality for the quantum Fisher information $\mathcal{F}_{\rm Q}[\varrho_{{\rm singlet},N},J_l]\leqslant 4 \va{J_l}$, we have
 $\mathcal{F}_{\rm Q}[\varrho_{{\rm singlet},N},J_y]=0.$ In other words,
 the singlet is not useful for metrology with linear interferometers.
 Let us now consider  the mixture
\be
\tilde{\varrho}_N=\left(1-\tfrac{1}{\gamma}\right)  \varrho_{{\rm singlet},N} + \tfrac{1}{\gamma} \varrho_{{\rm sym},N},
\ee
where $\varrho_{{\rm sym},N}$ is a symmetric state having the second moments $\exs{J_l^2}_{{\rm sym}, N}$.
We can easily see from \EQ{eq:symconst} that for the state $\tilde{\varrho}_N,$ we have $\exs{J_l^2}_{\tilde{\varrho}_N}=\exs{J_l^2}_{N}.$
In other words,  $\tilde{\varrho}_N$ has the same values for the second moments  that
have been measured experimentally.

We can relate the bound for general systems to the quantum Fisher information for symmetric systems as
\bea\label{eq:ineq}
\mathcal{B}_N
\leqslant \mathcal{F}_{\rm Q}[{\tilde{\varrho}_N},J_y]
= \tfrac{1}{\gamma}\mathcal{F}_{\rm Q}[\varrho_{{\rm sym},N},J_y].\eea
The inequality in \EQ{eq:ineq} holds because our bound cannot be larger than
the quantum Fisher information of state $\tilde{\varrho}_N$ having the expectation values $\exs{J_l^2}_N$.
The equality in \EQ{eq:ineq} is due to the fact
that both $\tilde{\varrho}_N$ and $J_y$ can be written
 as a block-diagonal matrix of blocks corresponding to different eigenvalues of
 $J_x^2+J_y^2+J_z^2$.
 Moreover, $\varrho_{{\rm singlet},N}$ and $\varrho_{{\rm sym},N}$ have nonzero elements in different blocks.
 Then we can use the general formula \cite{Toth2014Quantum}
\begin{equation}\label{eq:directsum}
{\mathcal F}_{\rm Q}[\bigoplus_k p_k \varrho_k, \bigoplus_k A_k ]=\sum_k p_k {\mathcal F}_{\rm Q}[\varrho_k,A_k],
\end{equation}
where $\varrho_k$ are density matrices with a unit trace and $\sum_k p_k=$\break1.
$\qed$

Extensive numerics for small systems show \EQ{eq:FQbound_ineq} is very close to an equality, hence it
can be used as a basis for making calculations for nonsymmetric states.
In this way, we arrive at  the bound
for the experimental system,
\be\label{eq:bounddicke}
\frac{\mathcal{B}_N}{N}\approx 2.94.
\ee
The "$\approx$" sign is used referring to the fact that we assume that
the inequality in \EQ{eq:ineq} is close to being saturated.
The details of the calculations are given in \APPENDIX{sec:dickeexp_appendix}.

It is instructive to compare the value, \eqref{eq:bounddicke}, to the one obtained in
\REF{Apellaniz2015Verifying}, where the metrological usefulness
has been estimated based on the second and fourth moments
of the collective angular momentum components, and
assuming that $\exs{J_z^2}$ is used for parameter estimation.
The result implies that $\mathcal{F}_{\rm Q}[\varrho_{N},J_y]/N\geqslant 3.3.$
Our result in \EQ{eq:bounddicke} is somewhat smaller, as we did not use the knowledge of the fourth moment, only the second moments.
The closeness of the two results
is a strong argument for the correctness of our calculations.

\section{Scaling of $\mathcal{F}_{\rm Q}[\varrho,J_l]$ with $N.$}
\label{sec:scaling}

Recent important
works  examine the scaling of the quantum Fisher information with the particle number for metrology under the presence of decoherence
\cite{{Escher2011General,*Demkowicz-Dobrzanski2012The}}.
They consider the quantum Fisher information defined for nonunitary, noisy evolution.
They find that for small $N$ it is close to the value obtained considering coherent dynamics. Hence,  even the Heisenberg scaling, $O(N^2),$  can be reached.
However, if $N$ is sufficiently large, then, due to the decoherence during the parameter estimation, the quantum Fisher information scales as $O(N).$

In contrast, we do not consider the usefulness of a quantum state
in some noisy metrological process, but we estimate the quantum
Fisher information assuming a perfect unitary dynamics.
Hence, the quantum Fisher information can be smaller than what we expect ideally only
due to imperfect state preparation \footnote{This is also relevant for \REF{Augusiak2016Asymptotic}, where $\mathcal{F}_{\rm Q}=O(N^2)$ is reached with weakly entangled states. }.
We can even find simple conditions for the state preparation
that lead to a Heisenberg scaling.
Based on \EQ{eq:FQFGHZ}, if one could realize quantum states $\varrho_N$ such that $F_{\rm GHZ}(\varrho_N)\geqslant 0.5+\epsilon$ for $N\rightarrow \infty$ for some $\epsilon>0,$
then we would reach $\mathcal{F}_{\rm Q}[\varrho_N,J_z]=O(N^2).$
Strong numerical evidence suggests that a similar relation holds for the fidelity $F_{\rm Dicke}$ and $\mathcal{F}_{\rm Q}[\varrho_N,J_y]$, but with a smaller threshold value for $F_{\rm Dicke}$ (see \APPENDIX{sec:fisherdicke_appendix}).
From another point of view, our method
can estimate $\mathcal{F}_{\rm Q}[\varrho,J_z]$ for large particle numbers, while a direct measurement of the metrological sensitivity
considerably underestimates it.

\section{Conclusions}

We have reported a general method to estimate the metrological usefulness of quantum states based on a few measurements, such as measurements of the fidelity or some collective observables.
We tested our approach on extensive experimental data from photonic and cold-gas experiments and demonstrated that it works
even for the case of thousands particles \cite{*[{For some of the programs used for this article, see the actual version of the QUBIT4MATLAB package at  {\tt http://www.mathworks.com/matlabcentral/}. The 3.0 version of the package is described in }] [{ .}] Toth2008QUBIT4MATLAB}.
In the future, it would be interesting to use our method to test the optimality of
various recent formulas giving a lower bound on the quantum Fisher information \cite{Oudot2015Two-mode,Zhang2014Quantum}. Another important question is how to improve the lower bounds on the quantum Fisher information in various experiments by using the knowledge of further operator expectation values.

\acknowledgments

We thank S. Altenburg, F. Fr\"owis, R. Demkowicz-Dobrza\'nski,
P. Hyllus, J. Ko{\l}odi\'nsky, O. Marty, M.W. Mitchell, M. Modugno, L. Pezze, L. Santos, A. Smerzi, I. Urizar-Lanz, and G. Vitagliano  for stimulating discussions. We thank the organizers, M. Oberthaler and P. Treutlein, and the participants of the 589.~Heraeus-Seminar on
``Continuous Variable Entanglement
in Atomic Systems'' for scientific exchange.
We acknowledge the support of the
EU (ERC Starting Grant 258647/GEDENTQOPT,
ERC Consolidator Grant 683107/TempoQ,
CHIST-ERA QUASAR, Marie Curie CIG 293993/ENFOQI, COST Action CA15220),
the Spanish Ministry of Economy, Industry and Competitiveness and the European Regional Development Fund FEDER through Grant No. FIS2015-67161-P (MINECO/FEDER), the Basque
Government (Project No. IT986-16), the OTKA (Contract No. K83858),
the UPV/EHU program UFI 11/55,
the FQXi Fund (Grant No. FQXi-RFP-1608), and the DFG (Forschungsstipendium KL 2726/2-1, Project ``Precise and efficient characterization of multi-qubit quantum states and gates with trapped ions'').

\appendix

\section{Proof of Observation 2}
\label{sec:proofobs2_app}

In this section, using \EQS{eq:LowerBoundOnF} and \eqref{eq:LegendreTransformFQ_H},
 we obtain analytically a tight lower bound on the quantum Fisher information based on the fidelity with respect to the GHZ state, $F_{\rm GHZ}$.

The calculation that we have to carry out is computing the bound,
\be\label{eq:BGHZ}
	\mathcal{B}(F_{\rm GHZ})=
	\sup_r \{rF_{\rm GHZ}-\sup_{\mu}[\lambda_{\max}(M_{\rm GHZ})]
	\},
\ee
where
\be\label{eq:MGHZ}
M_{\rm GHZ}=r\ketbra{{\rm GHZ}}-4(J_z-\mu)^2\openone.
\ee
We make our calculations in the $J_z$ basis, which is defined with the $2^N$ basis vectors
$b_0=\ket{00\dots000}$, $b_1=\ket{00\dots001}$, $b_2=\ket{00\dots010}$, \dots, $b_{(2^N-2)}=\ket{11\dots110}$,
and $b_{(2^N-1)}=\ket{11\dots111}$.
It is easy to see that the matrix, \eqref{eq:MGHZ},
 is almost diagonal in the $J_z$ basis.
 To be more specific, it can then be written as
\be
M_{\rm GHZ}=M_{2\times2} \oplus D,
\ee
where $\oplus$ denotes the direct sum and
\be
M_{2\times2}=
\left(\begin{array}{cc}
\frac{r}{2} -4 (\frac{N}{2}-\mu)^2 & \frac{r}{2}\\
\frac{r}{2} & \frac{r}{2} -4 (\frac{N}{2}+\mu)^2
\end{array}\right)
\ee
is given in the $\{b_0,b_{(2^N-1)}\}$ basis,
while $D$ is a diagonal matrix given in the basis of the rest of the $b_k$ vectors as
\be\label{eq:D}
D_{k}=-4(\langle b_k \vert J_z \vert b_k\rangle-\mu)^2
\ee
for $k=1,2,\dots,(2^{N}-2).$
This means that $M_{\rm GHZ}$ can be diagonalized as
\be\label{eq:diag}
{\rm diag}[\lambda_+,\lambda_-,D_1,D_2,...,D_{(2^N-2)}],
\ee
where the two eigenvalues of $M_{2\times2}$ are
\be
\lambda_{\pm}=\frac{r}{2}-N^2-4\mu^2\pm\sqrt{16\mu^2N^2+\frac{r^2}{4}}.
\ee

Next, we show a way that can simplify our calculations considerably.
As indicated in \EQ{eq:BGHZ}, we have to look for the maximal eigenvalue of $M_{\rm GHZ}$ and then optimize it over $\mu.$
We exchange the order of the two steps, that is, we
look for the maximum of each eigenvalue over $\mu$
and then find the maximal one.
Clearly, based on \EQ{eq:D} we obtain
\be
\sup_{\mu} D_k = 0,
\ee
since we can always choose a value for $\mu$ that makes $D_k$ 0, while
it is clear that it cannot be positive.
Thus, the maximal eigenvalue, maximized also over $\mu,$ can be obtained as
\bea\label{eq:LLL}
\sup_\mu [ \lambda_{\max}(M_{\rm GHZ})]&:=&\max[0,\sup_\mu (\lambda_+)]\nonumber\\
&=&
	\begin{cases}
		0 ,&  \mbox{if } r < 0,\\
		\frac{r}{2} + \tfrac{r^2}{16 N^2}, &   \mbox{if } 0\leqslant r\leqslant 4N^2,\\
		-N^2+r ,&   \mbox{if } r>4N^2,
	\end{cases}\nonumber\\
\eea
where we did not have to look for the maximum of $\lambda_-$ over $\mu$ since clearly $\lambda_+\ge\lambda_-.$
Finally, we have to substitute \EQ{eq:LLL} into \EQ{eq:BGHZ}, and carry
out the optimization over $r,$ considering $F_{\rm GHZ}\in[0,1].$
This way we arrive at \EQ{eq:FQFGHZ}. $\qed$

\section{Calculations in the symmetric subspace}
\label{sec:symm_subspace_appendix}

In this section, we prove an important fact, which can be used to simplify our calculations.

{\it Observation 4.} If a permutationally invariant $N$-qubit Hamiltonian $H$ has a nondegenerate ground state, then the ground state is in the symmetric subspace if $N>2.$ An analogous statement holds for the
maximal eigenvalue.

{\it Proof.} This is a well-known fact; we give a proof only for completeness.
Let $\vert\Psi\rangle$ denote the nondegenerate ground state.
This is at the same time the $T=0$ thermal ground state,
hence it must be a permutationally invariant pure state. For such states
%\begin{equation}
$S_{kl}\vert\Psi\rangle\langle\Psi\vert S_{kl}=\vert\Psi\rangle\langle\Psi\vert$,
%\end{equation}
where $S_{kl}$ is the swap operator exchanging qubits $k$ and $l.$
Based on this, it follows that
%\begin{equation}
$S_{kl}\vert\Psi\rangle=c_{kl}\vert\Psi\rangle,$
%\end{equation}
and $c_{kl}\in\{-1,+1\}$. There are three possible
cases to consider.

(i) All $c_{kl}=+1.$ In this case, for all permutation operators
$\Pi_j$ we have
\begin{equation}
\Pi_j\vert\Psi\rangle=\vert\Psi\rangle,\label{eq:sym}
\end{equation}
since any permutation operator $\Pi_j$ can be constructed as
%\begin{equation}
$\Pi_j=S_{k_{1}l_{1}}S_{k_{2}l_{2}}S_{k_{3}l_{3}}\dots S_{k_{m}l_{m}},$
%\end{equation}
where $m\geqslant1.$ Equation~(\ref{eq:sym}) means that the state $\vert\Psi\rangle$
is symmetric.

(ii) All $c_{kl}=-1.$ This means that the state is antisymmetric,
however, such a state exists only for $N=2$ qubits.

(iii) Not all $c_{kl}$ are identical to each other. In this case, there must be
$k_{+},l_{+},k_{-},l_{-}$ such that
\begin{eqnarray}
S_{k_{+}l_{+}}\vert\Psi\rangle & = & +\vert\Psi\rangle,\nonumber \\
S_{k_{-}l_{-}}\vert\Psi\rangle & = & -\vert\Psi\rangle.\label{eq:Feq1}
\end{eqnarray}
Let us assume that $k_{+}$, $l_{+}$, $k_{-}$ and $l_{-}$
are indices different from each other. In this case,
%\begin{equation}
$\vert\Psi'\rangle=S_{k_{+}k_{-}}S_{l_{+}l_{-}}\vert\Psi\rangle $
%\end{equation}
 is another ground state of Hamiltonian $H$ such that
\begin{eqnarray}
S_{k_{+}l_{+}}\vert\Psi'\rangle & = & -\vert\Psi'\rangle,\nonumber \\
S_{k_{-}l_{-}}\vert\Psi'\rangle & = & +\vert\Psi'\rangle.\label{eq:Feq2}
\end{eqnarray}
Comparing Eq.~(\ref{eq:Feq1}) and Eq.~(\ref{eq:Feq2}) we can conclude
that
%\begin{equation}
$\vert\Psi'\rangle\ne\vert\Psi\rangle$,
%\end{equation}
while due to the permutational invariance of $H$
we must have $\langle \Psi \vert H \vert \Psi\rangle=\langle \Psi' \vert H \vert \Psi'\rangle$.
Thus, $\vert\Psi\rangle$ is not a nongenerate ground state.
Let us now see what happens if $k_{+}$, $l_{+}$, $k_{-}$, and $l_{-}$
are not all different from each other.
The proof works in an analogous way for the only nontrivial case, $k_{+}=k_{-}$,
when $S_{k_{+}k_{-}}=\eins$.

Hence, if $N>2$, then only (i) is possible and $\vert\Psi\rangle$ must
be symmetric.  $\qed$

\section{Estimating the quantum Fisher information based on the fidelity with respect to Dicke states}
\label{sec:fisherdicke_appendix}

In this section, we show that if the fidelity with respect to the
Dicke state, \eqref{eq:FQ_F_cond}, is larger than a bound, then
$\mathcal{F}_{\rm Q}[\varrho,J_y]>0.$
Moreover, \FIG{fig:GHZ}(b) shows that the lower bound on $\mathcal{F}_{\rm Q}[\varrho,J_y]$ as a function of the fidelity $F_{\rm Dicke}$ normalized by $N^2$ is not the same curve for all $N.$
In this section, we demonstrate with numerical evidence
that the lower bound %on $\mathcal{F}_{\rm Q}[\varrho,J_y]$
normalized by $N^2$ collapses to a nontrivial curve for
large $N.$

As the first step, let us consider the state completely polarized in the $y$ direction,
\be\label{eq:psiy}
\ket{\Psi_y}=\ket{1}^{\otimes N}_y.
\ee
State \eqref{eq:psiy} does not change under a rotation around the $y$ axis,
hence ${\mathcal F}_{\rm Q}[\varrho,J_y]=0$.
Its fidelity with respect to the Dicke state, \eqref{eq:Dicke}, is
\be\label{eq:FDickey}
F_{\rm Dicke}(\ket{\Psi_y})=\frac{1}{2^N}\binom{N}{N/2}\approx \sqrt{\frac{2}{\pi N}}.
\ee
From the convexity of the bound on the quantum Fisher information
in $F_{\rm Dicke}$, it immediately follows that for $F_{\rm Dicke}$ smaller than \EQ{eq:FDickey} the optimal lower bound on ${\mathcal F}_{\rm Q}[\varrho,J_y]$
will give 0. For the examples shown in \FIG{fig:GHZ}(b), this fidelity limit is
$0.3125$ and $0.1254$ for $N=6$ and $N=40,$ respectively.

Next, we examine what happens if the fidelity is larger than \EQ{eq:FDickey}.

{\it Observation 5.} If for some state $\varrho$ we have
\be\label{eq:FQ_F_cond}
F_{\rm Dicke}(\varrho)\equiv{\rm Tr}(\ketbra{D_N}\varrho) > F_{\rm Dicke}(\ket{\Psi_y}),
\ee
then $\mathcal{F}_{\rm Q}[\varrho,J_y]>0.$ [The state $\ket{D_{N}}$ is given in \EQ{eq:Dicke}, and $F_{\rm Dicke}(\ket{\Psi_y})$ is given in \EQ{eq:FDickey}.]

{\it Proof.} We have to determine the
maximum for $F_{\rm Dicke}(\varrho)$
 for states that are
not useful for metrology, i.e., $\mathcal{F}_{\rm Q}[\varrho,J_y]=0.$
We know that $\mathcal{F}_{\rm Q}[\varrho,J_y]$ is the convex roof of
$4\va{J_y}$ \cite{Toth2013Extremal}. Hence, if we have a mixed state for which
$\mathcal{F}_{\rm Q}[\varrho,J_y]=0$, then it can always be decomposed into the
mixture of pure states $\ket{\Psi_k}$ for which
$\mathcal{F}_{\rm Q}[\Psi_k,J_y]=0$. As a consequence,
the extremal states of the set of states for which
$\mathcal{F}_{\rm Q}[\varrho,J_y]=0$ are pure states,
and we can restrict our search for pure states.
The optimization problem we have to solve can be given as
\be\label{eq:overlap}
\max_{\ket{\Psi}: \mathcal{F}_{\rm Q}[\ket{\Psi},J_y]=0}
\left\vert \langle \Psi \vert D_{N}
\rangle\right\vert^2.
\ee
Pure states $\ket{\Psi}$ for which
$\mathcal{F}_{\rm Q}[\ket{\Psi},J_y]=0$ must be invariant under $U_\phi=\exp(-iJ_y\phi)$ for any $\phi.$
Such states are the eigenstates of $J_y$. In order to maximize the overlap with the symmetric Dicke state
$\vert D_{N}
\rangle$
in \EQ{eq:overlap}, we have to look for symmetric eigenstates of
$J_y.$ These are the symmetric Dicke states in the $y$ basis
$\ket{{\rm D}_N^{(m)}}_y$. [See \EQ{eq:DickeNm}.]
In order to proceed, we have to write down $\ket{{\rm D}_N^{(m)}}_y$ in the $z$ basis.
Then, using the formula
\be
\sum_{k} \binom{n}{k} \binom{n}{q-k} (-1)^{k}
=\begin{cases} \binom{n}{q/2}(-1)^{q/2} &\text{for even }N,  \\ 0 &\text{for odd }N, \end{cases}
\ee
one finds
that the squared overlap is given by
\be
\vert \langle {\rm D}_N^{(N/2)}\vert {\rm D}_N^{(m)}\rangle_y \vert ^2
=\begin{cases}\frac{\binom{N/2}{m/2}^2 \binom{N}{N/2}}{2^{N}\binom{N}{m}} &\text{for even }N,\\0 & \text{for odd }N,\end{cases}
\ee
which is maximal for $m=0.\qed$

\begin{figure}
% \vskip-1cm
\vskip0.2cm

\centerline{
\epsfxsize6cm \epsffile{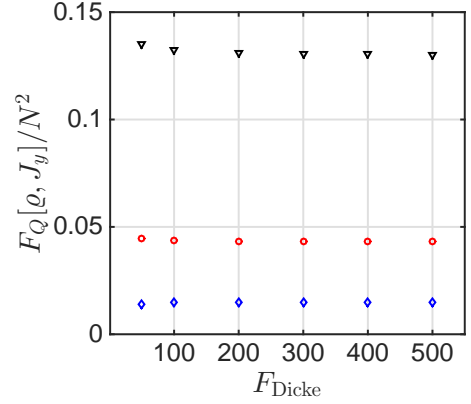}}
\vskip-0.2cm

\caption{(Color online)
The lower bound on ${\mathcal F}_{\rm Q}[\varrho,J_y],$ denoted $\mathcal{B}(F_{\rm Dicke}),$ for various particle numbers,
for $F_{\rm Dicke}=0.2$ (diamonds), $0.5$ (circles), and $0.7$ (triangles).
For $F_{\rm Dicke}=0.2,$ $10$ the calculated values times 10are shown, for better visibility.
 }
\label{fig:dicke_scaling}
\end{figure}

Next, we examine the behavior of our lower bound on
${\mathcal F}_{\rm Q}[\varrho,J_y]$ based on $F_{\rm Dicke}$ for large $N.$
In \FIG{fig:dicke_scaling}, the calculations up to $N=500$ present strong evidence that
for the fidelity values $F_{\rm Dicke}=0.2$, $0.5$, and $0.8$
the lower bound on the quantum Fisher information ${\mathcal F}_{\rm Q}[\varrho,J_y]$
has an $O(N^2)$ scaling.
If this is correct, then reaching a fidelity larger than a certain bound
for large $N$ would imply Heisenberg scaling for the bound on the quantum Fisher information. Note that it is difficult to present similar numerical evidence for small values of $F_{\rm Dicke},$ since in that case the bound for the quantum Fisher information
is nonzero only for large $N$ due to Observation 5.

\section{Boundary of physical states in the $(\exs{J_z},\exs{J_x^2})$-plane. }
\label{sec:physicalboundry_appendix}

In this section, we discuss how to find the physical region in the
$(\exs{J_z},\exs{J_x^2})$ plane, which was used to prepare
\FIG{fig:spin-squeezing}(a).

The physical region must be a convex one, since the set of quantum
states is convex and the coordinates depend linearly on the density matrix. Hence, we look for the minimal or maximal
$\exs{J_x^2}$ for a given $\exs{J_z}$ by looking for the ground states of
the Hamiltonians \cite{Sorensen2001Entanglement},
\be
H_{{\rm bnd}}^{(\pm)}(\mu)=\pm J_x^2-\mu J_z.\label{eq:Hpm}
\ee
The points of the boundary can be obtained
by evaluating $\exs{J_x^2}$ and $\exs{J_z}$
 for the ground states of \EQ{eq:Hpm}.
In particular, the ground states of $H_{{\rm bnd}}^{(+)}$ correspond to boundary points below point P corresponding to the fully polarized state
in \FIG{fig:spin-squeezing}(a).
The  ground states of $H_{{\rm bnd}}^{(-)}$ correspond to boundary points above point P.

For $0<\mu<\infty,$ the Hamiltonian $H_{{\rm bnd}}^{(+)}$ has nondegenerate ground states
with $\exs{J_x}=0.$ For even $N,$  the ground state of $H_{{\rm bnd}}^{(+)}$ minimizes both $\exs{J_x^2}$ and $\va{J_x}$ for a given $\exs{J_z}.$ For odd $N,$ this is not the case for small $\mu$ \cite{Sorensen2001Entanglement}.

On the other hand, $H_{{\rm bnd}}^{(-)}(\mu)$ has doubly degenerate ground states.
For the ground-state subspace, we have $\exs{J_x}=0.$ Hence,
for both even $N$ and odd $N,$ the ground state of $H_{{\rm bnd}}^{(-)}$ maximizes both $\exs{J_x^2}$ and $\va{J_x}$ for a given $\exs{J_z}.$

\section{Quantum Fisher information for states at the boundary of the physical region in the $(\exs{J_z},\exs{J_x^2})$-plane }

\label{sec:QFI_at_lboundry_appendix}

We show that, for even $N$, the ground states of $H_{{\rm bnd}}^{(+)}(\mu)$ defined in \EQ{eq:Hpm} are close to saturating \EQ{eq:archetypical}. As a consequence, for the boundary of the physical region in the $(\exs{J_z},\exs{J_x^2})$ plane below point P in \FIG{fig:spin-squeezing}, bound \eqref{eq:archetypical} is close to the optimal lower bound.

We carry out numerical calculations. Let us denote by $\ket{\Psi_{\mu}}$ the ground state of $H_{{\rm bnd}}^{(+)}(\mu)$.
Moreover, let us denote the relevant expectation values for this state $\exs{J_x^2}_{\mu}$ and $\exs{J_z}_{\mu}$.
We know that under the constraint $\exs{J_z}=\exs{J_z}_{\mu}$,
the state $\ket{\Psi_{\mu}}$ minimizes $\exs{J_x^2}.$ For $H_{{\rm bnd}}^{(+)}(\mu),$ the ground state is unique for
$0<\mu<\infty.$ Thus, there is no other quantum state with the same value for $\exs{J_z}$ and $\exs{J_x^2}.$

There is a very important consequence of the
uniqueness of the ground state of $H_{{\rm bnd}}^{(+)}(\mu)$ for the lower bound on the quantum Fisher information. We have discussed that our method based on the Legendre transform
gives the optimal lower bound for the quantum Fisher information
\be
\mathcal{F}_{\rm Q}[\varrho,J_y]\ge\mathcal{B}(\exs{J_z}_{\varrho},\ex{J_x^2}_{\varrho}),
\ee
where $\mathcal{B}$ denotes the optimal bound.
Since there is a unique state
corresponding to the boundary points, we must have for the states at the boundary
\be\label{eq:FQmu}
\mathcal{B}(\exs{J_z}_{\mu},\ex{J_x^2}_{\mu})=\mathcal{F}_{\rm Q}[\Psi_{\mu},J_y].
\ee
Thus, for the boundary points we do not have to compute
the lower bound with the method based on the Legendre transform.
We can just calculate the right-hand side of \EQ{eq:FQmu} instead.
Since we have a pure state, the quantum Fisher information is proportional to the variance
$\mathcal{F}_{\rm Q}[\varrho,J_y]=4\va{J_y}$
\cite{{Giovannetti2004Quantum-Enhanced,*Paris2009QUANTUM,*Demkowicz-Dobrzanski2014Quantum,*Pezze2014Quantum}}.

We add that, for even $N$, state $\ket{\Psi_\mu}$ not only minimizes $\exs{J_x^2}$ for a given value of  $\exs{J_z},$
but
also minimizes $\va{J_x}$, and this state is unique  \cite{Sorensen2001Entanglement}. Hence,
for the points on the boundary of physical states in the $(\exs{J_z},\va{J_x})$-space we have
\be\label{eq:FQmu2}
\mathcal{B}(\exs{J_z}_{\mu},\va{J_x}_{\mu})=\mathcal{F}_{\rm Q}[\Psi_{\mu},J_y],
\ee
where $\mathcal{B}$ denotes the optimal bound if the expectation value $\exs{J_z}$
and the variance $\va{J_x}$ are constrained.
Note that bound \eqref{eq:FQmu2} is monotonous in
$\va{J_x}_{\mu}$ \cite{Sorensen2001Entanglement}.

\begin{figure}
% \vskip-1cm
\centerline{
      \epsfxsize6cm \epsffile{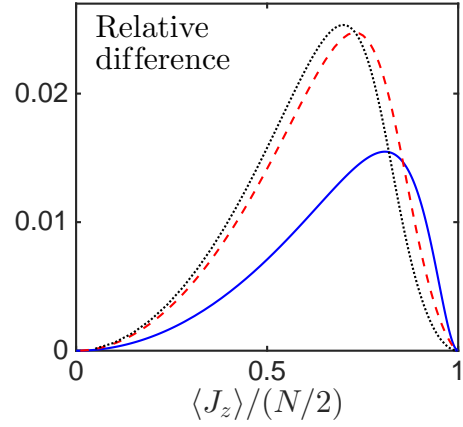}
}\vskip-0.3cm
\caption{Behavior of the bound in \EQ{eq:archetypical} for points at the boundary of physical states.
The relative difference
with respect to the optimal lower bound is plotted for
 $N=4$ (solid line), $N=20$ (dashed line), and $N=1000$ (dotted line).
}
\label{fig:spin-squeezing_boundary}
\end{figure}

In \FIG{fig:spin-squeezing_boundary}, we plot the relative difference between  the quantum Fisher information of $\ket{\Psi_{\mu}}$ and the lower bound \eqref{eq:archetypical} given as
\be\label{eq:PS_minus_FQ}
\frac{\mathcal{F}_{\rm Q}[\Psi_{\mu},J_y]-\frac{\exs{J_z}^2_{\mu}}{\va{J_x}_{\mu}}}{\mathcal{F}_{\rm Q}[\Psi_{\mu},J_y]}
\ee
for various particle numbers.
It can be seen that for an almost fully polarized state the difference is small,
but even for a state that is not fully polarized the relative difference is smaller than $3\%$ for the particle numbers considered.

\section {Why we can assume $\exs{J_x}=0$ for the discussion of spin-squeezed states}
\label{sec:assuming_Jx0_appendix}

We show that for the state minimizing $\mathcal{F}_{\rm Q}[\varrho,J_y]$ for given $\exs{J_z}$ and $\exs{J_x^2}$ we have $\exs{J_x}=0.$
Hence, if we constrain only $\exs{J_z}$ and $\exs{J_x^2},$
then we get the same bound as if we constrained $\exs{J_z}$ and $\exs{J_x^2},$ and we used an additional constraint $\exs{J_x}=0.$

For spin-squeezed states, we have to solve the following optimization task.
 We have to find a tight lower bound on the quantum Fisher information
\be \mathcal{F}_{\rm Q}[\varrho,J_y]\geqslant \mathcal{B}(\vec{w}_{\varrho})\ee where
$\vec{w}_{\varrho}=(\exs{J_z}_\varrho, \exs{J_x^2}_\varrho,\exs{J_x}_\varrho).$
For any $\varrho,$ we can define\break a state $\varrho_{-}=\sigma_z^{\otimes N} \varrho \sigma_z^{\otimes N}$,
for which $\vec{w}_{\varrho_{-}}=(\exs{J_z}_\varrho, \exs{J_x^2}_\varrho,$ $-\exs{J_x}_\varrho)$.
The metrological usefulness of $\varrho$ and $\varrho_{-}$ are the same, i.e., $\mathcal{F}_{\rm Q}[\varrho,J_y]=\mathcal{F}_{\rm Q}[\varrho_{-},J_y].$
Then, for any $\varrho,$ we can define a state
$
\varrho_{0}=\frac{1}{2}(\varrho+\varrho_{-}),
$
for which we have
$
\vec{w}_{\varrho_{0}}=(\exs{J_z}_\varrho, \exs{J_x^2}_\varrho,0).
$
Due to the convexity of the quantum Fisher information,
$\varrho_{0}$ cannot be better metrologically than $\varrho$ or
$\varrho_{-},$ that is, $\mathcal{F}_{\rm Q}[\varrho,J_y]=\mathcal{F}_{\rm Q}[\varrho_{-},J_y]\geqslant \mathcal{F}_{\rm Q}[\varrho_{0},J_y].$

Since for any $\varrho$ there is a corresponding $\varrho_0$ with the above properties, it follows that
$
\mathcal{B}(\vec{v}_{\varrho})= \mathcal{B}(\vec{v}_{\varrho_-})
\geqslant \mathcal{B}(\vec{w}_{\varrho_0})=
\mathcal{B}(\exs{J_z}_\varrho, \exs{J_x^2}_\varrho,0).
$
Thus, the worst-case bound for given $\exs{J_z}$
and $\exs{J_x^2}$ is $\mathcal{B}(\exs{J_z},\exs{J_x^2},0).$
Hence,
\be
\mathcal{B}(\exs{J_z}, \exs{J_x^2})=
\mathcal{B}(\exs{J_z}, \exs{J_x^2},\exs{J_x}=0),
\ee
and our claim is proved.

%[htdp]

\section {Many-particle experiments}

In this section, we consider cold-gas experiments creating
spin-squeezed states and Dicke states.

\subsection {Spin-squeezing experiment}
\label{sec:spinsq_exp_appendix}

\begin{figure}
\centerline{
\epsfxsize4.3cm \epsffile{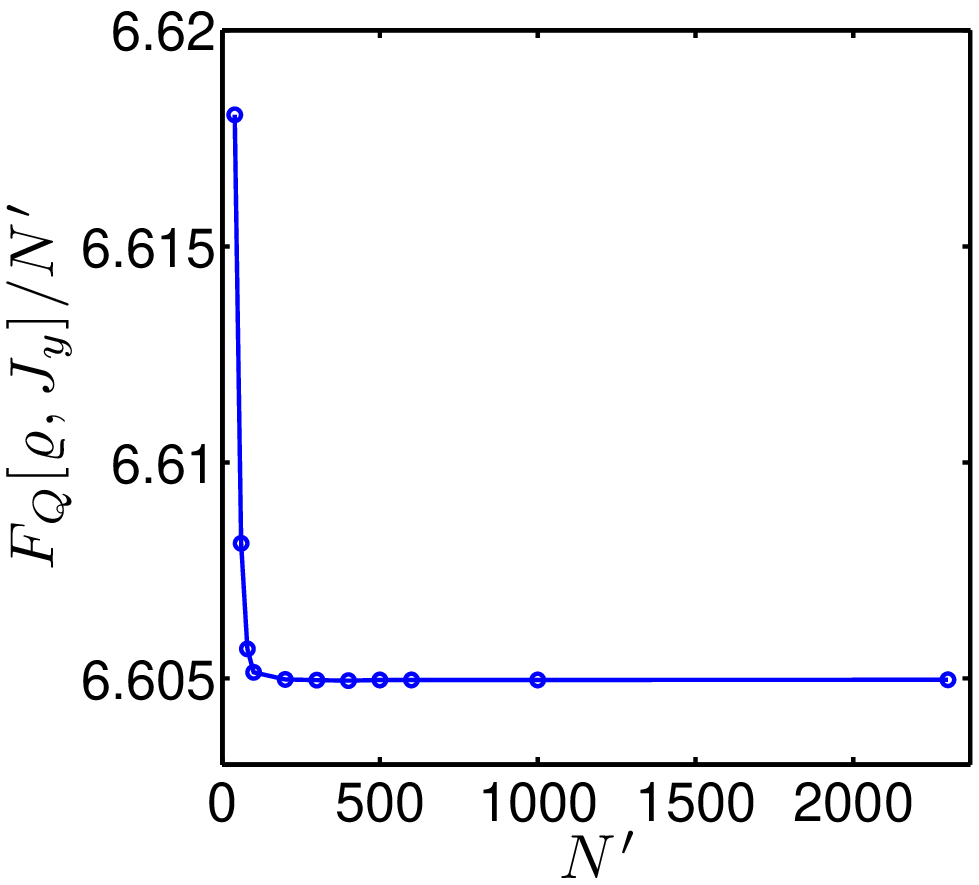}\hskip0.2cm
\epsfxsize4.45cm\epsffile{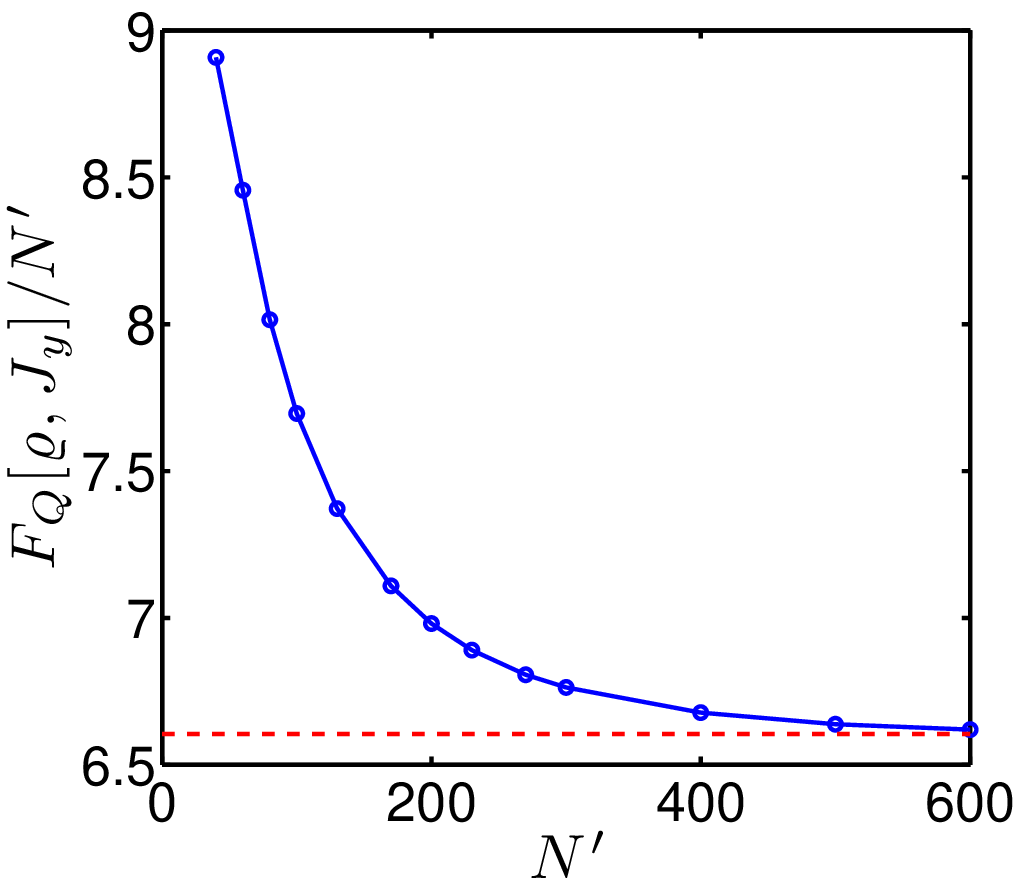}
}
\hskip0.7cm (a) \hskip3.8cm (b)
\caption{(Color online)
Lower bound on the quantum Fisher information based on
$\exs{J_z}$ and $\va{J_x}$ obtained for different particle numbers
making calculations in the symmetric subspace. $N=2300$ corresponds to the spin-squeezing experiment in \REF{Gross2010Nonlinear}. (a) Almost fully polarized spin-squeezed state.
Even for a moderate $N',$ the bound  is practically identical to the right-hand side of \EQ{eq:PS_exp}.
 (b) Spin-squeezed state that is not fully polarized.
 For large $N',$ the bound converges to the right-hand side of \EQ{eq:PS_exp},
 represented by the dashed line. In both panels,
 circles correspond to the results of our calculations,
 which are connected by straight lines to guide the eye.}
\label{fig:spinsqueezing}
\end{figure}

We now give the details of the calculations described in
\SEC{sec:spinsqueezingexp}.
We present a simple scheme that we need
to handle large systems.
We do not make calculations
directly for $N=2300$, but we start with smaller systems
and make calculations for larger and larger system sizes.
This is motivated as follows.
First, we can use the output of an optimization for a smaller particle number
as an initial guess for a larger particle number.
Thus, we need fewer steps for the numerical optimization
for large system sizes, which makes our computations faster.
Second, while we are able to carry out the calculation
for the particle number of the experiment, we
also see that we could even extrapolate the results from
the results obtained
for lower particle numbers.
This is useful for future application of our
method to very large systems.

The basic idea is that we transform
the collective quantities from $N$ to a smaller particle number $N'$
using the scaling relation
\bea
\label{eq:cond_exp_spsq}
\exs{J_z}&=& \frac{N'}{2}\alpha,\nonumber\\
\va{J_x}&=&\xi_s^2\frac{N'}{4}\alpha^2.
\eea
We see that for the scaling we consider, for
all $N'$ the bound in \EQ{eq:archetypical} is obtained as \be\label{eq:PS_exp_2}
\frac{\mathcal{F}_{\rm Q}[\varrho_{N'},J_y]}{N'}\geqslant \frac{1}{\xi_s^2}= 6.605 .
\ee
where $\varrho_{N'}$ is a state satisfying Eq. \eqref{eq:cond_exp_spsq}.
Let us first take $\alpha=0.85,$ which is somewhat lower than the experimental value, however,
it helps us to see various characteristics of the method.
At the end of the section we also discuss
the results for other values of $\alpha.$
Based on these ideas, we compute the bound $\mathcal{B}_{\rm sym}(\exs{J_z}_{\varrho_{N'}},\exs{J_x^2}_{\varrho_{N'}}),$ described in \SEC{sec:spinsqueezingexp}, for the quantum Fisher information for
an increasing system size $N'.$

The results are shown in \FIG{fig:spinsqueezing}(a).
The bound obtained in this way is close to the bound
in \EQ{eq:PS_exp} even for small $N'.$
For a larger particle number, i.e., $N'>200,$
it is constant and coincides with the bound in \EQ{eq:PS_exp}.
This also strongly supports the idea
that we could have used the results from small particle numbers to
extrapolate the bound for $N.$
Since for the experimental particle numbers we obtain that
$\mathcal{B}_{\rm sym}(\exs{J_z}_{\varrho_{N}},\exs{J_x^2}_{\varrho_{N}})$ equals the bound in \eqref{eq:archetypical},
we find that for $N'=N$ all three lower bounds in \EQ{eq:bounds} must be equal.
Hence, \EQ{eq:archetypical} is optimal for the experimental
system considered in this section.
Besides, these results also present a strong argument for the correctness
of our approach.

We now give more details of the calculation.
We were able to carry out the optimization up to $N'=2300$
with a usual laptop computer using the MATLAB programming language \footnote{For MATLAB R2015a, see {\tt http://www.mathworks.com}.}.
We started the calculation for each given particle number with the $r_k$ parameters
obtained for the previous simulation with a smaller particle number.

Let us consider a spin-squeezed state that is not fully polarized and
$\alpha=0.5.$ In \FIG{fig:spinsqueezing}(b), we can see that
for small particle numbers we have a bound on $\mathcal{F}_{\rm Q}[\varrho,J_y]$ larger than the one
obtained from \EQ{eq:archetypical}.
Thus for this case we could improve bound \eqref{eq:archetypical}
by assuming symmetry.
On the other hand, for large particle numbers
we approach \EQ{eq:archetypical}.

After seeing the results of the calculations for $\alpha=0.85$ and $\alpha=0.5,$
the question arises, what would the result be for larger $\alpha$,
that is, for even more polarized states? It turns out that
if we choose $\alpha$ larger than $0.85,$ then the convergence
of $\mathcal{F}_{\rm Q}[\varrho_{N'},J_y]/N'$ will be even faster than in \FIG{fig:spinsqueezing}(a),
and for the particle number of the experiment we obtain again
that \EQ{eq:archetypical} is saturated.

Finally, we add a note on a technical detail.
We carried out our calculations with the constraints on $\va{J_x},$ and $\ex{J_z},$
with the additional constraint $\exs{J_x}=0.$
For the experimental particle numbers, one can show that our results are valid even if we
constrain only $\va{J_x}$ and $\ex{J_z},$ and
do not use the $\exs{J_x}=0$ constraint.
This way, in principle, we can only get a bound that is equal to or lower than one we obtained
before. However,
we previously obtained a value identical to the analytical bound, \eqref{eq:archetypical}.
The optimal bound cannot be below the analytic bound, since then the analytic bound would overestimate
the quantum Fisher information, and it would not be a valid bound.
Hence, even an optimization without the
$\exs{J_x}=0$ constraint could not obtain a smaller value than our results.

\subsection{ Experiment creating Dicke states}
\label{sec:dickeexp_appendix}

% SMALL SYSTEM EXPERIMENT MOVED AHEAD

We now give the details for the calculations described in
\SEC{sec:dickeexp_large}. As in \APPENDIX{sec:spinsq_exp_appendix},
 we compute the bound for quantum Fisher information for
an increasing system size $N'$. However, now
we are not able to do the calculation for the experimental particle number, and
we use extrapolation from the results obtained for smaller particle numbers.

First, we transform the
measured second moments to values corresponding to a symmetric system using \EQ{eq:symconst} and \EQ{eq:gamma}.
For our case, $\gamma=1.301.$
In this way, we obtain
\bea
\exs{J_z^2}_{{\rm sym},N}&=&145.69,\nonumber\\
\exs{J_x^2}_{{\rm sym},N}&=&\exs{J_y^2}_{{\rm sym},N}=7.8\times 10^6.%\nonumber\\
\eea

Next, we carry out calculations for symmetric systems.
We consider a scaling that keeps expectation values such that the corresponding
quantum state must be symmetric.
Hence, we use the relations
\bea\label{eq:J2scaling}
\exs{J_z^2}_{{\rm sym},N'}&=&\exs{J_z^2}_{{\rm sym},N},\nonumber\\
\exs{J_x^2}_{{\rm sym},N'}&=&\exs{J_y^2}_{{\rm sym},N'}=\tfrac{1}{2}(\mathcal{J}_{N'}-\exs{J_z^2}_{{\rm sym},N'}),
\eea
where $\mathcal{J}_{N'}$ is defined in \EQ{eq:WJN}.
Note that with \EQ{eq:J2scaling}, $\exs{J_x^2+J_y^2+J_z^2}_{{\rm sym},N'}=\mathcal{J}_{N'}$ holds for all
$N',$ hence the state must be symmetric.
The main characteristics of the scaling relation,
\EQ{eq:J2scaling}, can be summarized as follows. $\exs{J_z^2}_{{\rm sym},N'}$ remains equal to $\exs{J_z^2}_{{\rm sym},N},$
while $\exs{J_x^2}_{{\rm sym},N'}$ and $\exs{J_y^2}_{{\rm sym},N'}$ are chosen such that
they are equal to each other and the state is symmetric.
For large $N$, \EQ{eq:J2scaling} implies a scaling of
$\exs{J_z^2}\sim{\rm const}.$ and $\exs{J_x^2}=\exs{J_z^2}\sim N(N+2)/8.$

Let us now turn to the central quantities of our paper, the lower bounds
on the quantum Fisher information. The quantum Fisher information for the experimentally obtained state $\varrho_{N}$ is bounded from below as
\be\label{eq:FQbound_ineq1}
\mathcal{F}_{\rm Q}[\varrho_{N},J_y]\geqslant \mathcal{B}_{N},
\ee
where $\mathcal{B}_{N}$ denotes a bound
based on $\exs{J_l^2}_{N}$ for $l=x,y,z.$
An analogous relation for the symmetric state $\varrho_{{\rm sym},N'}$ is
\be\label{eq:FQbound_ineq2}
\mathcal{F}_{\rm Q}[\varrho_{{\rm sym},N'},J_y]\geqslant \mathcal{B}_{{\rm sym},N'},
\ee
where $\mathcal{B}_{{\rm sym},N'}$ denotes a bound
based on $\exs{J_l^2}_{{\rm sym},N'}$ for $l=x,y,z.$

A central point in our scheme is that due to
the scaling properties of the system
 we can obtain the value for the particle number $N$ from the value for a smaller particle number $N'$\break as
 \cite{Zhang2014Quantum}
\bea\label{eq:FQsymscaling}
&&\mathcal{B}_{{\rm sym},N}
\approx \frac{\mathcal{J}_N}{\mathcal{J}_{N'}}\mathcal{B}_{{\rm sym},N'},
\eea
which we verify numerically.
Note that for large $N,$ we have
${\mathcal{J}_N}/{\mathcal{J}_{N'}}\sim N^2/(N')^2.$

\begin{figure}[t]
% \vskip0.7cm

\vskip0.5cm
\centerline{
{\epsfxsize6cm \epsffile{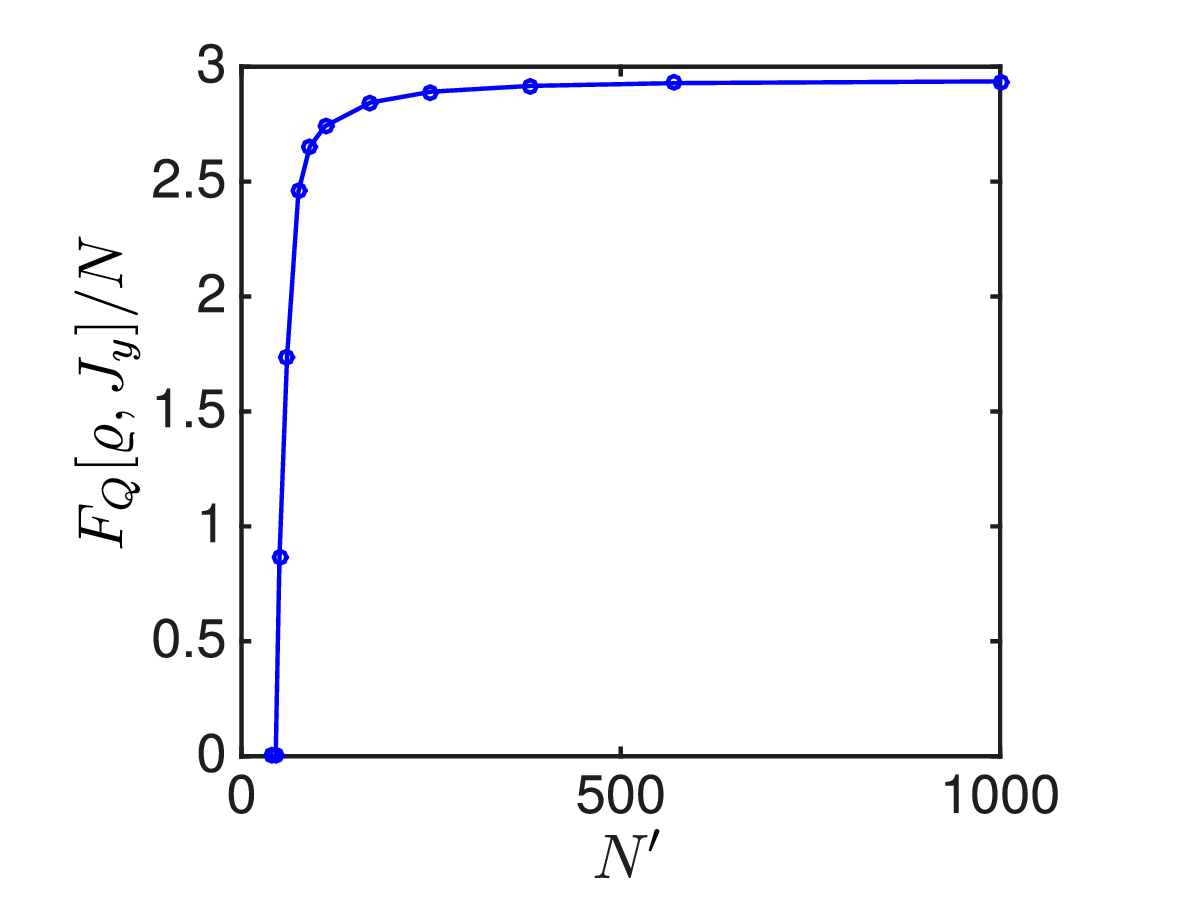}}
}\vskip-.2cm

\caption{(Color online)
 Quantum Fisher information extrapolated to $N=7900$ from calculations with different particle numbers $N'$ in an experiment creating Dicke states. Circles correspond to the results of our calculations,
 which are connected
by straight lines to guide the eye.}
\label{fig:dickeexp}
\end{figure}

As the last step, we have to return from the symmetric system to our real, not fully symmetric one.
Based on \EQ{eq:FQsymscaling}, and assuming that \EQ{eq:FQbound_ineq} is close to being saturated, a relation for the lower bound for the original problem can be obtained from the bound on the
symmetric problem with $N'$ particles
as
\bea\label{eq:FQbound_dicke_final}
\mathcal{B}_N
\approx
\frac{1}{\gamma}
\frac{\mathcal{J}_N}{\mathcal{J}_{N'}}
\mathcal{B}_{{\rm sym},N'}.
\eea
In \FIG{fig:dickeexp}, we plot the right-hand side of \EQ{eq:FQbound_dicke_final} as a function of
$N'.$
We can see that
$\mathcal{B}_{N'}$
is constant
or slightly increasing for $N'>400.$
This is strong evidence that \EQ{eq:FQsymscaling} is valid for large particle numbers.
With this, we arrive at \EQ{eq:bounddicke}.

%\bibliography{Bibliography2}
\bibliography{Bibliography2_fishwerwitness}

\begin{thebibliography}{90}%
\makeatletter
\providecommand \@ifxundefined [1]{%
 \@ifx{#1\undefined}
}%
\providecommand \@ifnum [1]{%
 \ifnum #1\expandafter \@firstoftwo
 \else \expandafter \@secondoftwo
 \fi
}%
\providecommand \@ifx [1]{%
 \ifx #1\expandafter \@firstoftwo
 \else \expandafter \@secondoftwo
 \fi
}%
\providecommand \natexlab [1]{#1}%
\providecommand \enquote  [1]{``#1''}%
\providecommand \bibnamefont  [1]{#1}%
\providecommand \bibfnamefont [1]{#1}%
\providecommand \citenamefont [1]{#1}%
\providecommand \href@noop [0]{\@secondoftwo}%
\providecommand \href [0]{\begingroup \@sanitize@url \@href}%
\providecommand \@href[1]{\@@startlink{#1}\@@href}%
\providecommand \@@href[1]{\endgroup#1\@@endlink}%
\providecommand \@sanitize@url [0]{\catcode `\\12\catcode `\$12\catcode
  `\&12\catcode `\#12\catcode `\^12\catcode `\_12\catcode `\%12\relax}%
\providecommand \@@startlink[1]{}%
\providecommand \@@endlink[0]{}%
\providecommand \url  [0]{\begingroup\@sanitize@url \@url }%
\providecommand \@url [1]{\endgroup\@href {#1}{\urlprefix }}%
\providecommand \urlprefix  [0]{URL }%
\providecommand \Eprint [0]{\href }%
\providecommand \doibase [0]{http://dx.doi.org/}%
\providecommand \selectlanguage [0]{\@gobble}%
\providecommand \bibinfo  [0]{\@secondoftwo}%
\providecommand \bibfield  [0]{\@secondoftwo}%
\providecommand \translation [1]{[#1]}%
\providecommand \BibitemOpen [0]{}%
\providecommand \bibitemStop [0]{}%
\providecommand \bibitemNoStop [0]{.\EOS\space}%
\providecommand \EOS [0]{\spacefactor3000\relax}%
\providecommand \BibitemShut  [1]{\csname bibitem#1\endcsname}%
\let\auto@bib@innerbib\@empty
%</preamble>
\bibitem [{\citenamefont {Horodecki}\ \emph {et~al.}(2009)\citenamefont
  {Horodecki}, \citenamefont {Horodecki}, \citenamefont {Horodecki},\ and\
  \citenamefont {Horodecki}}]{Horodecki2009Quantum}%
  \BibitemOpen
  \bibfield  {author} {\bibinfo {author} {\bibfnamefont {R.}~\bibnamefont
  {Horodecki}}, \bibinfo {author} {\bibfnamefont {P.}~\bibnamefont
  {Horodecki}}, \bibinfo {author} {\bibfnamefont {M.}~\bibnamefont
  {Horodecki}}, \ and\ \bibinfo {author} {\bibfnamefont {K.}~\bibnamefont
  {Horodecki}},\ }\href {\doibase 10.1103/RevModPhys.81.865} {\bibfield
  {journal} {\bibinfo  {journal} {Rev. Mod. Phys.}\ }\textbf {\bibinfo {volume}
  {81}},\ \bibinfo {pages} {865} (\bibinfo {year} {2009})}\BibitemShut
  {NoStop}%
\bibitem [{\citenamefont {G{\"u}hne}\ and\ \citenamefont
  {T{\'o}th}(2009)}]{Guhne2009Entanglement}%
  \BibitemOpen
  \bibfield  {author} {\bibinfo {author} {\bibfnamefont {O.}~\bibnamefont
  {G{\"u}hne}}\ and\ \bibinfo {author} {\bibfnamefont {G.}~\bibnamefont
  {T{\'o}th}},\ }\href {http://doi.org/10.1016/j.physrep.2009.02.004}
  {\bibfield  {journal} {\bibinfo  {journal} {Phys. Rep.}\ }\textbf {\bibinfo
  {volume} {474}},\ \bibinfo {pages} {1} (\bibinfo {year} {2009})}\BibitemShut
  {NoStop}%
\bibitem [{\citenamefont {Pezz\'e}\ and\ \citenamefont
  {Smerzi}(2009)}]{Pezze2009Entanglement}%
  \BibitemOpen
  \bibfield  {author} {\bibinfo {author} {\bibfnamefont {L.}~\bibnamefont
  {Pezz\'e}}\ and\ \bibinfo {author} {\bibfnamefont {A.}~\bibnamefont
  {Smerzi}},\ }\href {\doibase 10.1103/PhysRevLett.102.100401} {\bibfield
  {journal} {\bibinfo  {journal} {Phys. Rev. Lett.}\ }\textbf {\bibinfo
  {volume} {102}},\ \bibinfo {pages} {100401} (\bibinfo {year}
  {2009})}\BibitemShut {NoStop}%
\bibitem [{\citenamefont {Louchet-Chauvet}\ \emph {et~al.}(2010)\citenamefont
  {Louchet-Chauvet}, \citenamefont {Appel}, \citenamefont {Renema},
  \citenamefont {Oblak}, \citenamefont {Kjaergaard},\ and\ \citenamefont
  {Polzik}}]{Louchet-Chauvet2010Entanglement-assisted}%
  \BibitemOpen
  \bibfield  {author} {\bibinfo {author} {\bibfnamefont {A.}~\bibnamefont
  {Louchet-Chauvet}}, \bibinfo {author} {\bibfnamefont {J.}~\bibnamefont
  {Appel}}, \bibinfo {author} {\bibfnamefont {J.~J.}\ \bibnamefont {Renema}},
  \bibinfo {author} {\bibfnamefont {D.}~\bibnamefont {Oblak}}, \bibinfo
  {author} {\bibfnamefont {N.}~\bibnamefont {Kjaergaard}}, \ and\ \bibinfo
  {author} {\bibfnamefont {E.~S.}\ \bibnamefont {Polzik}},\ }\href
  {http://stacks.iop.org/1367-2630/12/i=6/a=065032} {\bibfield  {journal}
  {\bibinfo  {journal} {New J. Phys.}\ }\textbf {\bibinfo {volume} {12}},\
  \bibinfo {pages} {065032} (\bibinfo {year} {2010})}\BibitemShut {NoStop}%
\bibitem [{\citenamefont {Appel}\ \emph {et~al.}(2009)\citenamefont {Appel},
  \citenamefont {Windpassinger}, \citenamefont {Oblak}, \citenamefont {Hoff},
  \citenamefont {Kj{\ae}rgaard},\ and\ \citenamefont
  {Polzik}}]{Appel2009Mesoscopic}%
  \BibitemOpen
  \bibfield  {author} {\bibinfo {author} {\bibfnamefont {J.}~\bibnamefont
  {Appel}}, \bibinfo {author} {\bibfnamefont {P.~J.}\ \bibnamefont
  {Windpassinger}}, \bibinfo {author} {\bibfnamefont {D.}~\bibnamefont
  {Oblak}}, \bibinfo {author} {\bibfnamefont {U.~B.}\ \bibnamefont {Hoff}},
  \bibinfo {author} {\bibfnamefont {N.}~\bibnamefont {Kj{\ae}rgaard}}, \ and\
  \bibinfo {author} {\bibfnamefont {E.~S.}\ \bibnamefont {Polzik}},\ }\href
  {http://doi.org/10.1073/pnas.0901550106} {\bibfield  {journal} {\bibinfo
  {journal} {PNAS}\ }\textbf {\bibinfo {volume} {106}},\ \bibinfo {pages}
  {10960} (\bibinfo {year} {2009})}\BibitemShut {NoStop}%
\bibitem [{\citenamefont {Riedel}\ \emph {et~al.}(2010)\citenamefont {Riedel},
  \citenamefont {B{\"o}hi}, \citenamefont {Li}, \citenamefont {H{\"a}nsch},
  \citenamefont {Sinatra},\ and\ \citenamefont
  {Treutlein}}]{Riedel2010Atom-chip-based}%
  \BibitemOpen
  \bibfield  {author} {\bibinfo {author} {\bibfnamefont {M.~F.}\ \bibnamefont
  {Riedel}}, \bibinfo {author} {\bibfnamefont {P.}~\bibnamefont {B{\"o}hi}},
  \bibinfo {author} {\bibfnamefont {Y.}~\bibnamefont {Li}}, \bibinfo {author}
  {\bibfnamefont {T.~W.}\ \bibnamefont {H{\"a}nsch}}, \bibinfo {author}
  {\bibfnamefont {A.}~\bibnamefont {Sinatra}}, \ and\ \bibinfo {author}
  {\bibfnamefont {P.}~\bibnamefont {Treutlein}},\ }\href {\doibase
  10.1038/nature08988} {\bibfield  {journal} {\bibinfo  {journal} {Nature}\
  }\textbf {\bibinfo {volume} {464}},\ \bibinfo {pages} {1170} (\bibinfo {year}
  {2010})}\BibitemShut {NoStop}%
\bibitem [{\citenamefont {Gross}\ \emph {et~al.}(2010)\citenamefont {Gross},
  \citenamefont {Zibold}, \citenamefont {Nicklas}, \citenamefont {Esteve},\
  and\ \citenamefont {Oberthaler}}]{Gross2010Nonlinear}%
  \BibitemOpen
  \bibfield  {author} {\bibinfo {author} {\bibfnamefont {C.}~\bibnamefont
  {Gross}}, \bibinfo {author} {\bibfnamefont {T.}~\bibnamefont {Zibold}},
  \bibinfo {author} {\bibfnamefont {E.}~\bibnamefont {Nicklas}}, \bibinfo
  {author} {\bibfnamefont {J.}~\bibnamefont {Esteve}}, \ and\ \bibinfo {author}
  {\bibfnamefont {M.~K.}\ \bibnamefont {Oberthaler}},\ }\href
  {http://www.nature.com/nature/journal/v464/n7292/abs/nature08919.html}
  {\bibfield  {journal} {\bibinfo  {journal} {Nature}\ }\textbf {\bibinfo
  {volume} {464}},\ \bibinfo {pages} {1165} (\bibinfo {year}
  {2010})}\BibitemShut {NoStop}%
\bibitem [{\citenamefont {L{\"u}cke}\ \emph {et~al.}(2011)\citenamefont
  {L{\"u}cke}, \citenamefont {Scherer}, \citenamefont {Kruse}, \citenamefont
  {Pezz\'e}, \citenamefont {Deuretzbacher}, \citenamefont {Hyllus},
  \citenamefont {Peise}, \citenamefont {Ertmer}, \citenamefont {Arlt},
  \citenamefont {Santos}, \citenamefont {Smerzi},\ and\ \citenamefont
  {Klempt}}]{Lucke2011Twin}%
  \BibitemOpen
  \bibfield  {author} {\bibinfo {author} {\bibfnamefont {B.}~\bibnamefont
  {L{\"u}cke}}, \bibinfo {author} {\bibfnamefont {M.}~\bibnamefont {Scherer}},
  \bibinfo {author} {\bibfnamefont {J.}~\bibnamefont {Kruse}}, \bibinfo
  {author} {\bibfnamefont {L.}~\bibnamefont {Pezz\'e}}, \bibinfo {author}
  {\bibfnamefont {F.}~\bibnamefont {Deuretzbacher}}, \bibinfo {author}
  {\bibfnamefont {P.}~\bibnamefont {Hyllus}}, \bibinfo {author} {\bibfnamefont
  {J.}~\bibnamefont {Peise}}, \bibinfo {author} {\bibfnamefont
  {W.}~\bibnamefont {Ertmer}}, \bibinfo {author} {\bibfnamefont
  {J.}~\bibnamefont {Arlt}}, \bibinfo {author} {\bibfnamefont {L.}~\bibnamefont
  {Santos}}, \bibinfo {author} {\bibfnamefont {A.}~\bibnamefont {Smerzi}}, \
  and\ \bibinfo {author} {\bibfnamefont {C.}~\bibnamefont {Klempt}},\ }\href
  {\doibase 10.1126/science.1208798} {\bibfield  {journal} {\bibinfo  {journal}
  {Science}\ }\textbf {\bibinfo {volume} {334}},\ \bibinfo {pages} {773}
  (\bibinfo {year} {2011})}\BibitemShut {NoStop}%
\bibitem [{\citenamefont {Strobel}\ \emph {et~al.}(2014)\citenamefont
  {Strobel}, \citenamefont {Muessel}, \citenamefont {Linnemann}, \citenamefont
  {Zibold}, \citenamefont {Hume}, \citenamefont {Pezz{\'e}}, \citenamefont
  {Smerzi},\ and\ \citenamefont {Oberthaler}}]{Strobel2014Fisher}%
  \BibitemOpen
  \bibfield  {author} {\bibinfo {author} {\bibfnamefont {H.}~\bibnamefont
  {Strobel}}, \bibinfo {author} {\bibfnamefont {W.}~\bibnamefont {Muessel}},
  \bibinfo {author} {\bibfnamefont {D.}~\bibnamefont {Linnemann}}, \bibinfo
  {author} {\bibfnamefont {T.}~\bibnamefont {Zibold}}, \bibinfo {author}
  {\bibfnamefont {D.~B.}\ \bibnamefont {Hume}}, \bibinfo {author}
  {\bibfnamefont {L.}~\bibnamefont {Pezz{\'e}}}, \bibinfo {author}
  {\bibfnamefont {A.}~\bibnamefont {Smerzi}}, \ and\ \bibinfo {author}
  {\bibfnamefont {M.~K.}\ \bibnamefont {Oberthaler}},\ }\href
  {http://doi.org/10.1126/science.1250147} {\bibfield  {journal} {\bibinfo
  {journal} {Science}\ }\textbf {\bibinfo {volume} {345}},\ \bibinfo {pages}
  {424} (\bibinfo {year} {2014})}\BibitemShut {NoStop}%
\bibitem [{\citenamefont {Hyllus}\ \emph {et~al.}(2010)\citenamefont {Hyllus},
  \citenamefont {G\"uhne},\ and\ \citenamefont {Smerzi}}]{Hyllus2010Not}%
  \BibitemOpen
  \bibfield  {author} {\bibinfo {author} {\bibfnamefont {P.}~\bibnamefont
  {Hyllus}}, \bibinfo {author} {\bibfnamefont {O.}~\bibnamefont {G\"uhne}}, \
  and\ \bibinfo {author} {\bibfnamefont {A.}~\bibnamefont {Smerzi}},\ }\href
  {\doibase 10.1103/PhysRevA.82.012337} {\bibfield  {journal} {\bibinfo
  {journal} {Phys. Rev. A}\ }\textbf {\bibinfo {volume} {82}},\ \bibinfo
  {pages} {012337} (\bibinfo {year} {2010})}\BibitemShut {NoStop}%
\bibitem [{\citenamefont {Giovannetti}\ \emph {et~al.}(2004)\citenamefont
  {Giovannetti}, \citenamefont {Lloyd},\ and\ \citenamefont
  {Maccone}}]{Giovannetti2004Quantum-Enhanced}%
  \BibitemOpen
  \bibfield  {author} {\bibinfo {author} {\bibfnamefont {V.}~\bibnamefont
  {Giovannetti}}, \bibinfo {author} {\bibfnamefont {S.}~\bibnamefont {Lloyd}},
  \ and\ \bibinfo {author} {\bibfnamefont {L.}~\bibnamefont {Maccone}},\ }\href
  {\doibase 10.1126/science.1104149} {\bibfield  {journal} {\bibinfo  {journal}
  {Science}\ }\textbf {\bibinfo {volume} {306}},\ \bibinfo {pages} {1330}
  (\bibinfo {year} {2004})}\BibitemShut {NoStop}%
\bibitem [{\citenamefont {Paris}(2009)}]{Paris2009QUANTUM}%
  \BibitemOpen
  \bibfield  {author} {\bibinfo {author} {\bibfnamefont {M.~G.~A.}\
  \bibnamefont {Paris}},\ }\href {\doibase 10.1142/S0219749909004839}
  {\bibfield  {journal} {\bibinfo  {journal} {Int. J. Quant. Inf.}\ }\textbf
  {\bibinfo {volume} {07}},\ \bibinfo {pages} {125} (\bibinfo {year}
  {2009})}\BibitemShut {NoStop}%
\bibitem [{\citenamefont {Demkowicz-Dobrzanski}\ \emph
  {et~al.}(2015)\citenamefont {Demkowicz-Dobrzanski}, \citenamefont {Jarzyna},\
  and\ \citenamefont {Kolodynski}}]{Demkowicz-Dobrzanski2014Quantum}%
  \BibitemOpen
  \bibfield  {author} {\bibinfo {author} {\bibfnamefont {R.}~\bibnamefont
  {Demkowicz-Dobrzanski}}, \bibinfo {author} {\bibfnamefont {M.}~\bibnamefont
  {Jarzyna}}, \ and\ \bibinfo {author} {\bibfnamefont {J.}~\bibnamefont
  {Kolodynski}},\ }\href {\doibase http://dx.doi.org/10.1016/bs.po.2015.02.003}
  {\bibfield  {journal} {\bibinfo  {journal} {Prog. Optics}\ }\textbf {\bibinfo
  {volume} {60}},\ \bibinfo {pages} {345 } (\bibinfo {year} {2015})},\ \Eprint
  {http://arxiv.org/abs/arXiv:1405.7703} {arXiv:1405.7703} \BibitemShut
  {NoStop}%
\bibitem [{\citenamefont {Pezze}\ and\ \citenamefont
  {Smerzi}(2014)}]{Pezze2014Quantum}%
  \BibitemOpen
  \bibfield  {author} {\bibinfo {author} {\bibfnamefont {L.}~\bibnamefont
  {Pezze}}\ and\ \bibinfo {author} {\bibfnamefont {A.}~\bibnamefont {Smerzi}},\
  }in\ \href@noop {} {\emph {\bibinfo {booktitle} {Atom Interferometry (Proc.
  Int. School of Physics 'Enrico Fermi', Course 188, Varenna)}}},\ \bibinfo
  {editor} {edited by\ \bibinfo {editor} {\bibfnamefont {G.}~\bibnamefont
  {Tino}}\ and\ \bibinfo {editor} {\bibfnamefont {M.}~\bibnamefont
  {Kasevich}}}\ (\bibinfo  {publisher} {IOS Press, Amsterdam},\ \bibinfo {year}
  {2014})\ pp.\ \bibinfo {pages} {691--741},\ \Eprint
  {http://arxiv.org/abs/arXiv:1411.5164} {arXiv:1411.5164} \BibitemShut
  {NoStop}%
\bibitem [{\citenamefont {Helstrom}(1976)}]{Helstrom1976Quantum}%
  \BibitemOpen
  \bibfield  {author} {\bibinfo {author} {\bibfnamefont {C.}~\bibnamefont
  {Helstrom}},\ }\href {http://books.google.es/books?id=fv9SAAAAMAAJ} {\emph
  {\bibinfo {title} {Quantum Detection and Estimation Theory}}}\ (\bibinfo
  {publisher} {Academic Press, New York},\ \bibinfo {year} {1976})\BibitemShut
  {NoStop}%
\bibitem [{\citenamefont {Holevo}(1982)}]{Holevo1982Probabilistic}%
  \BibitemOpen
  \bibfield  {author} {\bibinfo {author} {\bibfnamefont {A.}~\bibnamefont
  {Holevo}},\ }\href@noop {} {\emph {\bibinfo {title} {Probabilistic and
  Statistical Aspects of Quantum Theory}}}\ (\bibinfo  {publisher}
  {North-Holland, Amsterdam},\ \bibinfo {year} {1982})\BibitemShut {NoStop}%
\bibitem [{\citenamefont {Braunstein}\ and\ \citenamefont
  {Caves}(1994)}]{Braunstein1994Statistical}%
  \BibitemOpen
  \bibfield  {author} {\bibinfo {author} {\bibfnamefont {S.~L.}\ \bibnamefont
  {Braunstein}}\ and\ \bibinfo {author} {\bibfnamefont {C.~M.}\ \bibnamefont
  {Caves}},\ }\href {\doibase 10.1103/PhysRevLett.72.3439} {\bibfield
  {journal} {\bibinfo  {journal} {Phys. Rev. Lett.}\ }\textbf {\bibinfo
  {volume} {72}},\ \bibinfo {pages} {3439} (\bibinfo {year}
  {1994})}\BibitemShut {NoStop}%
\bibitem [{\citenamefont {Petz}(2008)}]{Petz2008Quantum}%
  \BibitemOpen
  \bibfield  {author} {\bibinfo {author} {\bibfnamefont {D.}~\bibnamefont
  {Petz}},\ }\href@noop {} {\emph {\bibinfo {title} {Quantum information theory
  and quantum statistics}}}\ (\bibinfo  {publisher} {Springer, Berlin,
  Heilderberg},\ \bibinfo {year} {2008})\BibitemShut {NoStop}%
\bibitem [{\citenamefont {Braunstein}\ \emph {et~al.}(1996)\citenamefont
  {Braunstein}, \citenamefont {Caves},\ and\ \citenamefont
  {Milburn}}]{Braunstein1996Generalized}%
  \BibitemOpen
  \bibfield  {author} {\bibinfo {author} {\bibfnamefont {S.~L.}\ \bibnamefont
  {Braunstein}}, \bibinfo {author} {\bibfnamefont {C.~M.}\ \bibnamefont
  {Caves}}, \ and\ \bibinfo {author} {\bibfnamefont {G.~J.}\ \bibnamefont
  {Milburn}},\ }\href {https://doi.org/10.1006/aphy.1996.0040} {\bibfield
  {journal} {\bibinfo  {journal} {Ann. Phys.}\ }\textbf {\bibinfo {volume}
  {247}},\ \bibinfo {pages} {135} (\bibinfo {year} {1996})}\BibitemShut
  {NoStop}%
\bibitem [{\citenamefont {Hyllus}\ \emph {et~al.}(2012)\citenamefont {Hyllus},
  \citenamefont {Laskowski}, \citenamefont {Krischek}, \citenamefont
  {Schwemmer}, \citenamefont {Wieczorek}, \citenamefont {Weinfurter},
  \citenamefont {Pezz\'e},\ and\ \citenamefont {Smerzi}}]{Hyllus2012Fisher}%
  \BibitemOpen
  \bibfield  {author} {\bibinfo {author} {\bibfnamefont {P.}~\bibnamefont
  {Hyllus}}, \bibinfo {author} {\bibfnamefont {W.}~\bibnamefont {Laskowski}},
  \bibinfo {author} {\bibfnamefont {R.}~\bibnamefont {Krischek}}, \bibinfo
  {author} {\bibfnamefont {C.}~\bibnamefont {Schwemmer}}, \bibinfo {author}
  {\bibfnamefont {W.}~\bibnamefont {Wieczorek}}, \bibinfo {author}
  {\bibfnamefont {H.}~\bibnamefont {Weinfurter}}, \bibinfo {author}
  {\bibfnamefont {L.}~\bibnamefont {Pezz\'e}}, \ and\ \bibinfo {author}
  {\bibfnamefont {A.}~\bibnamefont {Smerzi}},\ }\href {\doibase
  10.1103/PhysRevA.85.022321} {\bibfield  {journal} {\bibinfo  {journal} {Phys.
  Rev. A}\ }\textbf {\bibinfo {volume} {85}},\ \bibinfo {pages} {022321}
  (\bibinfo {year} {2012})}\BibitemShut {NoStop}%
\bibitem [{\citenamefont {T\'oth}(2012)}]{Toth2012Multipartite}%
  \BibitemOpen
  \bibfield  {author} {\bibinfo {author} {\bibfnamefont {G.}~\bibnamefont
  {T\'oth}},\ }\href {\doibase 10.1103/PhysRevA.85.022322} {\bibfield
  {journal} {\bibinfo  {journal} {Phys. Rev. A}\ }\textbf {\bibinfo {volume}
  {85}},\ \bibinfo {pages} {022322} (\bibinfo {year} {2012})}\BibitemShut
  {NoStop}%
\bibitem [{\citenamefont {Krischek}\ \emph {et~al.}(2011)\citenamefont
  {Krischek}, \citenamefont {Schwemmer}, \citenamefont {Wieczorek},
  \citenamefont {Weinfurter}, \citenamefont {Hyllus}, \citenamefont {Pezz\'e},\
  and\ \citenamefont {Smerzi}}]{Krischek2011Useful}%
  \BibitemOpen
  \bibfield  {author} {\bibinfo {author} {\bibfnamefont {R.}~\bibnamefont
  {Krischek}}, \bibinfo {author} {\bibfnamefont {C.}~\bibnamefont {Schwemmer}},
  \bibinfo {author} {\bibfnamefont {W.}~\bibnamefont {Wieczorek}}, \bibinfo
  {author} {\bibfnamefont {H.}~\bibnamefont {Weinfurter}}, \bibinfo {author}
  {\bibfnamefont {P.}~\bibnamefont {Hyllus}}, \bibinfo {author} {\bibfnamefont
  {L.}~\bibnamefont {Pezz\'e}}, \ and\ \bibinfo {author} {\bibfnamefont
  {A.}~\bibnamefont {Smerzi}},\ }\href {\doibase
  10.1103/PhysRevLett.107.080504} {\bibfield  {journal} {\bibinfo  {journal}
  {Phys. Rev. Lett.}\ }\textbf {\bibinfo {volume} {107}},\ \bibinfo {pages}
  {080504} (\bibinfo {year} {2011})}\BibitemShut {NoStop}%
\bibitem [{\citenamefont {Hauke}\ \emph {et~al.}(2016)\citenamefont {Hauke},
  \citenamefont {Heyl}, \citenamefont {Tagliacozzo},\ and\ \citenamefont
  {Zoller}}]{Hauke2016Measuring}%
  \BibitemOpen
  \bibfield  {author} {\bibinfo {author} {\bibfnamefont {P.}~\bibnamefont
  {Hauke}}, \bibinfo {author} {\bibfnamefont {M.}~\bibnamefont {Heyl}},
  \bibinfo {author} {\bibfnamefont {L.}~\bibnamefont {Tagliacozzo}}, \ and\
  \bibinfo {author} {\bibfnamefont {P.}~\bibnamefont {Zoller}},\ }\href
  {\doibase 10.1038/nphys3700} {\bibfield  {journal} {\bibinfo  {journal} {Nat.
  Phys.}\ }\textbf {\bibinfo {volume} {12}},\ \bibinfo {pages} {778} (\bibinfo
  {year} {2016})}\BibitemShut {NoStop}%
\bibitem [{\citenamefont {Shitara}\ and\ \citenamefont
  {Ueda}(2016)}]{Shitara2016Determining}%
  \BibitemOpen
  \bibfield  {author} {\bibinfo {author} {\bibfnamefont {T.}~\bibnamefont
  {Shitara}}\ and\ \bibinfo {author} {\bibfnamefont {M.}~\bibnamefont {Ueda}},\
  }\href {\doibase 10.1103/PhysRevA.94.062316} {\bibfield  {journal} {\bibinfo
  {journal} {Phys. Rev. A}\ }\textbf {\bibinfo {volume} {94}},\ \bibinfo
  {pages} {062316} (\bibinfo {year} {2016})}\BibitemShut {NoStop}%
\bibitem [{\citenamefont {Pezze}\ \emph {et~al.}(2016)\citenamefont {Pezze},
  \citenamefont {Li}, \citenamefont {Li},\ and\ \citenamefont
  {Smerzi}}]{Pezze2016Witnessing}%
  \BibitemOpen
  \bibfield  {author} {\bibinfo {author} {\bibfnamefont {L.}~\bibnamefont
  {Pezze}}, \bibinfo {author} {\bibfnamefont {Y.}~\bibnamefont {Li}}, \bibinfo
  {author} {\bibfnamefont {W.}~\bibnamefont {Li}}, \ and\ \bibinfo {author}
  {\bibfnamefont {A.}~\bibnamefont {Smerzi}},\ }\href {\doibase
  10.1073/pnas.1603346113} {\bibfield  {journal} {\bibinfo  {journal} {PNAS}\
  }\textbf {\bibinfo {volume} {113}},\ \bibinfo {pages} {11459} (\bibinfo
  {year} {2016})}\BibitemShut {NoStop}%
\bibitem [{\citenamefont {Fr\"owis}\ \emph {et~al.}(2016)\citenamefont
  {Fr\"owis}, \citenamefont {Sekatski},\ and\ \citenamefont
  {D\"ur}}]{Frowis2016Detecting}%
  \BibitemOpen
  \bibfield  {author} {\bibinfo {author} {\bibfnamefont {F.}~\bibnamefont
  {Fr\"owis}}, \bibinfo {author} {\bibfnamefont {P.}~\bibnamefont {Sekatski}},
  \ and\ \bibinfo {author} {\bibfnamefont {W.}~\bibnamefont {D\"ur}},\ }\href
  {\doibase 10.1103/PhysRevLett.116.090801} {\bibfield  {journal} {\bibinfo
  {journal} {Phys. Rev. Lett.}\ }\textbf {\bibinfo {volume} {116}},\ \bibinfo
  {pages} {090801} (\bibinfo {year} {2016})}\BibitemShut {NoStop}%
\bibitem [{\citenamefont {Dicke}(1954)}]{Dicke1954Coherence}%
  \BibitemOpen
  \bibfield  {author} {\bibinfo {author} {\bibfnamefont {R.~H.}\ \bibnamefont
  {Dicke}},\ }\href {\doibase 10.1103/PhysRev.93.99} {\bibfield  {journal}
  {\bibinfo  {journal} {Phys. Rev.}\ }\textbf {\bibinfo {volume} {93}},\
  \bibinfo {pages} {99} (\bibinfo {year} {1954})}\BibitemShut {NoStop}%
\bibitem [{\citenamefont {Zhang}\ and\ \citenamefont
  {Duan}(2014)}]{Zhang2014Quantum}%
  \BibitemOpen
  \bibfield  {author} {\bibinfo {author} {\bibfnamefont {Z.}~\bibnamefont
  {Zhang}}\ and\ \bibinfo {author} {\bibfnamefont {L.~M.}\ \bibnamefont
  {Duan}},\ }\href {http://stacks.iop.org/1367-2630/16/i=10/a=103037}
  {\bibfield  {journal} {\bibinfo  {journal} {New J. Phys.}\ }\textbf {\bibinfo
  {volume} {16}},\ \bibinfo {pages} {103037} (\bibinfo {year}
  {2014})}\BibitemShut {NoStop}%
\bibitem [{\citenamefont {Fr\"owis}\ \emph {et~al.}(2015)\citenamefont
  {Fr\"owis}, \citenamefont {Schmied},\ and\ \citenamefont
  {Gisin}}]{Frowis2014Tighter}%
  \BibitemOpen
  \bibfield  {author} {\bibinfo {author} {\bibfnamefont {F.}~\bibnamefont
  {Fr\"owis}}, \bibinfo {author} {\bibfnamefont {R.}~\bibnamefont {Schmied}}, \
  and\ \bibinfo {author} {\bibfnamefont {N.}~\bibnamefont {Gisin}},\ }\href
  {\doibase 10.1103/PhysRevA.92.012102} {\bibfield  {journal} {\bibinfo
  {journal} {Phys. Rev. A}\ }\textbf {\bibinfo {volume} {92}},\ \bibinfo
  {pages} {012102} (\bibinfo {year} {2015})}\BibitemShut {NoStop}%
\bibitem [{\citenamefont {Apellaniz}\ \emph {et~al.}(2015)\citenamefont
  {Apellaniz}, \citenamefont {L\"ucke}, \citenamefont {Peise}, \citenamefont
  {Klempt},\ and\ \citenamefont {T\'oth}}]{Apellaniz2015Verifying}%
  \BibitemOpen
  \bibfield  {author} {\bibinfo {author} {\bibfnamefont {I.}~\bibnamefont
  {Apellaniz}}, \bibinfo {author} {\bibfnamefont {B.}~\bibnamefont {L\"ucke}},
  \bibinfo {author} {\bibfnamefont {J.}~\bibnamefont {Peise}}, \bibinfo
  {author} {\bibfnamefont {C.}~\bibnamefont {Klempt}}, \ and\ \bibinfo {author}
  {\bibfnamefont {G.}~\bibnamefont {T\'oth}},\ }\href
  {http://stacks.iop.org/1367-2630/17/i=8/a=083027} {\bibfield  {journal}
  {\bibinfo  {journal} {New J. Phys.}\ }\textbf {\bibinfo {volume} {17}},\
  \bibinfo {pages} {083027} (\bibinfo {year} {2015})}\BibitemShut {NoStop}%
\bibitem [{\citenamefont {Oudot}\ \emph {et~al.}(2015)\citenamefont {Oudot},
  \citenamefont {Sekatski}, \citenamefont {Fr\"{o}wis}, \citenamefont {Gisin},\
  and\ \citenamefont {Sangouard}}]{Oudot2015Two-mode}%
  \BibitemOpen
  \bibfield  {author} {\bibinfo {author} {\bibfnamefont {E.}~\bibnamefont
  {Oudot}}, \bibinfo {author} {\bibfnamefont {P.}~\bibnamefont {Sekatski}},
  \bibinfo {author} {\bibfnamefont {F.}~\bibnamefont {Fr\"{o}wis}}, \bibinfo
  {author} {\bibfnamefont {N.}~\bibnamefont {Gisin}}, \ and\ \bibinfo {author}
  {\bibfnamefont {N.}~\bibnamefont {Sangouard}},\ }\href {\doibase
  10.1364/JOSAB.32.002190} {\bibfield  {journal} {\bibinfo  {journal} {J. Opt.
  Soc. Am. B}\ }\textbf {\bibinfo {volume} {32}},\ \bibinfo {pages} {2190}
  (\bibinfo {year} {2015})}\BibitemShut {NoStop}%
\bibitem [{\citenamefont {Kitagawa}\ and\ \citenamefont
  {Ueda}(1993)}]{Kitagawa1993Squeezed}%
  \BibitemOpen
  \bibfield  {author} {\bibinfo {author} {\bibfnamefont {M.}~\bibnamefont
  {Kitagawa}}\ and\ \bibinfo {author} {\bibfnamefont {M.}~\bibnamefont
  {Ueda}},\ }\href {\doibase 10.1103/PhysRevA.47.5138} {\bibfield  {journal}
  {\bibinfo  {journal} {Phys. Rev. A}\ }\textbf {\bibinfo {volume} {47}},\
  \bibinfo {pages} {5138} (\bibinfo {year} {1993})}\BibitemShut {NoStop}%
\bibitem [{\citenamefont {Wineland}\ \emph {et~al.}(1994)\citenamefont
  {Wineland}, \citenamefont {Bollinger}, \citenamefont {Itano},\ and\
  \citenamefont {Heinzen}}]{Wineland1994Squeezed}%
  \BibitemOpen
  \bibfield  {author} {\bibinfo {author} {\bibfnamefont {D.~J.}\ \bibnamefont
  {Wineland}}, \bibinfo {author} {\bibfnamefont {J.~J.}\ \bibnamefont
  {Bollinger}}, \bibinfo {author} {\bibfnamefont {W.~M.}\ \bibnamefont
  {Itano}}, \ and\ \bibinfo {author} {\bibfnamefont {D.~J.}\ \bibnamefont
  {Heinzen}},\ }\href {\doibase 10.1103/PhysRevA.50.67} {\bibfield  {journal}
  {\bibinfo  {journal} {Phys. Rev. A}\ }\textbf {\bibinfo {volume} {50}},\
  \bibinfo {pages} {67} (\bibinfo {year} {1994})}\BibitemShut {NoStop}%
\bibitem [{\citenamefont {Greenberger}\ \emph {et~al.}(1990)\citenamefont
  {Greenberger}, \citenamefont {Horne}, \citenamefont {Shimony},\ and\
  \citenamefont {Zeilinger}}]{Greenberger1990Bells}%
  \BibitemOpen
  \bibfield  {author} {\bibinfo {author} {\bibfnamefont {D.~M.}\ \bibnamefont
  {Greenberger}}, \bibinfo {author} {\bibfnamefont {M.~A.}\ \bibnamefont
  {Horne}}, \bibinfo {author} {\bibfnamefont {A.}~\bibnamefont {Shimony}}, \
  and\ \bibinfo {author} {\bibfnamefont {A.}~\bibnamefont {Zeilinger}},\ }\href
  {http://doi.org/10.1119/1.16243} {\bibfield  {journal} {\bibinfo  {journal}
  {Am. J. Phys.}\ }\textbf {\bibinfo {volume} {58}},\ \bibinfo {pages} {1131}
  (\bibinfo {year} {1990})}\BibitemShut {NoStop}%
\bibitem [{\citenamefont {Bouwmeester}\ \emph {et~al.}(1999)\citenamefont
  {Bouwmeester}, \citenamefont {Pan}, \citenamefont {Daniell}, \citenamefont
  {Weinfurter},\ and\ \citenamefont {Zeilinger}}]{Bouwmeester1999Observation}%
  \BibitemOpen
  \bibfield  {author} {\bibinfo {author} {\bibfnamefont {D.}~\bibnamefont
  {Bouwmeester}}, \bibinfo {author} {\bibfnamefont {J.-W.}\ \bibnamefont
  {Pan}}, \bibinfo {author} {\bibfnamefont {M.}~\bibnamefont {Daniell}},
  \bibinfo {author} {\bibfnamefont {H.}~\bibnamefont {Weinfurter}}, \ and\
  \bibinfo {author} {\bibfnamefont {A.}~\bibnamefont {Zeilinger}},\ }\href
  {\doibase 10.1103/PhysRevLett.82.1345} {\bibfield  {journal} {\bibinfo
  {journal} {Phys. Rev. Lett.}\ }\textbf {\bibinfo {volume} {82}},\ \bibinfo
  {pages} {1345} (\bibinfo {year} {1999})}\BibitemShut {NoStop}%
\bibitem [{\citenamefont {Pan}\ \emph {et~al.}(2000)\citenamefont {Pan},
  \citenamefont {Bouwmeester}, \citenamefont {Daniell}, \citenamefont
  {Weinfurter},\ and\ \citenamefont {Zeilinger}}]{Pan2000Experimental}%
  \BibitemOpen
  \bibfield  {author} {\bibinfo {author} {\bibfnamefont {J.-W.}\ \bibnamefont
  {Pan}}, \bibinfo {author} {\bibfnamefont {D.}~\bibnamefont {Bouwmeester}},
  \bibinfo {author} {\bibfnamefont {M.}~\bibnamefont {Daniell}}, \bibinfo
  {author} {\bibfnamefont {H.}~\bibnamefont {Weinfurter}}, \ and\ \bibinfo
  {author} {\bibfnamefont {A.}~\bibnamefont {Zeilinger}},\ }\href
  {http://www.nature.com/nature/journal/v403/n6769/abs/403515a0.html}
  {\bibfield  {journal} {\bibinfo  {journal} {Nature}\ }\textbf {\bibinfo
  {volume} {403}},\ \bibinfo {pages} {515} (\bibinfo {year}
  {2000})}\BibitemShut {NoStop}%
\bibitem [{\citenamefont {Zhao}\ \emph {et~al.}(2003)\citenamefont {Zhao},
  \citenamefont {Yang}, \citenamefont {Chen}, \citenamefont {Zhang},
  \citenamefont {\ifmmode~\dot{Z}\else \.{Z}\fi{}ukowski},\ and\ \citenamefont
  {Pan}}]{Zhao2003Experimental}%
  \BibitemOpen
  \bibfield  {author} {\bibinfo {author} {\bibfnamefont {Z.}~\bibnamefont
  {Zhao}}, \bibinfo {author} {\bibfnamefont {T.}~\bibnamefont {Yang}}, \bibinfo
  {author} {\bibfnamefont {Y.-A.}\ \bibnamefont {Chen}}, \bibinfo {author}
  {\bibfnamefont {A.-N.}\ \bibnamefont {Zhang}}, \bibinfo {author}
  {\bibfnamefont {M.}~\bibnamefont {\ifmmode~\dot{Z}\else \.{Z}\fi{}ukowski}},
  \ and\ \bibinfo {author} {\bibfnamefont {J.-W.}\ \bibnamefont {Pan}},\ }\href
  {\doibase 10.1103/PhysRevLett.91.180401} {\bibfield  {journal} {\bibinfo
  {journal} {Phys. Rev. Lett.}\ }\textbf {\bibinfo {volume} {91}},\ \bibinfo
  {pages} {180401} (\bibinfo {year} {2003})}\BibitemShut {NoStop}%
\bibitem [{\citenamefont {Lu}\ \emph {et~al.}(2007)\citenamefont {Lu},
  \citenamefont {Zhou}, \citenamefont {G{\"u}hne}, \citenamefont {Gao},
  \citenamefont {Zhang}, \citenamefont {Yuan}, \citenamefont {Goebel},
  \citenamefont {Yang},\ and\ \citenamefont {Pan}}]{Lu2007Experimental}%
  \BibitemOpen
  \bibfield  {author} {\bibinfo {author} {\bibfnamefont {C.-Y.}\ \bibnamefont
  {Lu}}, \bibinfo {author} {\bibfnamefont {X.-Q.}\ \bibnamefont {Zhou}},
  \bibinfo {author} {\bibfnamefont {O.}~\bibnamefont {G{\"u}hne}}, \bibinfo
  {author} {\bibfnamefont {W.-B.}\ \bibnamefont {Gao}}, \bibinfo {author}
  {\bibfnamefont {J.}~\bibnamefont {Zhang}}, \bibinfo {author} {\bibfnamefont
  {Z.-S.}\ \bibnamefont {Yuan}}, \bibinfo {author} {\bibfnamefont
  {A.}~\bibnamefont {Goebel}}, \bibinfo {author} {\bibfnamefont
  {T.}~\bibnamefont {Yang}}, \ and\ \bibinfo {author} {\bibfnamefont {J.-W.}\
  \bibnamefont {Pan}},\ }\href {\doibase 10.1038/nphys507} {\bibfield
  {journal} {\bibinfo  {journal} {Nat. Phys.}\ }\textbf {\bibinfo {volume}
  {3}},\ \bibinfo {pages} {91} (\bibinfo {year} {2007})}\BibitemShut {NoStop}%
\bibitem [{\citenamefont {Gao}\ \emph {et~al.}(2010)\citenamefont {Gao},
  \citenamefont {Lu}, \citenamefont {Yao}, \citenamefont {Xu}, \citenamefont
  {G{\"u}hne}, \citenamefont {Goebel}, \citenamefont {Chen}, \citenamefont
  {Peng}, \citenamefont {Chen},\ and\ \citenamefont
  {Pan}}]{Gao2010Experimental}%
  \BibitemOpen
  \bibfield  {author} {\bibinfo {author} {\bibfnamefont {W.-B.}\ \bibnamefont
  {Gao}}, \bibinfo {author} {\bibfnamefont {C.-Y.}\ \bibnamefont {Lu}},
  \bibinfo {author} {\bibfnamefont {X.-C.}\ \bibnamefont {Yao}}, \bibinfo
  {author} {\bibfnamefont {P.}~\bibnamefont {Xu}}, \bibinfo {author}
  {\bibfnamefont {O.}~\bibnamefont {G{\"u}hne}}, \bibinfo {author}
  {\bibfnamefont {A.}~\bibnamefont {Goebel}}, \bibinfo {author} {\bibfnamefont
  {Y.-A.}\ \bibnamefont {Chen}}, \bibinfo {author} {\bibfnamefont {C.-Z.}\
  \bibnamefont {Peng}}, \bibinfo {author} {\bibfnamefont {Z.-B.}\ \bibnamefont
  {Chen}}, \ and\ \bibinfo {author} {\bibfnamefont {J.-W.}\ \bibnamefont
  {Pan}},\ }\href {\doibase 10.1038/nphys1603} {\bibfield  {journal} {\bibinfo
  {journal} {Nat. Phys.}\ }\textbf {\bibinfo {volume} {6}},\ \bibinfo {pages}
  {331} (\bibinfo {year} {2010})}\BibitemShut {NoStop}%
\bibitem [{\citenamefont {Leibfried}\ \emph {et~al.}(2004)\citenamefont
  {Leibfried}, \citenamefont {Barrett}, \citenamefont {Schaetz}, \citenamefont
  {Britton}, \citenamefont {Chiaverini}, \citenamefont {Itano}, \citenamefont
  {Jost}, \citenamefont {Langer},\ and\ \citenamefont
  {Wineland}}]{Leibfried2004Toward}%
  \BibitemOpen
  \bibfield  {author} {\bibinfo {author} {\bibfnamefont {D.}~\bibnamefont
  {Leibfried}}, \bibinfo {author} {\bibfnamefont {M.}~\bibnamefont {Barrett}},
  \bibinfo {author} {\bibfnamefont {T.}~\bibnamefont {Schaetz}}, \bibinfo
  {author} {\bibfnamefont {J.}~\bibnamefont {Britton}}, \bibinfo {author}
  {\bibfnamefont {J.}~\bibnamefont {Chiaverini}}, \bibinfo {author}
  {\bibfnamefont {W.}~\bibnamefont {Itano}}, \bibinfo {author} {\bibfnamefont
  {J.}~\bibnamefont {Jost}}, \bibinfo {author} {\bibfnamefont {C.}~\bibnamefont
  {Langer}}, \ and\ \bibinfo {author} {\bibfnamefont {D.}~\bibnamefont
  {Wineland}},\ }\href
  {http://www.sciencemag.org/content/304/5676/1476.abstract} {\bibfield
  {journal} {\bibinfo  {journal} {Science}\ }\textbf {\bibinfo {volume}
  {304}},\ \bibinfo {pages} {1476} (\bibinfo {year} {2004})}\BibitemShut
  {NoStop}%
\bibitem [{\citenamefont {Sackett}\ \emph {et~al.}(2000)\citenamefont
  {Sackett}, \citenamefont {Kielpinski}, \citenamefont {King}, \citenamefont
  {Langer}, \citenamefont {Meyer}, \citenamefont {Myatt}, \citenamefont {Rowe},
  \citenamefont {Turchette}, \citenamefont {Itano}, \citenamefont {Wineland},\
  and\ \citenamefont {Monroe}}]{Sackett2000Experimental}%
  \BibitemOpen
  \bibfield  {author} {\bibinfo {author} {\bibfnamefont {C.}~\bibnamefont
  {Sackett}}, \bibinfo {author} {\bibfnamefont {D.}~\bibnamefont {Kielpinski}},
  \bibinfo {author} {\bibfnamefont {B.}~\bibnamefont {King}}, \bibinfo {author}
  {\bibfnamefont {C.}~\bibnamefont {Langer}}, \bibinfo {author} {\bibfnamefont
  {V.}~\bibnamefont {Meyer}}, \bibinfo {author} {\bibfnamefont
  {C.}~\bibnamefont {Myatt}}, \bibinfo {author} {\bibfnamefont
  {M.}~\bibnamefont {Rowe}}, \bibinfo {author} {\bibfnamefont {Q.}~\bibnamefont
  {Turchette}}, \bibinfo {author} {\bibfnamefont {W.}~\bibnamefont {Itano}},
  \bibinfo {author} {\bibfnamefont {D.}~\bibnamefont {Wineland}}, \ and\
  \bibinfo {author} {\bibfnamefont {C.}~\bibnamefont {Monroe}},\ }\href
  {\doibase 10.1038/35005011} {\bibfield  {journal} {\bibinfo  {journal}
  {Nature}\ }\textbf {\bibinfo {volume} {404}},\ \bibinfo {pages} {256}
  (\bibinfo {year} {2000})}\BibitemShut {NoStop}%
\bibitem [{\citenamefont {Monz}\ \emph {et~al.}(2011)\citenamefont {Monz},
  \citenamefont {Schindler}, \citenamefont {Barreiro}, \citenamefont {Chwalla},
  \citenamefont {Nigg}, \citenamefont {Coish}, \citenamefont {Harlander},
  \citenamefont {H\"ansel}, \citenamefont {Hennrich},\ and\ \citenamefont
  {Blatt}}]{Monz201114-Qubit}%
  \BibitemOpen
  \bibfield  {author} {\bibinfo {author} {\bibfnamefont {T.}~\bibnamefont
  {Monz}}, \bibinfo {author} {\bibfnamefont {P.}~\bibnamefont {Schindler}},
  \bibinfo {author} {\bibfnamefont {J.~T.}\ \bibnamefont {Barreiro}}, \bibinfo
  {author} {\bibfnamefont {M.}~\bibnamefont {Chwalla}}, \bibinfo {author}
  {\bibfnamefont {D.}~\bibnamefont {Nigg}}, \bibinfo {author} {\bibfnamefont
  {W.~A.}\ \bibnamefont {Coish}}, \bibinfo {author} {\bibfnamefont
  {M.}~\bibnamefont {Harlander}}, \bibinfo {author} {\bibfnamefont
  {W.}~\bibnamefont {H\"ansel}}, \bibinfo {author} {\bibfnamefont
  {M.}~\bibnamefont {Hennrich}}, \ and\ \bibinfo {author} {\bibfnamefont
  {R.}~\bibnamefont {Blatt}},\ }\href {\doibase 10.1103/PhysRevLett.106.130506}
  {\bibfield  {journal} {\bibinfo  {journal} {Phys. Rev. Lett.}\ }\textbf
  {\bibinfo {volume} {106}},\ \bibinfo {pages} {130506} (\bibinfo {year}
  {2011})}\BibitemShut {NoStop}%
\bibitem [{\citenamefont {Kiesel}\ \emph {et~al.}(2007)\citenamefont {Kiesel},
  \citenamefont {Schmid}, \citenamefont {T\'oth}, \citenamefont {Solano},\ and\
  \citenamefont {Weinfurter}}]{Kiesel2007Experimental}%
  \BibitemOpen
  \bibfield  {author} {\bibinfo {author} {\bibfnamefont {N.}~\bibnamefont
  {Kiesel}}, \bibinfo {author} {\bibfnamefont {C.}~\bibnamefont {Schmid}},
  \bibinfo {author} {\bibfnamefont {G.}~\bibnamefont {T\'oth}}, \bibinfo
  {author} {\bibfnamefont {E.}~\bibnamefont {Solano}}, \ and\ \bibinfo {author}
  {\bibfnamefont {H.}~\bibnamefont {Weinfurter}},\ }\href {\doibase
  10.1103/PhysRevLett.98.063604} {\bibfield  {journal} {\bibinfo  {journal}
  {Phys. Rev. Lett.}\ }\textbf {\bibinfo {volume} {98}},\ \bibinfo {pages}
  {063604} (\bibinfo {year} {2007})}\BibitemShut {NoStop}%
\bibitem [{\citenamefont {Wieczorek}\ \emph {et~al.}(2009)\citenamefont
  {Wieczorek}, \citenamefont {Krischek}, \citenamefont {Kiesel}, \citenamefont
  {Michelberger}, \citenamefont {T\'oth},\ and\ \citenamefont
  {Weinfurter}}]{Wieczorek2009Experimental}%
  \BibitemOpen
  \bibfield  {author} {\bibinfo {author} {\bibfnamefont {W.}~\bibnamefont
  {Wieczorek}}, \bibinfo {author} {\bibfnamefont {R.}~\bibnamefont {Krischek}},
  \bibinfo {author} {\bibfnamefont {N.}~\bibnamefont {Kiesel}}, \bibinfo
  {author} {\bibfnamefont {P.}~\bibnamefont {Michelberger}}, \bibinfo {author}
  {\bibfnamefont {G.}~\bibnamefont {T\'oth}}, \ and\ \bibinfo {author}
  {\bibfnamefont {H.}~\bibnamefont {Weinfurter}},\ }\href {\doibase
  10.1103/PhysRevLett.103.020504} {\bibfield  {journal} {\bibinfo  {journal}
  {Phys. Rev. Lett.}\ }\textbf {\bibinfo {volume} {103}},\ \bibinfo {pages}
  {020504} (\bibinfo {year} {2009})}\BibitemShut {NoStop}%
\bibitem [{\citenamefont {Prevedel}\ \emph {et~al.}(2009)\citenamefont
  {Prevedel}, \citenamefont {Cronenberg}, \citenamefont {Tame}, \citenamefont
  {Paternostro}, \citenamefont {Walther}, \citenamefont {Kim},\ and\
  \citenamefont {Zeilinger}}]{Prevedel2009Experimental}%
  \BibitemOpen
  \bibfield  {author} {\bibinfo {author} {\bibfnamefont {R.}~\bibnamefont
  {Prevedel}}, \bibinfo {author} {\bibfnamefont {G.}~\bibnamefont
  {Cronenberg}}, \bibinfo {author} {\bibfnamefont {M.~S.}\ \bibnamefont
  {Tame}}, \bibinfo {author} {\bibfnamefont {M.}~\bibnamefont {Paternostro}},
  \bibinfo {author} {\bibfnamefont {P.}~\bibnamefont {Walther}}, \bibinfo
  {author} {\bibfnamefont {M.~S.}\ \bibnamefont {Kim}}, \ and\ \bibinfo
  {author} {\bibfnamefont {A.}~\bibnamefont {Zeilinger}},\ }\href {\doibase
  10.1103/PhysRevLett.103.020503} {\bibfield  {journal} {\bibinfo  {journal}
  {Phys. Rev. Lett.}\ }\textbf {\bibinfo {volume} {103}},\ \bibinfo {pages}
  {020503} (\bibinfo {year} {2009})}\BibitemShut {NoStop}%
\bibitem [{\citenamefont {Chiuri}\ \emph {et~al.}(2012)\citenamefont {Chiuri},
  \citenamefont {Greganti}, \citenamefont {Paternostro}, \citenamefont
  {Vallone},\ and\ \citenamefont {Mataloni}}]{Chiuri2012Experimental}%
  \BibitemOpen
  \bibfield  {author} {\bibinfo {author} {\bibfnamefont {A.}~\bibnamefont
  {Chiuri}}, \bibinfo {author} {\bibfnamefont {C.}~\bibnamefont {Greganti}},
  \bibinfo {author} {\bibfnamefont {M.}~\bibnamefont {Paternostro}}, \bibinfo
  {author} {\bibfnamefont {G.}~\bibnamefont {Vallone}}, \ and\ \bibinfo
  {author} {\bibfnamefont {P.}~\bibnamefont {Mataloni}},\ }\href {\doibase
  10.1103/PhysRevLett.109.173604} {\bibfield  {journal} {\bibinfo  {journal}
  {Phys. Rev. Lett.}\ }\textbf {\bibinfo {volume} {109}},\ \bibinfo {pages}
  {173604} (\bibinfo {year} {2012})}\BibitemShut {NoStop}%
\bibitem [{\citenamefont {Schindler}\ \emph {et~al.}(2013)\citenamefont
  {Schindler}, \citenamefont {M\"{u}ller}, \citenamefont {Nigg}, \citenamefont
  {Barreiro}, \citenamefont {Martinez}, \citenamefont {Hennrich}, \citenamefont
  {Monz}, \citenamefont {Diehl}, \citenamefont {Zoller},\ and\ \citenamefont
  {Blatt}}]{Schindler2013Quantum}%
  \BibitemOpen
  \bibfield  {author} {\bibinfo {author} {\bibfnamefont {P.}~\bibnamefont
  {Schindler}}, \bibinfo {author} {\bibfnamefont {M.}~\bibnamefont
  {M\"{u}ller}}, \bibinfo {author} {\bibfnamefont {D.}~\bibnamefont {Nigg}},
  \bibinfo {author} {\bibfnamefont {J.~T.}\ \bibnamefont {Barreiro}}, \bibinfo
  {author} {\bibfnamefont {E.}~\bibnamefont {Martinez}}, \bibinfo {author}
  {\bibfnamefont {M.}~\bibnamefont {Hennrich}}, \bibinfo {author}
  {\bibfnamefont {T.}~\bibnamefont {Monz}}, \bibinfo {author} {\bibfnamefont
  {S.}~\bibnamefont {Diehl}}, \bibinfo {author} {\bibfnamefont
  {P.}~\bibnamefont {Zoller}}, \ and\ \bibinfo {author} {\bibfnamefont
  {R.}~\bibnamefont {Blatt}},\ }\href {\doibase 10.1038/nphys2630} {\bibfield
  {journal} {\bibinfo  {journal} {Nat. Phys.}\ }\textbf {\bibinfo {volume}
  {9}},\ \bibinfo {pages} {361} (\bibinfo {year} {2013})}\BibitemShut {NoStop}%
\bibitem [{\citenamefont {Gross}(2012)}]{Gross2012Spin}%
  \BibitemOpen
  \bibfield  {author} {\bibinfo {author} {\bibfnamefont {C.}~\bibnamefont
  {Gross}},\ }\href {http://stacks.iop.org/0953-4075/45/i=10/a=103001}
  {\bibfield  {journal} {\bibinfo  {journal} {J. Phys. B: At. Mol. Opt. Phys.}\
  }\textbf {\bibinfo {volume} {45}},\ \bibinfo {pages} {103001} (\bibinfo
  {year} {2012})}\BibitemShut {NoStop}%
\bibitem [{\citenamefont {{Ma}}\ \emph {et~al.}(2011)\citenamefont {{Ma}},
  \citenamefont {{Wang}}, \citenamefont {{Sun}},\ and\ \citenamefont
  {{Nori}}}]{Ma2011Quantum}%
  \BibitemOpen
  \bibfield  {author} {\bibinfo {author} {\bibfnamefont {J.}~\bibnamefont
  {{Ma}}}, \bibinfo {author} {\bibfnamefont {X.}~\bibnamefont {{Wang}}},
  \bibinfo {author} {\bibfnamefont {C.~P.}\ \bibnamefont {{Sun}}}, \ and\
  \bibinfo {author} {\bibfnamefont {F.}~\bibnamefont {{Nori}}},\ }\href
  {\doibase 10.1016/j.physrep.2011.08.003} {\bibfield  {journal} {\bibinfo
  {journal} {Phys. Rep.}\ }\textbf {\bibinfo {volume} {509}},\ \bibinfo {pages}
  {89} (\bibinfo {year} {2011})}\BibitemShut {NoStop}%
\bibitem [{\citenamefont {Hald}\ \emph {et~al.}(1999)\citenamefont {Hald},
  \citenamefont {S\o{}rensen}, \citenamefont {Schori},\ and\ \citenamefont
  {Polzik}}]{Hald1999Spin}%
  \BibitemOpen
  \bibfield  {author} {\bibinfo {author} {\bibfnamefont {J.}~\bibnamefont
  {Hald}}, \bibinfo {author} {\bibfnamefont {J.~L.}\ \bibnamefont
  {S\o{}rensen}}, \bibinfo {author} {\bibfnamefont {C.}~\bibnamefont {Schori}},
  \ and\ \bibinfo {author} {\bibfnamefont {E.~S.}\ \bibnamefont {Polzik}},\
  }\href {\doibase 10.1103/PhysRevLett.83.1319} {\bibfield  {journal} {\bibinfo
   {journal} {Phys. Rev. Lett.}\ }\textbf {\bibinfo {volume} {83}},\ \bibinfo
  {pages} {1319} (\bibinfo {year} {1999})}\BibitemShut {NoStop}%
\bibitem [{\citenamefont {de~Echaniz}\ \emph {et~al.}(2005)\citenamefont
  {de~Echaniz}, \citenamefont {Mitchell}, \citenamefont {Kubasik},
  \citenamefont {Koschorreck}, \citenamefont {Crepaz}, \citenamefont
  {Eschner},\ and\ \citenamefont {Polzik}}]{Echaniz2005Conditions}%
  \BibitemOpen
  \bibfield  {author} {\bibinfo {author} {\bibfnamefont {S.~R.}\ \bibnamefont
  {de~Echaniz}}, \bibinfo {author} {\bibfnamefont {M.~W.}\ \bibnamefont
  {Mitchell}}, \bibinfo {author} {\bibfnamefont {M.}~\bibnamefont {Kubasik}},
  \bibinfo {author} {\bibfnamefont {M.}~\bibnamefont {Koschorreck}}, \bibinfo
  {author} {\bibfnamefont {H.}~\bibnamefont {Crepaz}}, \bibinfo {author}
  {\bibfnamefont {J.}~\bibnamefont {Eschner}}, \ and\ \bibinfo {author}
  {\bibfnamefont {E.~S.}\ \bibnamefont {Polzik}},\ }\href
  {http://stacks.iop.org/1464-4266/7/i=12/a=016} {\bibfield  {journal}
  {\bibinfo  {journal} {J. Opt. B: Quantum and Semiclassical Opt.}\ }\textbf
  {\bibinfo {volume} {7}},\ \bibinfo {pages} {S548} (\bibinfo {year}
  {2005})}\BibitemShut {NoStop}%
\bibitem [{\citenamefont {Sewell}\ \emph {et~al.}(2012)\citenamefont {Sewell},
  \citenamefont {Koschorreck}, \citenamefont {Napolitano}, \citenamefont
  {Dubost}, \citenamefont {Behbood},\ and\ \citenamefont
  {Mitchell}}]{Sewell2012Magnetic}%
  \BibitemOpen
  \bibfield  {author} {\bibinfo {author} {\bibfnamefont {R.~J.}\ \bibnamefont
  {Sewell}}, \bibinfo {author} {\bibfnamefont {M.}~\bibnamefont {Koschorreck}},
  \bibinfo {author} {\bibfnamefont {M.}~\bibnamefont {Napolitano}}, \bibinfo
  {author} {\bibfnamefont {B.}~\bibnamefont {Dubost}}, \bibinfo {author}
  {\bibfnamefont {N.}~\bibnamefont {Behbood}}, \ and\ \bibinfo {author}
  {\bibfnamefont {M.~W.}\ \bibnamefont {Mitchell}},\ }\href {\doibase
  10.1103/PhysRevLett.109.253605} {\bibfield  {journal} {\bibinfo  {journal}
  {Phys. Rev. Lett.}\ }\textbf {\bibinfo {volume} {109}},\ \bibinfo {pages}
  {253605} (\bibinfo {year} {2012})}\BibitemShut {NoStop}%
\bibitem [{\citenamefont {L\"ucke}\ \emph {et~al.}(2014)\citenamefont
  {L\"ucke}, \citenamefont {Peise}, \citenamefont {Vitagliano}, \citenamefont
  {Arlt}, \citenamefont {Santos}, \citenamefont {T\'oth},\ and\ \citenamefont
  {Klempt}}]{Lucke2014Detecting}%
  \BibitemOpen
  \bibfield  {author} {\bibinfo {author} {\bibfnamefont {B.}~\bibnamefont
  {L\"ucke}}, \bibinfo {author} {\bibfnamefont {J.}~\bibnamefont {Peise}},
  \bibinfo {author} {\bibfnamefont {G.}~\bibnamefont {Vitagliano}}, \bibinfo
  {author} {\bibfnamefont {J.}~\bibnamefont {Arlt}}, \bibinfo {author}
  {\bibfnamefont {L.}~\bibnamefont {Santos}}, \bibinfo {author} {\bibfnamefont
  {G.}~\bibnamefont {T\'oth}}, \ and\ \bibinfo {author} {\bibfnamefont
  {C.}~\bibnamefont {Klempt}},\ }\href {\doibase
  10.1103/PhysRevLett.112.155304} {\bibfield  {journal} {\bibinfo  {journal}
  {Phys. Rev. Lett.}\ }\textbf {\bibinfo {volume} {112}},\ \bibinfo {pages}
  {155304} (\bibinfo {year} {2014})}\BibitemShut {NoStop}%
\bibitem [{\citenamefont {Hamley}\ \emph {et~al.}(2012)\citenamefont {Hamley},
  \citenamefont {Gerving}, \citenamefont {Hoang}, \citenamefont {Bookjans},\
  and\ \citenamefont {Chapman}}]{Hamley2012Spin-nematic}%
  \BibitemOpen
  \bibfield  {author} {\bibinfo {author} {\bibfnamefont {C.}~\bibnamefont
  {Hamley}}, \bibinfo {author} {\bibfnamefont {C.}~\bibnamefont {Gerving}},
  \bibinfo {author} {\bibfnamefont {T.}~\bibnamefont {Hoang}}, \bibinfo
  {author} {\bibfnamefont {E.}~\bibnamefont {Bookjans}}, \ and\ \bibinfo
  {author} {\bibfnamefont {M.}~\bibnamefont {Chapman}},\ }\href {\doibase
  10.1038/nphys2245} {\bibfield  {journal} {\bibinfo  {journal} {Nat. Phys.}\
  }\textbf {\bibinfo {volume} {8}},\ \bibinfo {pages} {305} (\bibinfo {year}
  {2012})}\BibitemShut {NoStop}%
\bibitem [{\citenamefont {Luo}\ \emph {et~al.}(2017)\citenamefont {Luo},
  \citenamefont {Zou}, \citenamefont {Wu}, \citenamefont {Liu}, \citenamefont
  {Han}, \citenamefont {Tey},\ and\ \citenamefont
  {You}}]{Luo2017Deterministic}%
  \BibitemOpen
  \bibfield  {author} {\bibinfo {author} {\bibfnamefont {X.-Y.}\ \bibnamefont
  {Luo}}, \bibinfo {author} {\bibfnamefont {Y.-Q.}\ \bibnamefont {Zou}},
  \bibinfo {author} {\bibfnamefont {L.-N.}\ \bibnamefont {Wu}}, \bibinfo
  {author} {\bibfnamefont {Q.}~\bibnamefont {Liu}}, \bibinfo {author}
  {\bibfnamefont {M.-F.}\ \bibnamefont {Han}}, \bibinfo {author} {\bibfnamefont
  {M.~K.}\ \bibnamefont {Tey}}, \ and\ \bibinfo {author} {\bibfnamefont
  {L.}~\bibnamefont {You}},\ }\href {\doibase 10.1126/science.aag1106}
  {\bibfield  {journal} {\bibinfo  {journal} {Science}\ }\textbf {\bibinfo
  {volume} {355}},\ \bibinfo {pages} {620} (\bibinfo {year}
  {2017})}\BibitemShut {NoStop}%
\bibitem [{\citenamefont {S{\o}rensen}\ \emph {et~al.}(2001)\citenamefont
  {S{\o}rensen}, \citenamefont {Duan}, \citenamefont {Cirac},\ and\
  \citenamefont {Zoller}}]{Sorensen2001Many-particle}%
  \BibitemOpen
  \bibfield  {author} {\bibinfo {author} {\bibfnamefont {A.}~\bibnamefont
  {S{\o}rensen}}, \bibinfo {author} {\bibfnamefont {L.-M.}\ \bibnamefont
  {Duan}}, \bibinfo {author} {\bibfnamefont {J.}~\bibnamefont {Cirac}}, \ and\
  \bibinfo {author} {\bibfnamefont {P.}~\bibnamefont {Zoller}},\ }\href
  {http://www.nature.com/nature/journal/v409/n6816/abs/409063a0.html}
  {\bibfield  {journal} {\bibinfo  {journal} {Nature}\ }\textbf {\bibinfo
  {volume} {409}},\ \bibinfo {pages} {63} (\bibinfo {year} {2001})}\BibitemShut
  {NoStop}%
\bibitem [{\citenamefont {Korbicz}\ \emph {et~al.}(2005)\citenamefont
  {Korbicz}, \citenamefont {Cirac},\ and\ \citenamefont
  {Lewenstein}}]{Korbicz2005Spin}%
  \BibitemOpen
  \bibfield  {author} {\bibinfo {author} {\bibfnamefont {J.~K.}\ \bibnamefont
  {Korbicz}}, \bibinfo {author} {\bibfnamefont {J.~I.}\ \bibnamefont {Cirac}},
  \ and\ \bibinfo {author} {\bibfnamefont {M.}~\bibnamefont {Lewenstein}},\
  }\href {\doibase 10.1103/PhysRevLett.95.120502} {\bibfield  {journal}
  {\bibinfo  {journal} {Phys. Rev. Lett.}\ }\textbf {\bibinfo {volume} {95}},\
  \bibinfo {pages} {120502} (\bibinfo {year} {2005})}\BibitemShut {NoStop}%
\bibitem [{\citenamefont {T\'oth}\ \emph {et~al.}(2007)\citenamefont {T\'oth},
  \citenamefont {Knapp}, \citenamefont {G\"uhne},\ and\ \citenamefont
  {Briegel}}]{Toth2007Optimal}%
  \BibitemOpen
  \bibfield  {author} {\bibinfo {author} {\bibfnamefont {G.}~\bibnamefont
  {T\'oth}}, \bibinfo {author} {\bibfnamefont {C.}~\bibnamefont {Knapp}},
  \bibinfo {author} {\bibfnamefont {O.}~\bibnamefont {G\"uhne}}, \ and\
  \bibinfo {author} {\bibfnamefont {H.~J.}\ \bibnamefont {Briegel}},\ }\href
  {\doibase 10.1103/PhysRevLett.99.250405} {\bibfield  {journal} {\bibinfo
  {journal} {Phys. Rev. Lett.}\ }\textbf {\bibinfo {volume} {99}},\ \bibinfo
  {pages} {250405} (\bibinfo {year} {2007})}\BibitemShut {NoStop}%
\bibitem [{\citenamefont {Doherty}\ \emph {et~al.}(2002)\citenamefont
  {Doherty}, \citenamefont {Parrilo},\ and\ \citenamefont
  {Spedalieri}}]{Doherty2002Distinguishing}%
  \BibitemOpen
  \bibfield  {author} {\bibinfo {author} {\bibfnamefont {A.~C.}\ \bibnamefont
  {Doherty}}, \bibinfo {author} {\bibfnamefont {P.~A.}\ \bibnamefont
  {Parrilo}}, \ and\ \bibinfo {author} {\bibfnamefont {F.~M.}\ \bibnamefont
  {Spedalieri}},\ }\href {\doibase 10.1103/PhysRevLett.88.187904} {\bibfield
  {journal} {\bibinfo  {journal} {Phys. Rev. Lett.}\ }\textbf {\bibinfo
  {volume} {88}},\ \bibinfo {pages} {187904} (\bibinfo {year}
  {2002})}\BibitemShut {NoStop}%
\bibitem [{\citenamefont {Wunderlich}\ and\ \citenamefont
  {Plenio}(2009)}]{Wunderlich2009Quantitative}%
  \BibitemOpen
  \bibfield  {author} {\bibinfo {author} {\bibfnamefont {H.}~\bibnamefont
  {Wunderlich}}\ and\ \bibinfo {author} {\bibfnamefont {M.~B.}\ \bibnamefont
  {Plenio}},\ }\href {\doibase 10.1080/09500340903184303} {\bibfield  {journal}
  {\bibinfo  {journal} {J. Mod. Opt.}\ }\textbf {\bibinfo {volume} {56}},\
  \bibinfo {pages} {2100} (\bibinfo {year} {2009})}\BibitemShut {NoStop}%
\bibitem [{\citenamefont {T\'oth}\ \emph {et~al.}(2015)\citenamefont {T\'oth},
  \citenamefont {Moroder},\ and\ \citenamefont {G\"uhne}}]{Toth2015Evaluating}%
  \BibitemOpen
  \bibfield  {author} {\bibinfo {author} {\bibfnamefont {G.}~\bibnamefont
  {T\'oth}}, \bibinfo {author} {\bibfnamefont {T.}~\bibnamefont {Moroder}}, \
  and\ \bibinfo {author} {\bibfnamefont {O.}~\bibnamefont {G\"uhne}},\ }\href
  {\doibase 10.1103/PhysRevLett.114.160501} {\bibfield  {journal} {\bibinfo
  {journal} {Phys. Rev. Lett.}\ }\textbf {\bibinfo {volume} {114}},\ \bibinfo
  {pages} {160501} (\bibinfo {year} {2015})}\BibitemShut {NoStop}%
\bibitem [{\citenamefont {Rockafellar}(1997)}]{Rockafellar2015Convex}%
  \BibitemOpen
  \bibfield  {author} {\bibinfo {author} {\bibfnamefont {R.~T.}\ \bibnamefont
  {Rockafellar}},\ }\href@noop {} {\emph {\bibinfo {title} {Convex analysis}}}\
  (\bibinfo  {publisher} {Princeton University Press, Princeton},\ \bibinfo
  {year} {1997})\BibitemShut {NoStop}%
\bibitem [{\citenamefont {G\"uhne}\ \emph {et~al.}(2007)\citenamefont
  {G\"uhne}, \citenamefont {Reimpell},\ and\ \citenamefont
  {Werner}}]{Guhne2007Estimating}%
  \BibitemOpen
  \bibfield  {author} {\bibinfo {author} {\bibfnamefont {O.}~\bibnamefont
  {G\"uhne}}, \bibinfo {author} {\bibfnamefont {M.}~\bibnamefont {Reimpell}}, \
  and\ \bibinfo {author} {\bibfnamefont {R.~F.}\ \bibnamefont {Werner}},\
  }\href {\doibase 10.1103/PhysRevLett.98.110502} {\bibfield  {journal}
  {\bibinfo  {journal} {Phys. Rev. Lett.}\ }\textbf {\bibinfo {volume} {98}},\
  \bibinfo {pages} {110502} (\bibinfo {year} {2007})}\BibitemShut {NoStop}%
\bibitem [{\citenamefont {Eisert}\ \emph {et~al.}(2007)\citenamefont {Eisert},
  \citenamefont {Brandao},\ and\ \citenamefont
  {Audenaert}}]{Eisert2007Quantitative}%
  \BibitemOpen
  \bibfield  {author} {\bibinfo {author} {\bibfnamefont {J.}~\bibnamefont
  {Eisert}}, \bibinfo {author} {\bibfnamefont {F.~G. S.~L.}\ \bibnamefont
  {Brandao}}, \ and\ \bibinfo {author} {\bibfnamefont {K.~M.~R.}\ \bibnamefont
  {Audenaert}},\ }\href {http://stacks.iop.org/1367-2630/9/i=3/a=046}
  {\bibfield  {journal} {\bibinfo  {journal} {New J. Phys.}\ }\textbf {\bibinfo
  {volume} {9}},\ \bibinfo {pages} {46} (\bibinfo {year} {2007})}\BibitemShut
  {NoStop}%
\bibitem [{\citenamefont {T\'oth}\ and\ \citenamefont
  {Petz}(2013)}]{Toth2013Extremal}%
  \BibitemOpen
  \bibfield  {author} {\bibinfo {author} {\bibfnamefont {G.}~\bibnamefont
  {T\'oth}}\ and\ \bibinfo {author} {\bibfnamefont {D.}~\bibnamefont {Petz}},\
  }\href {\doibase 10.1103/PhysRevA.87.032324} {\bibfield  {journal} {\bibinfo
  {journal} {Phys. Rev. A}\ }\textbf {\bibinfo {volume} {87}},\ \bibinfo
  {pages} {032324} (\bibinfo {year} {2013})}\BibitemShut {NoStop}%
\bibitem [{\citenamefont {Yu}()}]{Yu2013Quantum}%
  \BibitemOpen
  \bibfield  {author} {\bibinfo {author} {\bibfnamefont {S.}~\bibnamefont
  {Yu}},\ }\href {http://arxiv.org/abs/1302.5311} {\bibinfo  {journal}
  {arXiv:1302.5311}\ }\BibitemShut {NoStop}%
\bibitem [{Note1()}]{Note1}%
  \BibitemOpen
\bibfield  {journal} {  }\bibinfo {note} {An alternative definition is
  $\protect \mathaccentV {hat}002{\protect \mathcal {F}}_{\protect \rm
  Q}(W)=\protect \qopname \relax m{sup}_{\nu } \protect \ensuremath {\delimiter
  "426830A {W}\delimiter "526930B }_{\psi _\nu }-4\protect \ensuremath {(\Delta
  J_l)^2}_{\psi _\nu },$ where $\ket {\psi _\nu }$ is the eigenstate with the
  maximal eigenvalue of the operator $W-4J_l^2-\nu J_l.$ In certain cases, this
  form can be calculated numerically more easily than Eq.~\protect \textup
  {\hbox {\mathsurround \z@ \protect \normalfont (\ignorespaces \ref
  {eq:LegendreTransformFQ_H}\unskip \@@italiccorr )}}.}\BibitemShut {Stop}%
\bibitem [{\citenamefont {Luis}(2004)}]{Luis2004Nonlinear}%
  \BibitemOpen
  \bibfield  {author} {\bibinfo {author} {\bibfnamefont {A.}~\bibnamefont
  {Luis}},\ }\href {\doibase http://dx.doi.org/10.1016/j.physleta.2004.06.080}
  {\bibfield  {journal} {\bibinfo  {journal} {Physics Letters A}\ }\textbf
  {\bibinfo {volume} {329}},\ \bibinfo {pages} {8} (\bibinfo {year}
  {2004})}\BibitemShut {NoStop}%
\bibitem [{\citenamefont {Boixo}\ \emph {et~al.}(2007)\citenamefont {Boixo},
  \citenamefont {Flammia}, \citenamefont {Caves},\ and\ \citenamefont
  {Geremia}}]{Boixo2007Generalized}%
  \BibitemOpen
  \bibfield  {author} {\bibinfo {author} {\bibfnamefont {S.}~\bibnamefont
  {Boixo}}, \bibinfo {author} {\bibfnamefont {S.~T.}\ \bibnamefont {Flammia}},
  \bibinfo {author} {\bibfnamefont {C.~M.}\ \bibnamefont {Caves}}, \ and\
  \bibinfo {author} {\bibfnamefont {J.}~\bibnamefont {Geremia}},\ }\href
  {\doibase 10.1103/PhysRevLett.98.090401} {\bibfield  {journal} {\bibinfo
  {journal} {Phys. Rev. Lett.}\ }\textbf {\bibinfo {volume} {98}},\ \bibinfo
  {pages} {090401} (\bibinfo {year} {2007})}\BibitemShut {NoStop}%
\bibitem [{\citenamefont {Choi}\ and\ \citenamefont
  {Sundaram}(2008)}]{Choi2008Bose-Einstein}%
  \BibitemOpen
  \bibfield  {author} {\bibinfo {author} {\bibfnamefont {S.}~\bibnamefont
  {Choi}}\ and\ \bibinfo {author} {\bibfnamefont {B.}~\bibnamefont
  {Sundaram}},\ }\href {\doibase 10.1103/PhysRevA.77.053613} {\bibfield
  {journal} {\bibinfo  {journal} {Phys. Rev. A}\ }\textbf {\bibinfo {volume}
  {77}},\ \bibinfo {pages} {053613} (\bibinfo {year} {2008})}\BibitemShut
  {NoStop}%
\bibitem [{\citenamefont {Roy}\ and\ \citenamefont
  {Braunstein}(2008)}]{Roy2008Exponentially}%
  \BibitemOpen
  \bibfield  {author} {\bibinfo {author} {\bibfnamefont {S.~M.}\ \bibnamefont
  {Roy}}\ and\ \bibinfo {author} {\bibfnamefont {S.~L.}\ \bibnamefont
  {Braunstein}},\ }\href {\doibase 10.1103/PhysRevLett.100.220501} {\bibfield
  {journal} {\bibinfo  {journal} {Phys. Rev. Lett.}\ }\textbf {\bibinfo
  {volume} {100}},\ \bibinfo {pages} {220501} (\bibinfo {year}
  {2008})}\BibitemShut {NoStop}%
\bibitem [{\citenamefont {Napolitano}\ \emph {et~al.}(2011)\citenamefont
  {Napolitano}, \citenamefont {Koschorreck}, \citenamefont {Dubost},
  \citenamefont {Behbood}, \citenamefont {Sewell},\ and\ \citenamefont
  {Mitchell}}]{Napolitano2011Interaction-based}%
  \BibitemOpen
  \bibfield  {author} {\bibinfo {author} {\bibfnamefont {M.}~\bibnamefont
  {Napolitano}}, \bibinfo {author} {\bibfnamefont {M.}~\bibnamefont
  {Koschorreck}}, \bibinfo {author} {\bibfnamefont {B.}~\bibnamefont {Dubost}},
  \bibinfo {author} {\bibfnamefont {N.}~\bibnamefont {Behbood}}, \bibinfo
  {author} {\bibfnamefont {R.}~\bibnamefont {Sewell}}, \ and\ \bibinfo {author}
  {\bibfnamefont {M.~W.}\ \bibnamefont {Mitchell}},\ }\href
  {http://www.nature.com/nature/journal/v471/n7339/abs/nature09778.html}
  {\bibfield  {journal} {\bibinfo  {journal} {Nature}\ }\textbf {\bibinfo
  {volume} {471}},\ \bibinfo {pages} {486} (\bibinfo {year}
  {2011})}\BibitemShut {NoStop}%
\bibitem [{\citenamefont {Hall}\ and\ \citenamefont
  {Wiseman}(2012)}]{Hall2012Does}%
  \BibitemOpen
  \bibfield  {author} {\bibinfo {author} {\bibfnamefont {M.~J.~W.}\
  \bibnamefont {Hall}}\ and\ \bibinfo {author} {\bibfnamefont {H.~M.}\
  \bibnamefont {Wiseman}},\ }\href {\doibase 10.1103/PhysRevX.2.041006}
  {\bibfield  {journal} {\bibinfo  {journal} {Phys. Rev. X}\ }\textbf {\bibinfo
  {volume} {2}},\ \bibinfo {pages} {041006} (\bibinfo {year}
  {2012})}\BibitemShut {NoStop}%
\bibitem [{\citenamefont {S\o{}rensen}\ and\ \citenamefont
  {M\o{}lmer}(2001)}]{Sorensen2001Entanglement}%
  \BibitemOpen
  \bibfield  {author} {\bibinfo {author} {\bibfnamefont {A.~S.}\ \bibnamefont
  {S\o{}rensen}}\ and\ \bibinfo {author} {\bibfnamefont {K.}~\bibnamefont
  {M\o{}lmer}},\ }\href {\doibase 10.1103/PhysRevLett.86.4431} {\bibfield
  {journal} {\bibinfo  {journal} {Phys. Rev. Lett.}\ }\textbf {\bibinfo
  {volume} {86}},\ \bibinfo {pages} {4431} (\bibinfo {year}
  {2001})}\BibitemShut {NoStop}%
\bibitem [{\citenamefont {Augusiak}\ \emph {et~al.}(2016)\citenamefont
  {Augusiak}, \citenamefont {Ko\l{}ody\ifmmode~\acute{n}\else \'{n}\fi{}ski},
  \citenamefont {Streltsov}, \citenamefont {Bera}, \citenamefont {Ac\'{\i}n},\
  and\ \citenamefont {Lewenstein}}]{Augusiak2016Asymptotic}%
  \BibitemOpen
  \bibfield  {author} {\bibinfo {author} {\bibfnamefont {R.}~\bibnamefont
  {Augusiak}}, \bibinfo {author} {\bibfnamefont {J.}~\bibnamefont
  {Ko\l{}ody\ifmmode~\acute{n}\else \'{n}\fi{}ski}}, \bibinfo {author}
  {\bibfnamefont {A.}~\bibnamefont {Streltsov}}, \bibinfo {author}
  {\bibfnamefont {M.~N.}\ \bibnamefont {Bera}}, \bibinfo {author}
  {\bibfnamefont {A.}~\bibnamefont {Ac\'{\i}n}}, \ and\ \bibinfo {author}
  {\bibfnamefont {M.}~\bibnamefont {Lewenstein}},\ }\href {\doibase
  10.1103/PhysRevA.94.012339} {\bibfield  {journal} {\bibinfo  {journal} {Phys.
  Rev. A}\ }\textbf {\bibinfo {volume} {94}},\ \bibinfo {pages} {012339}
  (\bibinfo {year} {2016})}\BibitemShut {NoStop}%
\bibitem [{\citenamefont {{T{\'o}th}}(2007)}]{Toth2007Detection}%
  \BibitemOpen
  \bibfield  {author} {\bibinfo {author} {\bibfnamefont {G.}~\bibnamefont
  {{T{\'o}th}}},\ }\href {\doibase 10.1364/JOSAB.24.000275} {\bibfield
  {journal} {\bibinfo  {journal} {J. Opt. Soc. Am. B}\ }\textbf {\bibinfo
  {volume} {24}},\ \bibinfo {pages} {275} (\bibinfo {year} {2007})}\BibitemShut
  {NoStop}%
\bibitem [{\citenamefont {T\'oth}\ \emph {et~al.}(2009)\citenamefont {T\'oth},
  \citenamefont {Wieczorek}, \citenamefont {Krischek}, \citenamefont {Kiesel},
  \citenamefont {Michelberger},\ and\ \citenamefont
  {Weinfurter}}]{Toth2009Practical}%
  \BibitemOpen
  \bibfield  {author} {\bibinfo {author} {\bibfnamefont {G.}~\bibnamefont
  {T\'oth}}, \bibinfo {author} {\bibfnamefont {W.}~\bibnamefont {Wieczorek}},
  \bibinfo {author} {\bibfnamefont {R.}~\bibnamefont {Krischek}}, \bibinfo
  {author} {\bibfnamefont {N.}~\bibnamefont {Kiesel}}, \bibinfo {author}
  {\bibfnamefont {P.}~\bibnamefont {Michelberger}}, \ and\ \bibinfo {author}
  {\bibfnamefont {H.}~\bibnamefont {Weinfurter}},\ }\href
  {http://stacks.iop.org/1367-2630/11/i=8/a=083002} {\bibfield  {journal}
  {\bibinfo  {journal} {New J. Phys.}\ }\textbf {\bibinfo {volume} {11}},\
  \bibinfo {pages} {083002} (\bibinfo {year} {2009})}\BibitemShut {NoStop}%
\bibitem [{\citenamefont {Holland}\ and\ \citenamefont
  {Burnett}(1993)}]{Holland1993Interferometric}%
  \BibitemOpen
  \bibfield  {author} {\bibinfo {author} {\bibfnamefont {M.~J.}\ \bibnamefont
  {Holland}}\ and\ \bibinfo {author} {\bibfnamefont {K.}~\bibnamefont
  {Burnett}},\ }\href {\doibase 10.1103/PhysRevLett.71.1355} {\bibfield
  {journal} {\bibinfo  {journal} {Phys. Rev. Lett.}\ }\textbf {\bibinfo
  {volume} {71}},\ \bibinfo {pages} {1355} (\bibinfo {year}
  {1993})}\BibitemShut {NoStop}%
\bibitem [{Note2()}]{Note2}%
  \BibitemOpen
  \bibinfo {note} {Due to symmetries of the problem, when minimizing ${\protect
  \mathcal F}_{\protect \rm Q}[\varrho ,J_y]$ with the constrains on $\protect
  \ensuremath {\delimiter "426830A {J_z}\delimiter "526930B }$ and $\protect
  \ensuremath {\delimiter "426830A {J_x^2}\delimiter "526930B }$, we do not
  have to add explicitly the constraint $\protect \ensuremath {\delimiter
  "426830A {J_x}\delimiter "526930B }=0.$ Optimization with only the first two
  constraints will give the same bound (see Appendix~\ref
  {sec:assuming_Jx0_appendix}).}\BibitemShut {Stop}%
\bibitem [{Note3()}]{Note3}%
  \BibitemOpen
  \bibinfo {note} {Outside the symmetric subspace, there are other states with
  $\protect \ensuremath {\delimiter "426830A {J_z}\delimiter "526930B
  }=\protect \ensuremath {\delimiter "426830A {J_x^2}\delimiter "526930B }=0,$
  which also correspond to point D. For example, such a state is the
  multiparticle singlet. However, usual spin-squeezing procedures remain in the
  symmetric subspace, thus we discuss only the Dicke state.}\BibitemShut
  {Stop}%
\bibitem [{\citenamefont {Zhao}\ \emph {et~al.}(2004)\citenamefont {Zhao},
  \citenamefont {Chen}, \citenamefont {Zhang}, \citenamefont {Yang},
  \citenamefont {Briegel},\ and\ \citenamefont {Pan}}]{Zhao2004Experimental}%
  \BibitemOpen
  \bibfield  {author} {\bibinfo {author} {\bibfnamefont {Z.}~\bibnamefont
  {Zhao}}, \bibinfo {author} {\bibfnamefont {Y.-A.}\ \bibnamefont {Chen}},
  \bibinfo {author} {\bibfnamefont {A.-N.}\ \bibnamefont {Zhang}}, \bibinfo
  {author} {\bibfnamefont {T.}~\bibnamefont {Yang}}, \bibinfo {author}
  {\bibfnamefont {H.~J.}\ \bibnamefont {Briegel}}, \ and\ \bibinfo {author}
  {\bibfnamefont {J.-W.}\ \bibnamefont {Pan}},\ }\href {\doibase
  10.1038/nature02643} {\bibfield  {journal} {\bibinfo  {journal} {Nature}\
  }\textbf {\bibinfo {volume} {430}},\ \bibinfo {pages} {54} (\bibinfo {year}
  {2004})}\BibitemShut {NoStop}%
\bibitem [{\citenamefont {Huang}\ \emph {et~al.}(2011)\citenamefont {Huang},
  \citenamefont {Liu}, \citenamefont {Peng}, \citenamefont {Li}, \citenamefont
  {Li}, \citenamefont {Li},\ and\ \citenamefont
  {Guo}}]{Huang2011Multi-partite}%
  \BibitemOpen
  \bibfield  {author} {\bibinfo {author} {\bibfnamefont {Y.-F.}\ \bibnamefont
  {Huang}}, \bibinfo {author} {\bibfnamefont {B.-H.}\ \bibnamefont {Liu}},
  \bibinfo {author} {\bibfnamefont {L.}~\bibnamefont {Peng}}, \bibinfo {author}
  {\bibfnamefont {Y.-H.}\ \bibnamefont {Li}}, \bibinfo {author} {\bibfnamefont
  {L.}~\bibnamefont {Li}}, \bibinfo {author} {\bibfnamefont {C.-F.}\
  \bibnamefont {Li}}, \ and\ \bibinfo {author} {\bibfnamefont {G.-C.}\
  \bibnamefont {Guo}},\ }\href {\doibase 10.1038/ncomms1556} {\bibfield
  {journal} {\bibinfo  {journal} {Nat. Commun.}\ }\textbf {\bibinfo {volume}
  {2}},\ \bibinfo {pages} {546} (\bibinfo {year} {2011})}\BibitemShut {NoStop}%
\bibitem [{\citenamefont {Leibfried}\ \emph {et~al.}(2005)\citenamefont
  {Leibfried}, \citenamefont {Knill}, \citenamefont {Seidelin}, \citenamefont
  {Britton}, \citenamefont {Blakestad}, \citenamefont {Chiaverini},
  \citenamefont {Hume}, \citenamefont {Itano}, \citenamefont {Jost},
  \citenamefont {Langer}, \citenamefont {Ozeri}, \citenamefont {Reichle},\ and\
  \citenamefont {Wineland}}]{Leibfried2005Creation}%
  \BibitemOpen
  \bibfield  {author} {\bibinfo {author} {\bibfnamefont {D.}~\bibnamefont
  {Leibfried}}, \bibinfo {author} {\bibfnamefont {E.}~\bibnamefont {Knill}},
  \bibinfo {author} {\bibfnamefont {S.}~\bibnamefont {Seidelin}}, \bibinfo
  {author} {\bibfnamefont {J.}~\bibnamefont {Britton}}, \bibinfo {author}
  {\bibfnamefont {R.~B.}\ \bibnamefont {Blakestad}}, \bibinfo {author}
  {\bibfnamefont {J.}~\bibnamefont {Chiaverini}}, \bibinfo {author}
  {\bibfnamefont {D.~B.}\ \bibnamefont {Hume}}, \bibinfo {author}
  {\bibfnamefont {W.~M.}\ \bibnamefont {Itano}}, \bibinfo {author}
  {\bibfnamefont {J.~D.}\ \bibnamefont {Jost}}, \bibinfo {author}
  {\bibfnamefont {C.}~\bibnamefont {Langer}}, \bibinfo {author} {\bibfnamefont
  {R.}~\bibnamefont {Ozeri}}, \bibinfo {author} {\bibfnamefont
  {R.}~\bibnamefont {Reichle}}, \ and\ \bibinfo {author} {\bibfnamefont
  {D.~J.}\ \bibnamefont {Wineland}},\ }\href {\doibase 10.1038/nature04251}
  {\bibfield  {journal} {\bibinfo  {journal} {Nature}\ }\textbf {\bibinfo
  {volume} {438}},\ \bibinfo {pages} {639} (\bibinfo {year}
  {2005})}\BibitemShut {NoStop}%
\bibitem [{\citenamefont {T\'oth}\ \emph {et~al.}(2010)\citenamefont {T\'oth},
  \citenamefont {Wieczorek}, \citenamefont {Gross}, \citenamefont {Krischek},
  \citenamefont {Schwemmer},\ and\ \citenamefont
  {Weinfurter}}]{Toth2010Permutationally}%
  \BibitemOpen
  \bibfield  {author} {\bibinfo {author} {\bibfnamefont {G.}~\bibnamefont
  {T\'oth}}, \bibinfo {author} {\bibfnamefont {W.}~\bibnamefont {Wieczorek}},
  \bibinfo {author} {\bibfnamefont {D.}~\bibnamefont {Gross}}, \bibinfo
  {author} {\bibfnamefont {R.}~\bibnamefont {Krischek}}, \bibinfo {author}
  {\bibfnamefont {C.}~\bibnamefont {Schwemmer}}, \ and\ \bibinfo {author}
  {\bibfnamefont {H.}~\bibnamefont {Weinfurter}},\ }\href {\doibase
  10.1103/PhysRevLett.105.250403} {\bibfield  {journal} {\bibinfo  {journal}
  {Phys. Rev. Lett.}\ }\textbf {\bibinfo {volume} {105}},\ \bibinfo {pages}
  {250403} (\bibinfo {year} {2010})}\BibitemShut {NoStop}%
\bibitem [{\citenamefont {T\'oth}\ and\ \citenamefont
  {Apellaniz}(2014)}]{Toth2014Quantum}%
  \BibitemOpen
  \bibfield  {author} {\bibinfo {author} {\bibfnamefont {G.}~\bibnamefont
  {T\'oth}}\ and\ \bibinfo {author} {\bibfnamefont {I.}~\bibnamefont
  {Apellaniz}},\ }\href {http://stacks.iop.org/1751-8121/47/i=42/a=424006}
  {\bibfield  {journal} {\bibinfo  {journal} {J. Phys. A: Math. Theor.}\
  }\textbf {\bibinfo {volume} {47}},\ \bibinfo {pages} {424006} (\bibinfo
  {year} {2014})}\BibitemShut {NoStop}%
\bibitem [{\citenamefont {Escher}\ \emph {et~al.}(2011)\citenamefont {Escher},
  \citenamefont {de~Matos~Filho},\ and\ \citenamefont
  {Davidovich}}]{Escher2011General}%
  \BibitemOpen
  \bibfield  {author} {\bibinfo {author} {\bibfnamefont {B.}~\bibnamefont
  {Escher}}, \bibinfo {author} {\bibfnamefont {R.}~\bibnamefont
  {de~Matos~Filho}}, \ and\ \bibinfo {author} {\bibfnamefont {L.}~\bibnamefont
  {Davidovich}},\ }\href {\doibase 10.1038/nphys1958} {\bibfield  {journal}
  {\bibinfo  {journal} {Nat. Phys.}\ }\textbf {\bibinfo {volume} {7}},\
  \bibinfo {pages} {406} (\bibinfo {year} {2011})}\BibitemShut {NoStop}%
\bibitem [{\citenamefont {Demkowicz-Dobrza{\'n}ski}\ \emph
  {et~al.}(2012)\citenamefont {Demkowicz-Dobrza{\'n}ski}, \citenamefont
  {Ko{\l}ody{\'n}ski},\ and\ \citenamefont
  {Gu{\c{t}}{\u{a}}}}]{Demkowicz-Dobrzanski2012The}%
  \BibitemOpen
  \bibfield  {author} {\bibinfo {author} {\bibfnamefont {R.}~\bibnamefont
  {Demkowicz-Dobrza{\'n}ski}}, \bibinfo {author} {\bibfnamefont
  {J.}~\bibnamefont {Ko{\l}ody{\'n}ski}}, \ and\ \bibinfo {author}
  {\bibfnamefont {M.}~\bibnamefont {Gu{\c{t}}{\u{a}}}},\ }\href {\doibase
  10.1038/ncomms2067} {\bibfield  {journal} {\bibinfo  {journal} {Nat.
  Commun.}\ }\textbf {\bibinfo {volume} {3}},\ \bibinfo {pages} {1063}
  (\bibinfo {year} {2012})}\BibitemShut {NoStop}%
\bibitem [{Note4()}]{Note4}%
  \BibitemOpen
  \bibinfo {note} {This is also relevant for Ref.~\cite
  {Augusiak2016Asymptotic}, where $\protect \mathcal {F}_{\protect \rm
  Q}=O(N^2)$ is reached with weakly entangled states.}\BibitemShut {Stop}%
\bibitem [{\citenamefont {{T{\'o}th}}(2008)}]{Toth2008QUBIT4MATLAB}%
  \BibitemOpen
  \bibfield  {author} {\bibinfo {author} {\bibfnamefont {G.}~\bibnamefont
  {{T{\'o}th}}},\ }\href {\doibase 10.1016/j.cpc.2008.03.007} {\bibfield
  {journal} {\bibinfo  {journal} {Comput. Phys. Commun.}\ }\textbf {\bibinfo
  {volume} {179}},\ \bibinfo {pages} {430} (\bibinfo {year}
  {2008})}\BibitemShut {NoStop}%
\bibitem [{Note5()}]{Note5}%
  \BibitemOpen
  \bibinfo {note} {For MATLAB R2015a, see {\protect \tt
  http://www.mathworks.com}.}\BibitemShut {Stop}%
\end{thebibliography}%

\end{document}